\documentclass[aps,prd,twocolumn,amsmath,showpacs,amssymb,superscriptaddress,nofootinbib,eqsecnum]{revtex4}
\newlength{\picwidth}
 
\makeatother
\usepackage{graphicx}
\usepackage{dcolumn}
\usepackage{bm}
\usepackage{amssymb,amsmath}
\usepackage{latexsym}
\usepackage{color}
\usepackage[normalem]{ulem} 
\usepackage{cancel}
\allowdisplaybreaks

\newcommand{\be}{\begin{equation}}
\newcommand{\ee}{\end{equation}}
\newcommand{\bes}{\begin{subequations}}
\newcommand{\ees}{\end{subequations}}
\newcommand{\ba}{\begin{eqnarray}}
\newcommand{\ea}{\end{eqnarray}}
\newcommand{\bma}{\begin{pmatrix}}
\newcommand{\ema}{\end{pmatrix}}
\newcommand{\ern}{\mathcal{E}}

\newcommand{\stell}{\affiliation{ Physics Department, Stellenbosch University,   Bag X1 Matieland, 7602, South Africa}}
\newcommand{\nithep}{\affiliation{National Institute for Theoretical Physics (NITheP), Western Cape, South Africa}}
\newcommand{\Caltech}{\affiliation{Theoretical Astrophysics 350-17, California Institute of Technology, Pasadena, California 91125, USA}}
\newcommand{\Maryland}{\affiliation{Maryland Center for Fundamental Physics \& Joint Space-Science Institute, Department of Physics, University of Maryland, College Park, MD 20742, USA}}

\begin{document}

\title{Avenues for Analytic exploration in Axisymmetric Spacetimes.\\
Foundations and the Triad Formalism. }

\author{Jeandrew Brink} \nithep \stell
\author{Aaron Zimmerman} \nithep \Caltech 
\author{Tanja Hinderer} \nithep \Maryland 
\date{\today}


\begin{abstract}
Axially symmetric spacetimes are the only vacuum models for isolated systems with continuous symmetries that also include dynamics. For such systems, we review the reduction of the vacuum Einstein field equations to their most concise form by dimensionally reducing to the three-dimensional space of orbits of the Killing vector, followed by a conformal rescaling. The resulting field equations can be written as a problem in three-dimensional gravity with a complex scalar field as  source. This scalar field, the Ernst potential is constructed from the norm and twist of the spacelike Killing field. In the case where the axial Killing vector is twist-free, we discuss the properties of the axis and simplify the field equations using a triad formalism. We study two physically motivated triad choices that further reduce the complexity of the equations and exhibit their hierarchical structure. The first choice is adapted to a harmonic coordinate that asymptotes to a cylindrical radius and leads to a 
simplification of the three-dimensional Ricci tensor and the boundary conditions on the axis. We illustrate its properties by explicitly solving the field equations in the case of static axisymmetric spacetimes. The other choice of triad is based on geodesic null coordinates adapted to null infinity as in the Bondi formalism. We then explore the solution space of the twist-free axisymmetric vacuum field equations, identifying the known (unphysical) solutions together with the assumptions made in each case. This singles out the necessary conditions for obtaining physical solutions to the equations. 
\end{abstract}
\maketitle

\section{Introduction}
\label{sec:Intro}

Numerical relativity has revolutionized our understanding of General Relativity (GR) in the last decade, allowing us to study situations of high curvature and strongly nonlinear dynamics (see \cite{Centrella2010} for a comprehensive review). In particular, numerical relativity has allowed for the solution of the two-body problem in GR, giving a description of the interaction and merger of compact objects. Successful numerical simulations of merging black holes have shown that these events can be well described by Post-Newtonian theory up until the black holes are quite near merger,  and after merger black hole perturbation theory  accurately describes the ring-down. Where perturbation theory fails, a simple transition between the regimes of the ``chirp'' waveform associated with Post-Newtonian theory and the exponential decay to a stationary black hole is observed. A primary focus of current research is to combine these computationally expensive simulations with analytical approximations to create full 
inspiral-merger-ringdown gravitational waveforms  \cite{Ohme2012, ninja12, phenom2008, phenom2007, caleob09, eobeurope, eobimr}. Such waveforms will serve as templates for the matched-filtering based signal detection methods that will be used in ground-based gravitational-wave detectors coming into operation within the next few years~\cite{Abbott:2007, Acernese:2008, Grote:2008zz, Kuroda:2010}.

Despite the success of perturbation and numerical methods in modeling binary merger waveforms, a detailed understanding of the nonlinear regime of a binary merger remains an open problem. It is in this stage of the merger that the black hole binary emits most of its radiated energy (see \cite{Barausse2012} and the references therein) and experiences a possibly strong kick due to beamed emission of radiation \cite{Bekenstein73,Cooperstock77, Fitchett83,Lousto2012}. As such, deeper analytic understanding of nonlinear dynamics in GR, including better insights into the two-body problem, gravitational wave generation, and black hole formation, remains a primary research goal.

The purpose of this paper is to review and expand on the analytic techniques involved in the study of the field equations in axisymmetry.  Along the way we will collect many known and useful results, placing them into a unified context and notation. We intend this comprehensive overview of the state of knowledge in the field to serve as a launching point for future analytic investigations and searches for physically relevant, exact dynamical solutions in an era where a wealth of numerical data from simulations is available to guide our intuition.

We will largely restrict the  discussion to the simplest case where there is no rotation about the axis of symmetry (so that the Killing vector of the symmetry is ``twist-free''). While specializing to such a great degree does limit the 
scope of our discussion,  at least two interesting scenarios are still included in the spacetimes under consideration. The first is the case of a head-on merger of two non-rotating black holes, the simplest instance of the two body problem in GR. The second is the critical collapse of axially symmetric gravitational waves, which gives insights into the formation of black holes (for a review see \cite{lrr-2007-5}). 

Our approach to exploring the Einstein field equations in this context closely follows the methods developed by Hoenselaers, Geroch, and Xanthopolous~\cite{Xanth1, Geroch1971,Geroch1972, Hoenselaers1977,Hoenselaers1978a,Hoenselaers1978b}. The basic idea is to reduce the number of equations to a minimum by applying a dimensional reduction and conformal rescaling to the axisymmetric field equations. The resulting equations are then expressed on a null basis, in the manner of the Newman-Penrose (NP) formalism~\cite{NewmanPenrose1962} but in only three dimensions. This triad formalism  imposes an additional structure on the equations to be solved, which can lead to valuable physical insights as in the NP formalism. As will be illustrated in the text, the resulting system of equations is simple enough to allow us to keep track of the assumptions made in trying to obtain a solution and to analyze the properties of a given solution. This approach may have the potential to make dynamical spacetime problems 
analytically tractable and to provide a consistent framework for systematically characterizing the results of axisymmetric numerical simulations. The formulation given here also has a close connection to that used to find solutions to the well-studied stationary axisymmetric vacuum (SAV) equations.

To place our work in a broader context and to motivate the approach to the field equations advocated here,  we now briefly review the development both of the field of exact solutions as well as aspects of the subsequent  development of numerical relativity. Symmetry has often played a primary role in arriving at a solution to the field equations (Ref.~\cite{Stephani2003} contains a comprehensive review). Famous solutions such as the Schwarzschild black hole, de Sitter, Anti de Sitter, and Friedmann-Robertson-Walker cosmological solutions all possess large numbers of symmetries. Relaxing the degree of symmetries present, but still imposing sufficient symmetry to make headway in solving the field equations, leads to the study of stationary, axisymmetric vacuum (SAV) spacetimes (equivalently, spacetimes with two commuting Killing vectors), which has been completely solved~\cite{Ernst1968, Harrison1968, KramerNeugebauer1968, Geroch1971, Geroch1972, Xanth1, ErisB, Neu, HarrisonWET, HKXI, HKXII, HKXIII, HKXIV}. It 
was shown that the task of solving the SAV equations can be reduced to seeking a solutions of Ernst's equation~\cite{Ernst1968, ErnstEQ} on a flat  manifold. Various techniques to generate new solutions from known ones were developed in~\cite{Harrison1968, KramerNeugebauer1968, Geroch1971, Geroch1972}, based on examining the integral extension (prolongation structure) of the SAV field equations. Over the ensuing decade a variety of additional techniques were explored, including the use of harmonic maps~\cite{Xanth1,ErisB}, B\"{a}ckland transformations~\cite{Neu,HarrisonWET}, soliton and inverse scattering techniques~\cite{GravitationalSolitons}, and the use of generating functions to exponentiate the infinitesimal Hoenselaers-Kinnersley-Xanthopoulos (HKX) transformations~\cite{HKXI,HKXII,HKXIII,HKXIV}, to name a few. These techniques are all interrelated~\cite{CosRelationships}, and each has in turn taught us about the structure and properties of the SAV field equations. They allow, for example, the 
generation of an SAV spacetime with any desired asymptotic mass- and current-multipole moments~\cite{Fodor1989}. Unfortunately, by their nature, SAV solutions cannot include gravitational radiation and tell us little about the dynamics of spacetime.

Building on the progress made in studying in spacetimes with two Killing vectors, in the early and mid-1970's triad methods were developed for spacetimes with a single symmetry, and applied to stationary spacetimes~\cite{Perjes1970} and to dynamical, axisymmetric spacetimes~\cite{Hoenselaers1977,Hoenselaers1978a,Hoenselaers1978b}. At this time, however, the availability of increasingly powerful computers offered a promising new approach to obtaining solutions of the Einstein field equations for fully generic spacetimes by numerical means. In the relativity community at large, the major focus of research on solving the field equations shifted from systematically exploring the analytic structure to attempting their solution numerically. However, the numerical integration of the field equations proved to be unexpectedly difficult, especially in the axisymmetric case. 

The advent of  strongly hyperbolic and stable formulations of the field equations (e.g. the commonly used BSSN~\cite{Shibata1995,Baumgarte1998} and generalized harmonic~\cite{Pretorius2005a} formulations)  made the long term simulations of binary black hole simulations an exciting reality. With the steady progress since the breakthrough by Pretorius~\cite{Pretorius2005prl}, the merger of compact objects has become routine ~\cite{ninja12,2012CQGra..29l4004P}, although still computationally limited in duration and mass ratio. The insights afforded by these successes can now serve to guide research efforts aimed at obtaining an analytical understanding of dynamical solutions to the field equations. The relative simplicity of the gravitational waveforms and other observables generated during the highly nonlinear phase of a binary coalescence indicate that even this phase of merger could potentially be amenable to analytic techniques. A renewed interest in analytic investigations of axisymmetric spacetimes \cite{
Dain2011} has already led to interesting results such as the discovery and use of geometric inequalities ~\cite{2012CQGra..29g3001D, 2011PhRvL.107e1101D, 2011CQGra..28j5014A, 2008CQGra..25n5021D, 2008IJMPD..17..519D, 2006PhRvL..96j1101D, 2012CQGra..29p5008G, 2011CQGra..28x5017C}, studies of the radiation in a head-on collision \cite{2002CQGra..19..811D, 1999PhRvD..59h4018M, 1996PhRvD..53.1745M}, models for understanding gravitational recoil \cite{2010PhRvL.104v1101R,2012PhRvD..85h4030J} and geometrical insights on gravitational radiation  \cite{1996CQGra..13.1155D,1969PhRv..187.1784N, 1999CQGra..16..611C}.

Initially, symmetries  played an important role in the development of numerical relativity because of the great reduction in computational cost in axisymmetry compared to a fully 4D simulation. Some of the first successful work in numerical relativity was done in axisymmetry~\cite{smarrcadez, bernst, anninosheadon, numbrillw}, following the initial attempt of \cite{Hahn1964304}. 
Coordinate singularities at the axis of symmetry \cite{cartesian, interpolate, conformal, 2006PhDT.......157R, maeda}, and growing constraint violations, even when using strongly hyperbolic formulations of the field equations~\cite{rinnestewart}, presented computational challenges in fully axisymmetric codes. 
Because of these difficulties, successful codes capable of long-term evolutions of axisymmetric systems have only recently been developed~\cite{rinnehyp,sorkin,Choptuik:2003as,numbrillw}.

The continued interest in axisymmetric simulations is driven mainly by the desire to understand the critical collapse of gravitational waves~\cite{PhysRevLett.70.2980,Sorkin:2010tm} and in the higher accuracy and lower computational cost of these simulations. In addition, similar dimensional reductions as in the axisymmetric case are also used in numerical simulations of spacetimes in theories with higher dimensions; see e.g. the mergers studied in~\cite{Witek2010a, Witek2010b}.

In addition to the simplicity of the nonlinear dynamics observed in numerical simulations of the merger event, there are other tantalizing indications that the axisymmetric problem could be solvable analytically. The field equations of axisymmetric spacetimes can be written in terms of a generalized Ernst potential on a curved background, given the appropriate dimensional reduction~\cite{Xanth1}, discussed in detail in Sec.~\ref{sec:Axisym}. In this formulation some of the techniques used to find solutions for the  SAV field equations, such as harmonic maps, have a straightforward generalization to the dynamic axisymmetric case. Viewing the numerical simulations in the context of the analytic techniques employed in the past may help provide new insight into questions such as
 the nature of initial junk radiation in numerical simulations, the reasons for the robustness of certain approaches such as the puncture method \cite{PhysRevLett.96.111101}, the nature of singularity formation during a collapse process, and the possible distinction between which features of initial data contribute to the the mass of the final black hole and which components are ultimately radiated away (similar to the way in which poles and scattering data can be differentiated in the nonlinear solution of the KdV equations~\cite{solitons}).

The intent of this work is to provide a framework that could be used in future work to explore and interact with the results of axisymmetric simulations, drawing on the accumulated analytic and numerical results available for these spacetimes to date. We now briefly outline the structure and contents of the paper.

\subsection{Overview of this paper}

We will begin our discussion in full generality, explicitly carrying out in Sec.~\ref{sec:Axisym} the series of reductions that ends in the field equations for vacuum, twist-free, axisymmetric spacetimes. Here we largely follow the discussion of \cite{Geroch1971}, although in Appendix~\ref{DimensionalReduction} we present the derivation in the familiar notation of the 3+1 decomposition used in numerical relativity.  The resulting set of equations is equivalent to 3D GR coupled to a complex scalar potential  $\mathcal E$ which obeys the Ernst equation. The manifold $ \mathcal{S}$ on which these fields are defined is obtained by conformally rescaling the metric on the quotient space $\bar{\mathcal{S}}$ with the norm of the axial KV. The space $\bar{\mathcal{S}}$ should be thought of as the physical 4D manifold $\mathcal M$ modulo the orbits of the Killing vector (KV) $\xi^\mu$, or $\bar {\mathcal S} = \mathcal M / \xi^\mu$. We then specialize to the case of non-rotating spacetimes, where the field equations 
are equivalent to 3D GR with a real harmonic scalar field source that obeys the Klein-Gordon equation.  In Sec.~\ref{sec:AxisCond} we discuss general considerations regarding the existence of an axis, and note that the problem of divergences at the axis is in principal easily handled analytically.

In Sec.~\ref{sec:TriadEqns} we express the 3D field equations in terms of a  triad formulation 
that was first developed by Hoenselaers~\cite{Hoenselaers1977, Hoenselaers1978a, Hoenselaers1978b}, but seemingly not used by other authors (it should be compared to a similar triad formulation presented by Perj\'{e}s \cite{Perjes1970} and used in the case of stationary spacetimes). This formulation is derived by the projection of the dimensionally-reduced field equations onto a 3D null basis, composed of two null and one orthonormal spatial vector. The field equations and Bianchi identities are then written out in full, in terms of the 3D rotation coefficients. Considering the success of the NP equations and the valuable insights they provide, this formulation of the axisymmetric field equations merits a more thorough investigation than it appears to have received. 

In Sec.~\ref{sec:3dto4d}, we relate the 3D rotation coefficients and curvatures quantities to the familiar NP quantities on $\mathcal M$, thus providing a dictionary between NP quantities and the quantities that arise in the triad formulation. This facilitates a connection to known results and an interpretation of the physical content of the triad equations. 

In the triad formulation, we have the freedom to specialize our choice of basis vectors. In Sec.~\ref{sec:TriadChoice}, we present two useful choices of triad vectors and accompanying coordinates which serve to simplify the field equations. The first choice is, to our knowledge, new, and is analogous to the use of Lagrangian coordinates in fluid mechanics. In this triad choice the spatial triad leg is adapted to the gradient of the scalar field that encodes the dynamical degree of freedom in the twist-free axisymmetric spacetime. By virtue of the field equations, this scalar field is a harmonic coordinate which asymptotically becomes a cylindrical radius. This first coordinate choice is well-suited for analyzing the behavior of the metric functions and rotation coefficients on and near the axis. The second triad choice is inspired by the tetrad commonly used in the NP formalism, where one null vector is taken to be geodesic and  orthogonal to null hypersurfaces. This choice is useful in that it connects 
directly 
to many known solutions of the field equations, and to the dynamics at asymptotic null infinity, where the peeling property~\cite{Sachs1961, Sachs1962, Penrose1963, Penrose1965, PenroseRindler2} holds.

Our purpose in Sec.~\ref{sec:Solutions} is twofold. The first is to catalog known axisymmetric vacuum solutions, 
together with the assumptions that lead to each solution in terms of the triad formalism. This isolates the conditions required for the spacetime to represent a physically relevant solution. Secondly, we provide two example derivations of (known) spacetimes in the context of the triad equations, to illustrate typical techniques used to find solutions in this formulation. While we generally do not say much about the extensively-studied SAV spacetimes (see e.g. \cite{Stephani2003}), in Sec.~\ref{sec:Split} we discuss the equations governing SAV spacetimes in the context of our new coordinate choice from Sec.~\ref{sec:Choice2}. We conclude in Sec.~\ref{sec:Conclusions}. Additional useful results are collected in a series of appendices.

Throughout this paper, we use geometrized units with $G = c =1$. We use Einstein summation conventions, with Greek indices indicating 4D coordinate indices (in practice these can be taken as abstract tensor indices). Latin indices from the middle of the alphabet $(i,j,k,\dots)$ run over 3D coordinates (two spatial coordinates and one time coordinate), and Latin indices from the beginning of the alphabet $(a,b,c,\dots)$ run over 3D triad indices. Indices with a hat correspond to tetrad components of  a tensor in the physical manifold $\mathcal{M}$ and run over $1,2,3,4$. Indices preceded by a comma either indicate a partial derivative with respect to the coordinates, as in $f_{,i}$, or the directional derivative of a scalar quantity with respect to the members of a null basis, as in $f_{,a}$. Similarly, we use semicolons to denote covariant differentiation in the coordinate basis, as in $V_{\mu;\nu}$. Indices preceded by a bar as in $v_{a|b}$ indicate the intrinsic derivative on the triad basis. 
Symmetrization of indices is denoted by enclosing them in parenthesis, and anti-symmetrization by using square brackets. We use a spacetime signature of $(-+++)$ on the 4D spacetime, and $(-++)$ on the 3D quotient space. Note that this modern convention differs from the signature used by many authors in the literature referenced here. An asterisk denotes complex conjugation.

\section{Reduction of the axisymmetric field equations}
\label{sec:Axisym}

In this section, we review a formalism for expressing the full four dimensional Einstein field equations in a simpler three dimensional form when there is a single continuous symmetry present in the spacetime. We then specialize the resulting equations to the vacuum case, and then to spacetimes that admit a twist-free Killing vector.

The formalism for the dimensional reduction was presented by Geroch \cite{Geroch1971} for a single symmetry, and extended by him to the case of two commuting symmetries \cite{Geroch1972} in order to study SAV spacetimes. This reduction has been extensively used, especially in the investigation of stationary spacetimes \cite{Stephani2003}. We closely follow Geroch's derivation and notation in what follows. We also compare the dimensional reduction to the familiar $3+1$ decomposition used in numerical relativity, for which \cite{BaumgarteShapiro2010,Gourgoulhon2005} provide excellent references. Finally, Dain's review of axisymmetric spacetimes \cite{Dain2011} complements the discussion provided here and throughout this paper.

The reduction in complexity when one studies the field equations for a 3D Lorentzian metric as opposed to a 4D metric becomes immediately apparent by counting the number of independent components of the  Weyl tensor, given by $N(N+1)(N+2)(N-3)/12$ in $N$ dimensions. That is, zero independent components in 3D, and ten in 4D. 

In the reduction to axisymmetric, vacuum spacetimes discussed in greater depth in Sec~\ref{sec:Reduction}, all of the gravitational field's dynamical degrees of freedom enter as two  scalar functions whose gradients serve as sources for the 3D Ricci curvature. In the twist-free case, one of these scalars vanishes. The fact that the gravitational field is determined by a single remaining scalar demonstrates the tremendous simplification over the full 4D case with no symmetries present.

The reduction proceeds in three steps. The first step is to derive the equations on the three manifold. Presented in Sec.~\ref{sec:Reduction}, this process is similar to the 3+1 spacetime split familiar to numerical relativists. As a second step, we specialize to vacuum spacetimes. The last step of the reduction is a conformal rescaling, discussed in Sec.~\ref{sec:ErnstPot}, which simplifies the 3D field equations further and makes apparent the existence of a generalized Ernst potential.

\subsection{The space of orbits and the general reduction of the field equations}
\label{sec:Reduction}

We begin by considering  a 4D manifold $\mathcal M$ that admits a metric $g_{\mu \nu}$ and a Killing vector (KV) field $\xi^\mu$. Throughout this paper, we will consider $\xi^\mu$ to be spacelike; however, the same formalism is easily extended to the case of a timelike symmetry \cite{Geroch1971,Stephani2003}. The KV field represents a continuous symmetry, and it defines a set of integral curves called the orbits of $\xi^\mu$. Motion along these orbits leaves the spacetime invariant and preserves the metric. This means that tensor fields on $\mathcal M$ have vanishing Lie derivative along $\xi^\mu$. For the case of the metric tensor $\mathcal L_{\xi} g_{\mu \nu} = 0$ leads to the Killing equation,
\begin{align}
\label{killingEQ}
\nabla_{(\mu} \xi_{\nu)}= 0 \, .
\end{align}
Intuitively, we see that one of the dimensions of $\mathcal M$ is redundant, and so we would like to reduce the study of this spacetime to the study of some 3D space. 

\begin{figure}[t]
 \includegraphics[width=\columnwidth ] {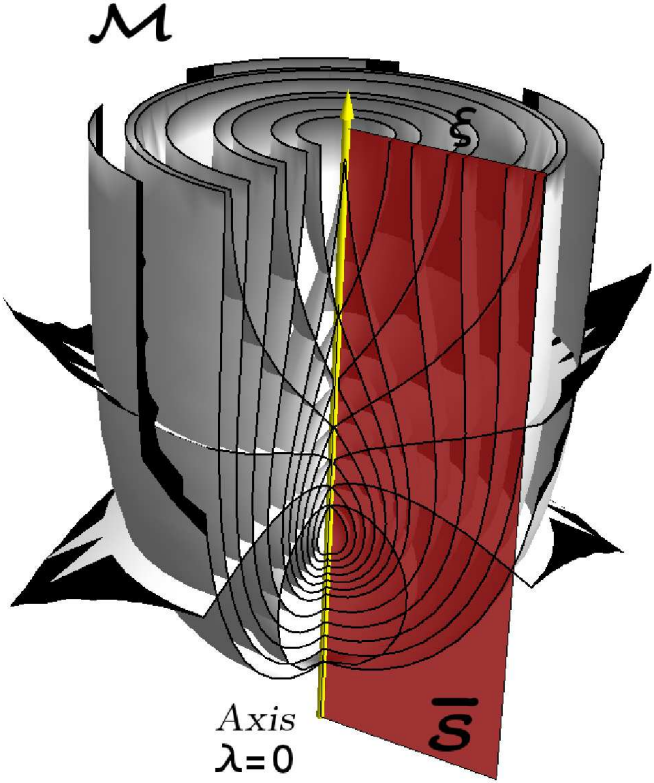}
\caption{Schematic illustration of the decomposition of a twist-free axisymmetric spacetime with closed orbits.  Since $\omega^\mu=0$, $\bar{\mathcal S}$, the quotient space of $\mathcal M$ that contains all orbits of $\xi^\mu$, is also a subspace of $\mathcal M$.  The fact that the orbits are closed implies that a set of fixed points, namely the axis, must exist if the spacetime is asymptotically flat.}
\label{MANI3}
\end{figure}

Naively, one would think of considering dynamics in $\mathcal M$ only on surfaces to which $\xi^\mu$ is orthogonal.
In practice however, $\xi^\mu$ is only orthogonal  to such a foliation of submanifolds of $\mathcal M$ if its twist $\omega^\mu$, given by
\begin{align}
 \label{twist}
\omega_\mu & = \epsilon_{\mu \nu \rho \sigma} \xi^\nu \nabla^\rho \xi^\sigma \,,
\end{align}
 vanishes. When $\omega_{\mu} =0$, the KV $\xi^\mu$ points in the same direction as the gradient of some scalar function $\phi$ on $\mathcal M$, but this is not true  in general~\cite{Wald1984}. Instead of considering some hypersurface in $\mathcal M$, we consider a new space, which we call $\bar{ \mathcal S}$ following Geroch \cite{Geroch1971}. The space $\bar{ \mathcal S }$ is defined as the collection of orbits of $\xi^\mu$ in $\mathcal M$; it is a 3D space that can be shown to posses all the properties of a manifold. The space $\bar{ \mathcal S}$ can be represented as a surface in $\mathcal M$ only if $\omega^\mu = 0$. Figure~\ref{MANI3} provides an illustration of the case of a twist-free symmetry with closed orbits. 

 We denote with an over-bar tensor fields on $\bar{\mathcal S}$. These fields are orthogonal to the KV on all their indices, e.g. $\bar T^\alpha_\beta \xi^\beta = \bar T^\alpha_\beta \xi_\alpha = 0$. 
A metric $\bar h_{\mu \nu}$ on $\bar{ \mathcal S}$ can be defined by ``subtracting'' 
the exterior product of two unit vectors  pointing in the direction of the KV from the metric $g_{\mu \nu}$. The resultant metric on $\bar{\mathcal{S}}$ is
\begin{align}
\label{eq:ThreeMetric}
\bar h_{\mu \nu} & = g_{\mu \nu} - \lambda^{-1} \xi_\mu \xi_\nu \,.
\end{align}
Note that $\xi^\mu \bar h_{\mu\nu} = 0$ and the Lie derivative of $\bar h_{\mu\nu}$ along $\xi^\mu$ vanishes.
The function $\lambda$ that appears in Eq.~\eqref{eq:ThreeMetric} is the norm of the spacelike KV,
\begin{align}
\label{norm}
\xi^\mu \xi_\mu = \lambda >0 \,,
\end{align}
and will play a key role in the reduction that follows.

By raising an index on $\bar h_{\mu \nu}$ using $g^{\mu \nu}$, we can define a projection operator $\bar h^\alpha_\nu$ which projects 4D fields onto $ \bar{\mathcal S}$. Arbitrary tensor fields can be projected into $\bar{\mathcal S}$ by contracting all of their indices onto the projector,
\begin{align}
\bar V^\alpha &= \bar h^\alpha_\mu V^\mu \,, \qquad {\rm and} \qquad \bar T_{\alpha \beta} = \bar h_\alpha^\mu \bar h_\beta^\nu T_{\mu \nu}\,,
\end{align}
and similarly for tensors of arbitrary rank. We also define the operator $\bar D_\alpha$ by contracting the usual 4D covariant derivative of a tensor field with the projector on all its indices, 
\begin{align}
\label{DERIVOP}
\bar D_\alpha \bar T_{\beta \gamma} & = \bar h_\alpha^\mu\bar h_\beta^\nu \bar h_\gamma^\rho (\nabla_\mu \bar T_{\nu \rho}) \,.
\end{align}
It can be shown that the operator $\bar D_\alpha$ obeys all the usual axioms associated with the unique covariant derivative operator on a manifold with metric $\bar h_{\mu \nu}$~\cite{Geroch1971}. 

Given the metric $\bar h_{\mu\nu}$ on $\bar{\mathcal S}$ and a compatible covariant derivative, we can compute the Riemann tensor on $\bar{\mathcal S}$, and relate it to the 4D Riemann tensor and the KV $\xi^\mu$. In doing so, the 4D field equations will be expressed entirely in terms of quantities on $\bar{\mathcal S}$. This projection of the 4D field equations is achieved by writing out Gauss-Codazzi equations generalized to the case of a timelike quotient space. This calculation, although computationally intensive, is only a slight modification of the standard techniques of the $3+1$ split often used in numerical relativity and is detailed in Appendix~\ref{DimensionalReduction}. Here, we summarize the key results that will be used later in the text.

The contracted Gauss equation expresses the 3D Ricci curvature $\bar R_{\alpha \beta}$ on $\bar{\mathcal{S}}$ in terms of the Ricci tensor $R_{\mu\nu}$ on the manifold $\mathcal{\mathcal M}$, derivatives of the norm $\lambda$ of the KV  and its twist $\omega^\mu$  as
\begin{align}
\label{RICCI3}
\bar R_{\alpha \beta}  =&  \bar h_\alpha^\mu \bar h_\beta^\nu R_{\mu \nu} + \frac{1}{2\lambda}\bar D_\alpha \bar D_\beta \lambda - \frac{1}{4 \lambda^2} \bar D_\alpha \lambda \bar D_\beta \lambda \notag
\\ &
 -\frac{1}{2 \lambda^2}\left( \bar h_{\alpha \beta} \omega_\gamma \omega^\gamma - \omega_\alpha \omega_\beta \right) \,.
\end{align} 
Since $\bar{\mathcal S}$ is a 3D  manifold all the curvature information on $\bar{\mathcal S}$ is contained in the Ricci tensor $\bar R_{\alpha \beta}$ associated with $\bar h_{\mu \nu}$, with the remaining geometric content of $\mathcal M$ given by the magnitude $\lambda$ and twist $\omega^\mu$ of $\xi^\mu$. Note that Eq.~\eqref{RICCI3} has the same form as the Einstein field equations on the three manifold ${\mathcal S}$ with additional source terms on the right hand side;  in the case where there are matter fields, we would re-express $R_{\mu \nu}$ in terms of the stress energy tensor $T_{\mu \nu}$. We are primarily interested in the vacuum field equations, in which case $R_{\mu \nu} =0$ and the geometry on the three manifold is entirely sourced by $\lambda$ and $\omega^\mu$. As such, we need equations governing the evolution of $\lambda$ and $\omega^\mu$ in order to complete our reduction of the field equations.

This second set of equations is analogous to the Codazzi equations~\cite{BaumgarteShapiro2010}, since they are derived by applying the Ricci identity to the unit vector tangent to the KV. They are detailed in Appendix~\ref{GaussCODAZZIK}. The resulting equation governing $\lambda$ is
\begin{align}
\label{laplambda}
\bar D^2 \lambda = & \frac{1}{2\lambda} \bar D_\alpha \lambda \bar D^\alpha \lambda - \frac{1}{\lambda} \omega_\mu \omega^\mu - 2 R_{\mu \nu} \xi^{\mu} \xi^{\nu} \,,
\end{align}
 where the 3D wave operator is defined using $\bar D^2 \equiv \bar D^\alpha \bar D_\alpha$. The twist $\omega^\mu$ obeys the equations
\begin{align}
\label{domega}
\bar D^\alpha \omega_\alpha =& \frac{3}{2\lambda} \omega_\alpha \bar D^\alpha \lambda \,, \\
 \label{twistpotential}
\bar D_{[\alpha} \omega_{\beta]}=&- \epsilon_{\alpha \beta \rho \sigma}  \xi^\rho R^{\sigma}_{\tau}\xi^{\tau} .
\end{align} 
Together, Eqs.~\eqref{RICCI3}--\eqref{twistpotential} can be solved on $\bar {\mathcal S}$ for $\bar h_{\alpha \beta}, \, \lambda$ and $\omega_\mu$. We can then find an expression for the KV $\xi^\mu$ using the identity, derived in Appendix~\ref{GaussCODAZZIK},
\begin{align}
\nabla_\mu \xi_\nu & = \frac{1}{2\lambda} \epsilon_{\mu \nu \rho \sigma} \xi^\rho \omega^\sigma - \frac{1}{\lambda}\xi_{[\mu} \nabla_{\nu]} \lambda
\end{align}
together with the fact that $\xi^\mu \bar h_{\mu \nu} = 0$. With the KV and $\bar h_{\mu \nu}$, we can finally reconstruct the full 4D metric $g_{\mu \nu}$ on $\mathcal M$, completing the solution of the field equations.

The field equations on $\bar{\mathcal S}$ are greatly simplified compared to the full Einstein field equations, but they are still formidable. As such, we will make a series of further specializations with the aim of rendering them tractable. In the past, the assumption of a second, timelike symmetry has resulted in the SAV equations and their solution. We will briefly discuss the SAV equations in Sec.~\ref{sec:Split}, in the context of a convenient coordinate system we introduce in Sec.~\ref{sec:Choice2}. Since our purpose is to pursue new solutions, outside of Sec.~\ref{sec:Split} we will not assume any further symmetries. Instead, we give the reductions of the field equations in the case of vacuum, and then twist-free spacetimes in the sections that follow.

\subsection{Coordinates adapted to the symmetry}
\label{sec:Coordinates}

In this section we detail the consequences of using a coordinate system adapted to the Killing symmetry. For a spacetime admitting a KV there exists coordinates $x^\mu = (x^i, \phi)$ on $\mathcal M$ such that $\xi^\mu = \delta^\mu_\phi$, where $\phi$ is a coordinate that does not appear in the metric, $\mathcal L_\xi g_{\mu \nu} = \partial g_{\mu \nu} /\partial \phi = 0$~\cite{MTW,Stephani2003}. To find the form of the metric $g_{\mu\nu}$ in coordinates adapted to an axial KV, we first note that 
\begin{align}
 g_{\phi \phi}=g_{\phi \mu}\xi^\mu
= g_{\mu \nu} \xi^\mu \xi^\nu = \lambda,
\end{align}
which also implies that $\xi_\phi=\lambda$. We denote the remaining covariant components of $\xi_\mu$ by $B_i$, so that $\xi_\mu = (B_i, \lambda)$. Since fully projected quantities on $\bar{\mathcal S}$ are orthogonal to $\xi^\mu$, e.g. $\bar V_\alpha \xi^\alpha = \bar V_\phi =0$, the $\phi$ components of projected tensors vanish, and the remaining components of $\bar h_{\mu \nu}$ are the $3\times 3$ block of components $\bar h_{ij}$. 
Using this in Eq.~\eqref{eq:ThreeMetric}, the metric $g_{\mu \nu}$ takes a simple form
\begin{align}
 \label{MetDecomp}
g_{\mu \nu} =& \left(\begin{array}{cc}
\bar h_{ij} +\lambda^{-1} B_i B_j & B_i \\
 B_j&\lambda\end{array}\right) \,.
\end{align}
Denoting the inverse of $\bar h_{ij}$ by $\bar h^{ij}$ and using it to raise and lower 3D indices, we can define $B^i = \bar h^{ij} B_j$ and $B^2 = \bar h^{ij} B_i B_j$. This allows us to write the inverse of the metric~\eqref{MetDecomp} as
\begin{align}
\label{InverseDecomp}
g^{\mu \nu} =& \left(\begin{array}{cc}
\bar h^{ij}& -\lambda^{-1} B^i \\
 -\lambda^{-1} B_j&(\lambda - B^2)^{-1}\end{array}\right) \,.
\end{align}
The determinant of $g_{\mu \nu}$ can be expressed as
\begin{align}
\det g_{\mu \nu} &=  g = \lambda \det \bar h_{ij} = \lambda \bar h \,.
\end{align}
Finally, in this basis the relationship between twist of the KV and $B_i$ can be found by defining the projected antisymmetric tensor $\bar \epsilon_{\alpha \beta \gamma} = \epsilon_{\alpha \beta \gamma \mu} \xi^\mu /\sqrt{\lambda}$. Using the definition of the twist~\eqref{twist} and projecting onto $\bar{\mathcal S}$ we have
\begin{align}
\omega_i = \sqrt{\lambda} \, \bar \epsilon_{i}{}^{jk} \bar D_j B_k \,,
\end{align}
from which we can see that if $B_i$ vanishes, so does the twist.

This decomposition of the 4D metric and its inverse in terms of the 3D metric and the KV should be compared to the analogous decompositions of the 4D metric into a spatial metric, lapse, and shift vector in a $3+1$ split, e.g. as found in \cite{BaumgarteShapiro2010}. For the remainder of this text, we will use coordinates adapted to the Killing symmetry, so that the decompositions~\eqref{MetDecomp} and~\eqref{InverseDecomp} hold. The most useful consequence of this choice is that all of the information contained in quantities projected onto $\bar{\mathcal S}$ is contained in the components on the coordinate basis $x^i$. As such, we will  write projected four dimensional indices $\alpha, \beta, \dots$ as Latin three dimensional indices such as $i, j, k, \dots$ which run over coordinates on $\bar{\mathcal S}$.

\subsection{The vacuum field equations}

We now consider the case of vacuum 4D spacetimes. This sets the 4D Ricci tensor to zero in the equations derived in Sec.~\ref{sec:Reduction}. Importantly, we see from Eq.~\eqref{twistpotential} that the curl of the twist vector vanishes. We can thus define a twist potential $\omega$ such that
\begin{align}
\omega_\mu = \nabla_\mu \omega \,,
\end{align}
From Eqs.~\eqref{laplambda},~\eqref{domega}, and~\eqref{RICCI3}, recalling that we may use 3D indices for quantities projected onto $\bar{\mathcal S}$, we have as our field equations
\begin{align}
\bar D^2 \lambda&= \frac{1}{2\lambda} \bar D^i\lambda \bar D_i\lambda-\frac{1}{\lambda} \bar D^i\omega \bar D_i\omega \,,
 \notag\\
\bar D^2 \omega&=\frac{3}{2\lambda}\bar D_i\omega \bar D^i\lambda \,, \notag\\
\bar R_{ij}&=\frac{1}{2\lambda^2}\left[ \bar D_i\omega \bar D_j\omega -\bar h_{ij}\bar D^k\omega \bar D_k\omega    \right] \notag\\
&+ \frac{1}{2\lambda} \bar D_i \bar D_j\lambda - \frac{1}{4\lambda^2} \bar D_i\lambda \bar D_j\lambda \,.
\label{Vac3DEQ}
\end{align}

\subsection{The conformally rescaled equations and the Ernst potential}
\label{sec:ErnstPot}

A further simplification to the reduced field equations~\eqref{Vac3DEQ} can be obtained by conformally rescaling the metric $\bar h_{ij}$. We define $h_{\mu\nu} $ to be
\begin{align}
h_{\mu \nu}= \lambda \bar h_{\mu\nu}= \lambda g_{\mu\nu}-\xi_\mu \xi_\nu , \label{hXanth2}
\end{align}
and investigate the conformally rescaled 3D manifold which we will call $\mathcal S$.
The vacuum field equations~\eqref{Vac3DEQ} can now be rewritten in terms of $h_{ij}$, bearing in mind that the Christoffel symbols associated with the two metrics are  related by 
\begin{align}
\Gamma^i_{kl} 
 &=\overline{\Gamma}^i_{jk}+ \frac{1}{2\lambda} \left(\delta^{i}_{j}\lambda_{,k}+\delta^i_{k}\lambda_{,j}-\bar h_{jk} \bar h^{il}   \lambda_{,l}\right).
\end{align}
The wave operator, $D^2$ associated with $h_{\mu\nu}$ is related to $\bar D^2$ as
\begin{align}
\bar D^2 f &= \lambda D^2 f -\frac{1}{2}  D_i\lambda D^i f  \,,
\end{align}
and further, $ \bar D^i f \bar D_i f = \lambda {D}^i f {D}_i f$. Substituting these identities into Eqs.~\eqref{Vac3DEQ}, the field equations can be expressed using the metric $h_{ij}$ \cite{Geroch1971,Dain2011}
\begin{align}
{D}^2 \lambda&= \frac{1}{\lambda} {D}^i\lambda {D}_i\lambda-\frac{1}{\lambda} {D}^i\omega {D}_i\omega \,,
 \notag\\
{D}^2 \omega&=\frac{2}{\lambda}\omega_i {D}^i\lambda \,,\notag\\
{R}_{ij}^{\rm 3D}&=\frac{1}{2\lambda^2}\left[ {D}_i\omega {D}_j\omega + {D}_i\lambda {D}_j\lambda \right] \,. \label{Vac3DEQrescaled}
\end{align}
The symbol ${R}_{ij}^{\rm 3D}$ denotes the Ricci curvature of the rescaled three manifold with metric $h_{ij}$.
There is additional structure in these equations which can be made more apparent by introducing the complex Ernst potential $\ern = \lambda + i \omega$~\cite{Ernst1968}. In terms of this potential, Eqs.~\eqref{Vac3DEQrescaled}  become
\begin{align}
 \label{EQERN} 
 D^2 \ern &= \frac{2 D_i\ern D^i \ern}{ (\ern+\ern^*) } \,, &
{R}_{ij}^{\rm 3D} &= \frac{2  D_{(i}\ern D_{j)} \ern^*   }{(\ern+\ern^*)^2}\,.
\end{align}
It is important to note that the Ernst potential usually discussed in the context of stationary spacetimes is based on the norm and twist of a timelike KV, rather than the spacelike KV as discussed in this section. This results in some sign differences in various definitions, c.f. the relevant chapters of \cite{Stephani2003}. The relationship between the Ernst potential defined here and the Ernst potential used in conjunction with SAV spacetimes is explained further in Sec.~\ref{sec:Split}. 

\subsection{Reduction to the case of twist-free Killing vectors}
\label{sec:NoTwist}

The axisymmetric field equations~\eqref{EQERN}, though much simplified from their full 4D form, remain intractable. For the remainder of this paper, we restrict our exploration to the situation depicted in Fig.~\ref{MANI3}, where $\xi^\mu$ is hypersurface orthogonal, so that $\omega=0$.  In doing so we eliminate the possibility of the study of rotating axisymmetric spacetimes, but we benefit from further simplifications to the field equations. A number of physically interesting dynamical spacetime solutions are twist free, these including the head-on collision of black holes and non-spinning, axisymmetric critical collapse.

 The twist-free assumption reduces the problem of finding solutions to the field equations to the study of a harmonic scalar  $\psi$ on the three manifold ${\mathcal{S}}$, where we define $\psi$ via
\begin{align}
\lambda = e^{2 \psi}\,.
\end{align}
The field equations~\eqref{EQERN} become
\begin{align}
{D}^2\psi  &= 0, &
{R}_{ij}^{\rm 3D}&=2 {D}_i\psi {D}_j\psi  ,
\label{fieldEQpsi}
\end{align}
and the Ricci scalar associated with the three metric, which we denote as $R$, is given by the contraction of~\eqref{fieldEQpsi},
\begin{align}
\label{ricci3D}
 R = 2  \psi_{;i}\psi^{;i}\, , 
\end{align}
where we used semicolons in place of $D_i$ to condense the notation for the covariant derivatives. 
The scalar $R$ is the only nonzero eigenvalue of $R_{ij}^{\rm 3D}$ and corresponds to the eigenvector $\psi_{;i}$.

Some general  properties of gravity in 3D are discussed in \cite{Giddings1984}. In particular, since the 3D gravitational field has no dynamics, due to the vanishing of the Weyl tensor, the only dynamical degree of freedom in the problem is the scalar $\psi$. This reduced number of variables drastically simplifies the calculations. In the sections that follow we present a systematic way of analyzing Eqs.~\eqref{fieldEQpsi} using a triad formalism, without immediately specializing to any given coordinate system. The fact that $\psi$ is harmonic makes it a convenient choice of coordinate on $\mathcal S$ that in addition greatly simplifies the components of the Ricci tensor $R_{ij}^{\rm 3D}$. The full implication of choosing $\psi$ as a coordinate, as well as another gauge choice adapted to geodesic null coordinates on $\mathcal S$, are discussed in the sections that follow.

\subsection{The axis}
\label{sec:AxisCond}

All of the previous results in this section hold for KVs with generic orbits. Here, we review some additional results which apply if the orbits are closed, as in the case of axisymmetry. Motion along the orbits of a KV maps the spacetime onto itself, and by the definition of the KV this map preserves the metric. This map may have fixed points, where it is simply the identity operator, and these fixed points comprise the axis of the spacetime. Much is known about the axis of an axisymmetric spacetime, see e.g.~\cite{Carter1970, MarsSenovilla1993, Carot2000}. A key result due to Carter~\cite{Carter1970} is that any vacuum spacetime with a KV that has closed orbits and is asymptotically flat admits fixed points and therefore isolated systems which possess an axial KV $\xi^\mu$ will have an axis.

This axis is 2D and timelike~\cite{Carter1970}, and will be denoted $W_2$. On the axis the magnitude of the axial KV vanishes,
\begin{align}
\xi^\mu \xi_\mu|_{W_2} =\lambda |_{W_2} = 0 \,.
\end{align}
Note that the derivative of the KV, $\xi^\mu{}_{; \nu}$ cannot vanish on the axis, or else $\xi^\mu$ would vanish everywhere (see e.g.~\cite{Wald1984} for further discussion).

When the axis is free of singularities, a condition known as elementary flatness holds in a neighborhood of the axis. 
This condition expresses the fact that in the local Lorentz frame of a small neighborhood about a point 
on $W_2$, we can make a loop around the axis, and the circumference of this loop must be equal to $2 \pi$ 
times its radius. If this is not true, then there is a conical singularity in this small neighborhood, and  traversing the circle around them results in a  deficit (or surplus) angle. One way to express elementary flatness is to find a set of coordinates in which the line element has the form $ds^2_0 = g_{\rho \rho}  d \rho^2 + \lambda d\phi^2$ near the axis, holding the third spatial coordinate fixed. Dividing the proper length around a circle by $2 \pi$ times the proper distance to the axis yields a constant $K_{\rm D}$ which, if different from unity, gives a measure of the deficit angle~\cite{Sladek2010}, 
\begin{align}
K_{\rm D} = \lim_{\lambda \to 0} \frac{\int^{2 \pi}_{0} \sqrt{\lambda}\ d \phi}{2 \pi \int_0^\rho \sqrt{g_{\rho \rho}}\ d\rho } \,.
\label{deficitangle}
\end{align}

A coordinate invariant form of this same condition that is more useful from our perspective was given by Mars and Senovilla \cite{MarsSenovilla1993},
\begin{align}
\label{ElemFlat}
 \lim_{\lambda \to 0}\frac{\lambda_{,\mu} \lambda^{,\mu} }{4 \lambda} = 1 \,.
\end{align} 
We derive this result using a specific coordinate system in Sec.~\ref{sec:Choice2}. Expressing Eq.~\eqref{ElemFlat} in terms of $\psi$ and the conformal three metric $h_{ij}$ we have
\begin{align}
 \lim_{\lambda \to 0}e^{4 \psi} h^{ij} \psi_{,i} \psi_{,j} = 1 \,, \label{psiAxis}
\end{align}
Equation~\eqref{psiAxis} provides explicit boundary conditions for quantities on $\mathcal S$ as the axis $\lambda=0$ is approached, if we wish our axis to be free of conical singularities.

\section{The twist-free field equations expressed using a  triad formalism}
\label{sec:TriadEqns}

To explore the field equations on the 3D manifold $\mathcal S$, we employ a triad formalism in which we choose a basis for the tangent bundle before further selecting coordinates on the manifold. In this section we follow Hoensaelers~\cite{Hoenselaers1977, Hoenselaers1978a, Hoenselaers1978b} in writing out the 3D field equations~\eqref{fieldEQpsi} and Bianchi identities on $\mathcal S$ in a manner similar to the Newman-Penrose (NP) equations \cite{NewmanPenrose1962,MathTheoryofBlackHoles}. This form of the equations is particularly convenient for the study of the exact solutions of the field equations, since it makes manifest what the various possible assumptions and simplifications might be for special and physically interesting cases. The procedure is to define a null (or orthonormal) triad  and write out in full the field equations expressed in this basis (we note that a similar formalism was developed by Perj\'{e}s in \cite{Perjes1970} in the context of stationary spacetimes, using a complex 
triad). Our approach largely follows the conventions for the tetrad formalism used in Chandrasekhar's text~\cite{MathTheoryofBlackHoles}, which also gives general background on the technique.

We begin by selecting a triad basis 
\begin{align}
\zeta^i_a= (l^i,n^i,c^i)  \,,
\label{Triad}
\end{align}
so that the metric expressed on this triad basis 
\begin{align}
\eta_{ab} = h_{i j} \zeta^i_a \zeta^j_b  
\label{ETAdef}
\end{align}
 contains only constant coefficients. A null triad choice that is particularly useful is one for which
\begin{align}
\label{NullNorm}
l_i l^i&=n_i n^i=l_i c^i=n_i c^i=0 \, \ \  \mbox{and} &
c_i c^i & = - l_i n^i = 1 \,, 
\end{align}
and the non-zero metric components are $\eta_{12}=\eta_{21}=-1$ and $\eta_{33}=1$. 
The orientation of the triad is fixed by the equations
\begin{align}
\epsilon_{ijk}l^j n^k = c_i,\ \
\epsilon_{ijk}c^j l^k = l_i, \ \
\epsilon_{ijk}n^j c^k = n_i. \
\end{align}
Given the normalization in Eq.~\eqref{NullNorm}, the metric on the coordinate basis is expressed in terms of the triad vectors as
\begin{align}
\label{Hmn}
h_{ij} = -l_i n_j-n_i l_j+ c_i c_j\,.
\end{align}
The fundamental variables in a triad formalism are the Ricci rotation coefficients $\gamma_{abc}$, which record how the basis vectors change as we traverse the manifold. They are defined by 
\begin{align}
\label{RotCoeff}
\gamma_{abc}= \zeta_{a j;k} \zeta_b^j \zeta_c^k \,.
\end{align}
The rotation coefficients are antisymmetric in the first two indices $\gamma_{abc} = \gamma_{[ab]c}$ (note our ordering of indices induces a sign change from Chandrasekhar's definition \cite{MathTheoryofBlackHoles}). In 3D there are nine independent real rotation coefficients, as opposed to the 24 real rotation coefficients that exist in 4D.
We adopt the following naming convention first introduced in~\cite{Hoenselaers1977},
\begin{align}
\label{ninerot}
\alpha&=\gamma_{121}=l_{i;j}n^{i}l^{j}\,,&
\beta&=\gamma_{311}= c_{i;j}l^{i}l^{j}\,,\notag\\
\gamma&=\gamma_{231}=n_{i;j}c^{i}l^{j}\,,&
\delta&=\gamma_{122}=l_{i;j}n^{i}n^{j}\,,\notag\\
\epsilon&=\gamma_{312}=c_{i;j}l^{i}n^{j}\,, &
\zeta&=\gamma_{232}=n_{i;j}c^{i}n^{j}\,,\notag\\
\eta&= \gamma_{123}=l_{i;j}n^{i}c^{j}\,,&
\theta&=\gamma_{313}=c_{i;j}l^{i}c^{j}\,,\notag\\
\iota &=\gamma_{233}=n_{i;j}c^{i}c^{j} \,. 
\end{align}
 The projection of the 3D Ricci tensor $R_{ij}$ (we drop the superscript 3D from here on) onto this basis gives us six curvature scalars, which we denote
\begin{align}
\label{phidef}
\phi_5 &= R_{11} = R_{ij} l^{i} l^{j},&
\phi_4 &= R_{12} = R_{ij} l^{i} n^{j}, \notag\\
\phi_3 &= R_{13} = R_{ij} l^{i} c^{j},&
\phi_2 &= R_{22} = R_{ij} n^{i} n^{j}\notag\\
\phi_1 &= R_{23} = R_{ij} n^{i} c^{j},&
\phi_0 &= R_{33} = R_{ij} c^{i} c^{j}.
\end{align}
The Ricci scalar is given by%
\footnote{Note that here we are using a different definition of $\phi_0$ and $\phi_4$ than \cite{Hoenselaers1977}. If we denote the scalars of~\cite{Hoenselaers1977} with a superscript $H$, the relationship between the two conventions is such that $\phi_4^H+ \phi_0^H = \phi_4$ and $2\phi_0^H = \phi_0$.}
\begin{align}
\label{RicciScalar}
R= R^i_{\ i} = \phi_0-2 \phi_4 \,.
\end{align}

The field equations describe how the rotation coefficients listed in Eqs.~\eqref{ninerot} change in a particular basis direction to ensure that Eqs.~\eqref{fieldEQpsi} are satisfied.  The change along a basis direction is a directional derivative given by
\begin{align}
V_{a_1 \cdots a_n,b} & =  (V_{i_1 \cdots i_n} \zeta_{a_1}{}^{i_1} \cdots\zeta_{a_n}{}^{i_n})_{;j} \zeta_b{}^j \,.
\label{DirectionalDer}
\end{align}
The basis-dependent directional derivative can be related to the intrinsic covariant derivative of a tensor projected onto the triad basis,
\ba
\label{IntrinsicDer}
V_{a_1 \cdots a_n|b} & =& (V_{i_1 \cdots i_n})_{;j} \zeta_{a_1}{}^{i_1} \cdots\zeta_{a_n}{}^{i_n} \zeta_b{}^j \, ,
\ea
by taking into account the manner in which the basis itself changes. Using Eqs.~\eqref{RotCoeff},~\eqref{DirectionalDer}, and~\eqref{IntrinsicDer} one can show that the relationship between the directional and intrinsic derivatives is
\begin{align}
V_{a_1 \cdots a_n|b}& = V_{a_1 \cdots a_n,b}+ \gamma^c_{\ a_1 b }V_{ c \cdots  a_n} + \cdots +\gamma^c_{ \ a_n b} V_{a_1  \cdots c}  \,.
\end{align}
Recall that triad indices are raised using the constant metric $\eta^{ab}$ defined by $\eta^{ab}\eta_{bc}=\delta^a_{ c}$, which has the same component form as $\eta_{ab}$. It is important to note that in a triad formalism, the directional derivatives of a scalar function do not commute, while intrinsic derivatives do.  The commutation relations for directional derivatives are
\be
\label{comm}
f_{,[ab]}=f_{,m}\gamma_{[a}{}^m{}_{b]} \,.
\ee
Using the relationship between intrinsic and directional derivatives, we now express the field equations on the triad basis in terms of the rotation coefficients~\cite{MathTheoryofBlackHoles}. The Ricci tensor obeys
\begin{align}
 \label{RICCIFIELD}
R_{ab} 
&= - \gamma_{a}{}^{m}{}_{m,b}-\gamma^{m}{}_{ab,m}- \gamma^{mn}{}_{n} \gamma_{mab}-\gamma_{a}{}^{mn}\gamma_{bnm}.
\end{align}
Writing out the field equations~\eqref{RICCIFIELD} in full leads to
\begin{widetext}
\ba
\label{Bondi0}
(\epsilon - \gamma)_{,3} + \iota_{,1} - \theta_{,2}
&=& \alpha  \iota -2 \beta  \zeta +\gamma ^2   +\epsilon ^2  +  \theta (\delta 
 +2   \iota) +\phi_0 \,,
\\
\label{Bondi1}
\eta_{,2} - \delta_{,3}
&= & \zeta  (\alpha -\theta)  -\delta 
  (\epsilon + \eta) +\iota(\epsilon -\eta) +\phi_1 \,,
\\
 \label{Bondi2}
\zeta_{,3}  -\iota_{,2}
& =& \iota(\iota - \delta)+  \zeta(\epsilon  + 2   \eta - \gamma) +\phi_2 \,,
\\
\label{Bondi3}
\alpha_{,3} - \eta_{,1}
&= &-\alpha ( \gamma + \eta) +\beta ( \delta -  \iota)
+\theta   (\gamma - \eta)   +\phi_3 \,,
\\
\label{Bondi4}
(\alpha+ \theta)_{\text{,2}} -\text{$\delta $}_{\text{,1}} -\text{$\epsilon
   $}_{\text{,3}}
&=& - \gamma ( \eta - \epsilon)    -\delta ( \theta +2 \alpha) -\epsilon(\epsilon+\eta)  -\theta  \iota +\phi_4 \,,
\\
\label{Bondi5}
\theta_{,1}-\beta_{,3}
&=&\theta(\theta - \alpha) +\beta( \gamma + 2 \eta - \epsilon) +\phi_5 \,,
\\
\label{BondiI1}
\iota_{,1} + \theta_{,2} - (\epsilon+\gamma)_{,3}
&=& \alpha  \iota +\gamma ^2-\delta 
   \theta -\epsilon ^2 \,,
\\
\label{BondiI2}
(\epsilon + \eta)_{,1} - \beta_{,2} - \alpha_{,3}
&=& \alpha 
   (\gamma +\eta) +\beta(  \delta +  \iota) +\theta(\epsilon  +\eta) \,,
\\
\label{BondiI3}
\zeta_{,1} + \delta_{,3} - (\gamma + \eta)_{,2}
&=& \zeta(\alpha + \theta) + \delta(\epsilon + \eta) + \iota(\gamma + \eta) \,.
\ea
In the twist-free case, the curvature scalars $\phi_i$ appearing in the above expressions are obtained from the Ricci tensor $R_{ab}$ computed by projecting Eqs.~\eqref{fieldEQpsi} onto the triad,
\begin{align}
\label{RICCITET}
R_{ab}= 2\psi_{,a}\psi_{,b} .
\end{align}
Of the above set of nine field equations there are six equations that contain Ricci curvature components and three that do not. These three equations constitute the 3D version of the eliminant relations (see Chandrasekhar~\cite{MathTheoryofBlackHoles}). As expected, there are fewer equations on $\mathcal S$ than in the 4D case (9 here versus 36 equations in 4D).

The three, 3D Bianchi identities
\begin{align}
R_a{}^b{}_{,b}-\frac{1}{2}R_{,a}+ \gamma_{bam}R^{bm}+\gamma_b{}^m{}_mR_a{}^b =0 \,
\end{align}
are written as\footnotemark
\begin{align}
\label{BIANCI1}
\frac{1}{2}\left(\phi _0\right)_{\text{,1}}+\left(\phi _5\right)_{\text{,2}}-\left(\phi_3\right)_{\text{,3}}
&=  \theta ( \phi _0  +  \phi _4)  - (\iota   +2 \delta ) \phi _5  
+(\eta + \gamma -2 \epsilon)  \phi_3  -  \beta  \phi _1 \,,\\
\label{BIANCI2}
\frac{1}{2}\left(\phi_0\right)_{\text{,2}}
+\left(\phi _2\right)_{\text{,1}}-
\left(\phi _1\right)_{\text{,3}}
&=-\iota(  \phi _0+  \phi _4)  +(2 \alpha   + \theta ) \phi _2  
+ (2 \gamma - \epsilon-\eta )\phi _1 +\zeta  \phi _3 \,, \\  
\label{BIANCI3}
\left(\frac{1}{2} \phi_0 + \phi_4 \right)_{,3} - (\phi_1)_{,1} - (\phi_3)_{,2} 
& =  -\phi _1 (2 \theta+ \alpha  )+\beta  \phi _2 
+( \epsilon - \gamma )(\phi_4+\phi _0)+ (\delta   + 2 \iota)  \phi_3  - \zeta  \phi _5 \,.
\end{align}
\end{widetext}

In terms of the rotation coefficients, the commutation relations~\eqref{comm} are 
\begin{align}
\label{COMM}
f_{,21}-f_{,12}&= \delta 
   f_{,1}+\alpha  f_{,2}+(\gamma  +\epsilon) 
   f_{,3},
\notag\\
f_{,31}
-f_{,13}&=  
   (\gamma +\eta )f_{,1}  - \beta  f_{,2}+\theta  f_{,3},
\notag\\
 f_{,23}-f_{,32}& =-\zeta f_{,1}+
   (\epsilon +\eta) f_{,2}+\iota  f_{,3} \,.
\end{align}
and must be used whenever interchanging the order of directional derivatives.
Finally,  it is useful to note that the operator $D^2f=f_{,a}{}^{|a}$ can be expressed as
\begin{align}
\label{LapOp}
D^2f&=
-2 f_{,12} +f_{,33}  - (\iota +2 \delta ) f_{,1}+\theta  f_{,2}-2
   \epsilon  f_{,3}\notag\\
&=-2f_{,21}+ f_{,33}- \iota  f_{,1}+  (2 \alpha +\theta ) f_{,2}+2 \gamma f_{,3} \,.
\end{align}
This concludes the general triad formulation using the rotation coefficients as fundamental variables. 
The equations given are valid for both twisting and twist-free spacetimes, with the only difference being the complexity of the 3D Ricci tensor. These equations can be simplified to a great degree by a judicious choice of triad. 
We explore two especially useful triad choices in Sec.~\ref{sec:Solutions}, where we also specialize to the case of twist-free spacetimes. 

\footnotetext{Note that Eqs.~\eqref{BIANCI1}-\eqref{BIANCI3} differ from those derived by Hoenselaers in \cite{Hoenselaers1977}, Eq.~(3.5c), with respect to the sign in front of his $d$ operator. In addition to this difference and the difference in notation the equations presented in this section differ from those of Hoenselaers in that our triad has a different normalization, and so some of the signs differ. Specifically, every factor of $l^i$ and derivative in the $l^i$ direction receives a sign change, which changes the signs in front of many of the rotation coefficients and some of the curvature scalars. With these considerations, the two sets of equations are identical.}

\section{Relating Physical 4d quantities to computed 3d quantities}
\label{sec:3dto4d}

In this section we provide the explicit correspondence between NP quantities on the physical 4D spacetime $\mathcal M$ and the computationally concise quantities on the conformal manifold $\mathcal S$. Knowledge of this correspondence is useful for various reasons: (i) Initial conditions for integrating the much simpler 3D field equations~\eqref{Bondi0}--\eqref{BondiI3} are most readily specified on $\mathcal M$. (ii) The boundary conditions on the axis, discussed in Sec.~\ref{sec:AxisCond} and Appendix~\ref{sec:ROTCOEFPSICOORD}, require information about the smoothness of the physical quantities on $\mathcal M$, since the 3D conformal metric $h_{ij}$ is singular on the axis. (iii) Searching for solutions to the field equations involves making choices of the triad and the gauge, and having a direct translation of the assumptions made in 3D to the implications for the physical quantities is advantageous. This relationship between specializations in 3D and 4D also identifies the conditions on the 3D 
quantities corresponding to known solutions.

To exhibit the correspondence, we first note that in the twist free case, the metric decomposition of Sec.~\ref{sec:Coordinates} simplifies to the case where $B_i =0$ and the 4D metric $g_{\mu \nu}$ can thus be expressed as
 \begin{align}
\label{MetDecomp2}
g_{\mu \nu}=
e^{-2\psi}\left(\begin{array}{cc}
h_{ij}&0\\
0&e^{4\psi}\end{array}\right) =e^{-2\psi} \tilde{g}_{\mu \nu} \,.
\end{align}
The metric $\tilde{g}_{\mu\nu}$, which is conformal to the physical metric, provides a useful intermediate step for the calculations that follow.

\subsection{Spin coefficients}
\label{sec:4dSpin}

We now define the relationship between the NP spin coefficients on $\mathcal M$ and the rotation coefficients defined in Eqs.~\eqref{ninerot}. There is some freedom in the choice of tetrad as we go between the 4D and 3D manifolds, which we fix by choosing the tetrad so that the directions of all the null basis vectors coincide, and so that the parametrization of the out-going null vectors are the same. In order to avoid confusion, all quantities on $\mathcal M$ such as spin coefficients, $\hat{\kappa}$, $\hat{\epsilon}$ and Weyl scalars, $\hat{\Psi}_i$ are given with a hat ($\hat{.}$). Quantities associated with the conformally rescaled 4-metric $\tilde{g}_{\mu\nu}$ are all indicated with a tilde ($\tilde{.}$), and 3D quantities will remain unadorned. 

The standard complex null tetrad on $\mathcal M$ is
\begin{align}
\hat{\zeta} ^\mu_{\hat a} =\left(\hat{l}^\mu,\hat{n}^\mu,\hat{m}^{\mu},\hat{m}^{*\mu} \right), 
\label{tetNP}  
\end{align}
with the non-zero metric components being $\hat{\eta}_{ln} = \hat{\eta}_{nl}=- \hat{\eta}_{mm^*}= -\hat{\eta}_{m^* m} = -1$. Now consider another tetrad constructed by augmenting the triad~\eqref{Triad} with the vector $d^{\mu} = e^{-2\psi}\delta^{\mu}_\phi$ which has the same direction as the KV $\xi^\mu$, to yield the tetrad
\begin{align}
\tilde{\zeta}_a^\mu = (l^\mu, n^\mu, c^\mu, d^\mu),
\label{tetaug}
\end{align}
where we have omitted the tilde's to emphasize that this tetrad is built from the same triad vectors that we use on $\mathcal S$ (although strictly speaking they are the lift of these vectors onto a conformal 4D space).
It can be verified directly using Eqs.~\eqref{Hmn} and~\eqref{MetDecomp2} that the conformal metric $\tilde{g}_{\mu\nu}$ can be expressed as 
\begin{align}
\label{Gmn}
\tilde{g}_{\mu\nu} =- l_\mu n_\nu- n_\mu l_\nu + c_\mu c_\nu+ d_\mu d_\nu  
\end{align}
where the covector $d_\mu$ is $d_\mu = e^{2\psi} \delta ^\phi_\mu$.

To find the relationship between the NP spin coefficients on $\mathcal M$ and the rotation coefficients on $\mathcal S$, we first calculate the rotation coefficients associated with the conformally related metric $\tilde{g}_{\mu \nu}$ on the basis in Eq.~\eqref{tetaug}. To do this expediently we introduce the quantities $\lambda_{abc}$ that are defined as~\cite{MathTheoryofBlackHoles}
\begin{align}
\lambda_{abc} &= \gamma_{bac} -\gamma_{bca} = (\zeta_{b \gamma,\beta}- \zeta_{b \beta,\gamma } )\zeta_c^\beta \zeta_a^\gamma ,
\label{lambdaabc}
\end{align}
and are antisymmetric in the first and third indices. The major advantage of working with the quantities $\lambda_{abc}$ is that they can be computed using coordinate derivatives rather than covariant derivatives. This property makes the comparison between quantities defined on different metrics given the same coordinate choice easy. Given a set of $\lambda_{abc}$'s the rotation coefficients can be constructed using the relation 
\begin{align}
\gamma_{abc}
=-\frac{1}{2}(\lambda_{abc}+\lambda_{cab}-\lambda_{bca})\,.
\label{rotrecon}
\end{align}

The 24 rotation coefficients associated with the conformal metric  $\tilde{g}_{\mu\nu}$ can be related to the nine rotation coefficients associated with $h_{ab}$ by noting that $\tilde{\lambda}_{abc} = \lambda_{abc}$ when $a,b,c$ run over $1,2,3$. The remaining 15 rotation coefficients can be subdivided into nine coefficients of the form $\tilde{\gamma}_{a4b} $, three $\tilde{\gamma}_{ab4} $ coefficients, and three $\tilde{\gamma}_{a 4 4} $ coefficients . From the definitions in Eq.~\eqref{lambdaabc} and the vector $d^\mu$, it is straightforward to verify that $\tilde{\lambda}_{ab4}=\tilde{\lambda}_{a4b}=0$, and so the 12 coefficients $\tilde{\gamma}_{a4b} $ and $\tilde{\gamma}_{ab4} $ vanish. There are then only three non-zero rotation coefficients,
\begin{align}
\tilde{\gamma}_{a 4 4} = -\tilde{\lambda}_{a4 4} = 2\psi_{,a},
\end{align} 
in addition to those in Eq.~\eqref{ninerot}.

Given the rotation coefficients associated with the augmented tetrad in Eq.~\eqref{tetaug}, the spin coefficients associated with the physical space tetrad in Eq.~\eqref{tetNP} can be obtained from a transformation of the form 
\begin{align}
\hat{\zeta} ^\mu_{\hat a} &= Q_{\hat a}^b(\psi) \tilde{\zeta}_b ^{\mu}, &
\hat{\zeta}_{\hat a\mu} &=P_{\hat a}^b (\psi) \tilde{\zeta}_{b\mu}  .
\label{tetrans}
\end{align}
Specifically, $P_{\hat a}^b= e^{-2\psi} Q_{\hat a}^b$ and the nonzero components of $Q_{\hat a}^b$ are
\begin{align}
Q_1^1&=1, &  Q_3^3 &= Q_4^3= \frac{e^\psi}{\sqrt{2}},\notag\\ 
Q_2^2&=e^{2\psi}, &  Q_3^4 &= - Q^4_4=  \frac{ ie^\psi  }{\sqrt{2}}. 
\end{align}
Note that the fact that $Q^1_1=1$ ensures that the parametrization of outgoing null 
vector $l^i$ on the three manifold coincides with the associated vector on the 4D spacetime.
The vectors on $\mathcal M$ are then given in terms of the tetrad~\eqref{tetaug} by
\begin{align}
\hat{l}^\mu&=l^\mu,& \hat{n}^\mu &= e^{2\psi} n^{\mu}, \notag\\ 
\hat{m}^\mu &= e^\psi( c^\mu + i  d^\mu)/\sqrt{2},& \hat {m}^{* \mu} &= e^\psi( c^\mu - i d^\mu)/\sqrt{2}.
\label{physicalvec}
\end{align}
By repeatedly using the definition~\eqref{lambdaabc} on the different tetrads, we find that the $\lambda_{abc}$ functions associated with the physical tetrad~\eqref{tetNP} [and thus the rotation coefficients via Eq.~\eqref{rotrecon}] are related to those on the augmented tetrad given in Eq.~\eqref{tetaug} by
\begin{align}
\label{hatrot}
\hat{\lambda}_{\hat a \hat b \hat c}
&=  Q^{c}_{\hat c}  Q^{a}_{\hat a}   P_{\hat b}^{b} \tilde{\lambda}_{abc}+   Q^{c}_{\hat c}  Q^{a}_{\hat a} 
\left[ \tilde{\eta}_{ba}(  P_{\hat b}^{b} )_{,c} -  \tilde{\eta}_{bc} ( P_{\hat b}^{b} )_{,a } \right] \,,
\end{align}
 where the constant metric $\tilde{\eta}_{ab}$ has the non-zero components, $\tilde{\eta}_{12} =\tilde{\eta}_{21} =-1$ and $\tilde{\eta}_{33}=\tilde{\eta}_{44}=1$.

Since the $P^a_{\hat b}$'s are functions only of $\psi$ all the physical rotation coefficients  reconstructed using Eq.~\eqref{rotrecon} given Eq.~\eqref{hatrot} can be written in terms of the nine rotation coefficients on the triad basis, the three directional derivatives of the scalar function $\psi$, and functions of $\psi$ itself. It can also be observed that all the physical rotation coefficients expressed on the basis in Eq.~\eqref{physicalvec} are real. The physical spin coefficients using the NP naming convention  \cite{MathTheoryofBlackHoles}, when  expressed in terms of the rotation coefficients defined on $\mathcal{S}$ are 
\begin{align}
\label{4spins}
\hat{\rho  }&= \frac{\theta }{2},&
\hat{\sigma }&= \frac{1}{2} \left(2 \psi
   _{\text{,1}}+\theta \right),\notag\\
\hat{\kappa }&= \frac{  e^{-\psi }}{\sqrt{2}} \beta ,&
\hat{\nu }&= \frac{  e^{3 \psi }}{\sqrt{2}}\zeta  , \notag\\
\hat{\tau }&= \frac{e^{\psi }}{\sqrt{2}} \left(\psi _{\text{,3}}+\epsilon \right)  ,&
\hat{\pi }&= \frac{e^{\psi }}{\sqrt{2}} \left(\gamma -\psi _{\text{,3}}\right)  ,\notag\\
\hat{\alpha }&= -\frac{e^{\psi }}{2
   \sqrt{2}} \left(2 \psi _{\text{,3}}+\eta \right)  ,&
\hat{\beta }&= -\frac{ e^{\psi }}{2 \sqrt{2}}\eta   , \notag\\
\hat{\lambda }&= \frac{1}{2} e^{2 \psi } \left(\iota -2 \psi _{\text{,2}}\right),&
\hat{\mu }&= \frac{1}{2} \iota  e^{2 \psi }, \notag\\
\hat{\epsilon }&= -\frac{1}{2} \left(2 \psi
   _{\text{,1}}+\alpha \right),&
\hat{\gamma }&= -\frac{1}{2} \delta  e^{2 \psi } .
\end{align}
The identifications in Eqs.~\eqref{4spins}  gives us the benefit of all the usual intuition regarding the spin coefficients in the 4D spacetime when computing quantities on the manifold $\mathcal{S}$. We will explore these relationships and their physical implications more fully in Sec.~\ref{sec:Solutions} when we review the exact solutions to the field equations.

\subsection{Curvature and Weyl scalars}
\label{sec:4dWeyl}
The second set of quantities that are useful for exploring the physical content of spacetime, such as gravitational radiation, are the Weyl scalars. In this section we will show that they have a particularly simple representation in terms of the 3D rotation coefficients and directional derivatives of $\psi$. 

The fact that the Weyl tensor is conformally invariant implies that on the coordinate basis $\hat C ^\alpha{}_{\beta \gamma \delta} = \tilde C^\alpha{}_{\beta \gamma \delta}$.
Lowering the index $\alpha$, expressing the tensor on the tetrad basis in Eq.~\eqref{physicalvec}, and subsequently using Eq.~\eqref{tetrans} to express it on the augmented basis in Eq.~\eqref{tetaug},  we obtain an expression for the physical Weyl tensor in terms of the Weyl tensor on the augmented basis,
\begin{align}
\hat{C}_{\hat a \hat b \hat c \hat d}
= e^{-2\psi}\tilde{C}_{abcd} Q_{\hat a}^a Q_{\hat b}^b Q_{\hat c}^c  Q_{\hat d}^d  \,.
\end{align}
The quantity $\tilde{C}_{abcd}$ is readily computed in terms of the rotation coefficients and directional derivatives of $\psi$ on $\mathcal S$ from the standard expression for the Riemann tensor~\cite{MathTheoryofBlackHoles}, which in vacuum is identical to the Weyl tensor: 
\begin{align}
\tilde{R}_{abcd} &=   \tilde{\gamma}_{a bc,d}-\tilde{\gamma}_{abd,c}+
\tilde{\eta}^{fg}\tilde{\gamma}_{baf}(\tilde{\gamma}_{cgd}-\tilde{\gamma}_{dgc})\notag\\
&+\tilde{\eta}^{fg}(\tilde{\gamma}_{fac}\tilde{\gamma}_{bgd}-\tilde{\gamma}_{fad}\tilde{\gamma}_{bgc}) \,.
 \label{RIEMANN}
\end{align}
Writing out Eqs.~\eqref{RIEMANN} in full, making use of the definitions of $\phi_i$ given in 
Eqs.~\eqref{phidef} and~\eqref{RICCITET}, and substituting in the field equations~\eqref{Bondi0}--\eqref{BondiI3} wherever necessary yields the following expressions for the  Weyl scalars on the physical manifold:
\begin{align}
\hat \Psi_0&=\hat C_{1313}= \left(\alpha  \psi _{,1}+3 (\psi _{,1})^2+\psi
   _{,11}+\beta  \psi _{,3}\right),
\notag\\
\hat \Psi_1&=\hat C_{1213}=\frac{e^{ \psi }}{\sqrt{2}} \left(\psi
   _{,1} \left(3 \psi _{,3} -\gamma \right)+\psi _{,31}+\beta  \psi
   _{,2}\right), \notag\\
\hat \Psi_2&=\hat C_{1342}=\frac{e^{2 \psi }}{2}  \left(\psi _{,1} \left(2 \psi
   _{,2}-\iota \right)+\theta  \psi _{,2}+2 (\psi _{,3})^2+\psi
   _{,33}\right) , \notag\\
\hat \Psi_3&=\hat C_{1242} =-\frac{e^{3 \psi }}{\sqrt{2}} \left(\zeta  \psi _{,1}-\psi _{,2}
   \left(3 \psi _{,3}+\epsilon \right)-\psi
   _{,32}\right),\notag\\
\hat \Psi_4&=\hat C_{2424}= -e^{4 \psi } \left(\delta  \psi _{,2}-3 (\psi
   _{,2})^2-\psi _{,22}+\zeta  \psi _{,3}\right). 
\label{p0}
\end{align}
It is important to note that the assumption of twist-free axisymmetry greatly decreases the number of independent functions to be considered: the NP spin coefficients and Weyl scalars which in general are complex are all real in the twist-free case, effectively cutting the problem of finding solutions in half. Further simplifications can be achieved with specific gauge and tetrad choices.

\section{Two triad choices}
\label{sec:TriadChoice}

In this section we discuss the implications of two physically-motivated triad choices which further simplify Eqs.~\eqref{Bondi0}--\eqref{BondiI3} and the Bianchi identities~\eqref{BIANCI1}-\eqref{BIANCI3}. The first choice is to use $\psi$ as a coordinate and to associate the triad direction $c_a$ with its gradient. This choice greatly simplifies the Ricci tensor on the three manifold and is suited to applying the boundary condition on the axis. The second is to use geodesic null coordinates. This allows us to make direct contact with the Bondi formalism and thus the emitted radiation reaching future null infinity $\mathcal I^+$ in asymptotically flat spacetimes. 

\subsection{Choosing $\psi$ as a coordinate}
\label{sec:Choice2}

The field equations~\eqref{fieldEQpsi} describe a gravitational field on a three manifold sourced by a harmonic scalar field $\psi$ which obeys ${D}^2\psi  = 0$. In 4D gravity, harmonic coordinates  have been successfully employed, e.g. for proving the well-posedness of the Cauchy problem for the Einstein equations~\cite{CB1962,Wald1984}. The usefulness of harmonic coordinates in 4D, together with the fact that the 3D Ricci tensor greatly simplifies if $\psi$ is chosen as a coordinate leads us to investigate this gauge choice further. 

We now specialize our triad so that $c_a$ points in the same direction as the gradient of $\psi$. The normalization condition \mbox{$c_a c^a=1$} implies that 
\be
\label{eqCA}
c_a = \sqrt{\frac 2R} \psi_{,a} \,,
\ee 
where $R=2\psi_{,a}\psi^{,a}$ is the 3D Ricci scalar defined in~\eqref{ricci3D}. Note that the sign of $R$ determines whether $\psi_{,a}$ is timelike, spacelike, or null. For Schwarzschild, $R>0$, and so we might expect this to be true of a physically reasonable spacetime, especially one that settles down to Schwarzschild after some dynamical evolution, and as such we will assume that $\psi_{,a}$ is spacelike. 

Given the definition of $c_a$ in Eq.~\eqref{eqCA}  we can express the Ricci tensor~\eqref{fieldEQpsi} as $R_{ij}= R c_{i} c_{j}$, and so the six curvature scalars defined in Eqs.~\eqref{phidef} are  $\phi_5= \phi_4= \phi_3= \phi_2= \phi_1 =0$ and $\phi_0 = R$. This greatly simplifies the Bianchi identities, which are
\begin{align}
\frac{R_{,1}}{R}&=  2 \theta ,
 & \frac{R_{,2}}{R} &=-2\iota , 
&   \frac{R_{,3}}{R} &=  2(\epsilon- \gamma) \,,
 \label{BIANCI3b}
\end{align}
and which gives the rotation coefficients appearing in Eq.~\eqref{BIANCI3b} the interpretation of being proportional to the rate of change of $\ln R$ in a particular direction. 

Because $R$ is a scalar, the curl of its gradient, $\epsilon^{abc}R_{|bc} =0$, must vanish. Equivalently, the commutator equations \eqref{COMM} with $f=R$ must hold. This augments the field equations with the following three equations,
\ba
\label{Rcomm}
\iota _{,1}+\theta _{,2} - \alpha  \iota +\delta  \theta +\epsilon ^2 - \gamma^2=0, \notag \\
(\epsilon -\gamma)_{,1}-\theta_{,3}- \beta \iota -\theta(\epsilon+ \eta)=0,\notag\\
(\epsilon - \gamma)_{,2}+\iota_{,3}- \zeta \theta - \iota( \gamma+ \eta)=0 \,.
\ea
The fact that $c_a$ points along the gradient of a scalar places additional conditions on the rotation coefficients. To see this, we compute the intrinsic derivative of $c_a$ and express the result on the triad basis to obtain
\begin{align}
c_{a|b} = \left\{
\begin{array}{ccc}
 \beta  & \epsilon  & \theta  \\
 -\gamma  & -\zeta  & -\iota  \\
 0 & 0 & 0
\end{array}
\right\} \,.
\end{align}
Now, noting that $c_{a|b} + \frac{1}{2}c_a (\ln R)_{,b}= \sqrt{2/R}\psi_{|ab}$, and using the directional derivatives of $R$ computed in~\eqref{BIANCI3b}, we have
\begin{align}
\sqrt{\frac{2}{R}}\psi_{|ab} =\left\{
\begin{array}{ccc}
 \beta  & \epsilon  & \theta  \\
 -\gamma  & -\zeta  & -\iota  \\
 \theta  & -\iota  & \epsilon- \gamma 
\end{array}
\right\} \, . 
\label{psitriad}
\end{align}
However since $\psi_{,a}$ is a gradient, this matrix should be symmetric. Thus $\gamma = -\epsilon$. Further, we note that $D^2 \psi = 0$ is automatically satisfied.

The Bianchi identities~\eqref{BIANCI3b}, in addition to the field equations~\eqref{Bondi0}--\eqref{BondiI3}, allow us to find a particularly simple expression for the wave operator of $\ln R$,
\begin{align}
D^2(\ln R) = 2(R-2\epsilon^2-2\zeta \beta) \,.
\label{LapopPsi}
\end{align}

It is interesting to note that if $\psi$ is chosen as a coordinate and the tetrad leg $c_a$ is fixed using \eqref{eqCA}, then the directional derivatives of $\psi$ that enter into the 4D expressions for the NP scalars become particularly simple. Explicitly $\psi_{,1}=\psi_{,2}=0$ and $\psi_{,3}= \sqrt{R/2}$. This implies that the expressions for the Weyl scalars~\eqref{p0} become
\begin{align}
\hat \Psi_0&= \sqrt{\frac{R}{2}}\beta , &
\hat \Psi_1&=\frac{e^{ \psi } \sqrt{R}}{2 } \theta, \notag\\
\hat \Psi_2&= e^{2 \psi } \left( \frac{R}{2}+\sqrt{\frac{R}{2}} \epsilon \right) , \notag\\
\hat \Psi_3&=-\frac{e^{3 \psi }\sqrt{R}}{2 } \iota,&\hat \Psi_4&= -e^{4 \psi }\sqrt{\frac{R}{2}}\zeta . 
\label{p0psi}
\end{align}
The rotation coefficients that enter these expressions are the same rotation coefficients that appear in the second derivative of $\psi$ expressed on the triad basis, Eq.~\eqref{psitriad}. This underscores the fact that the scalar $\psi$ sources the gravitational field. Another important consequence of Eqs.~\eqref{p0psi} is that for this tetrad choice, if $l^a$ is geodesic, i.e. $\beta=0$, then the geodesic is a principal null geodesic of the spacetime, $\hat \Psi_0 = 0$.

Thus far  the other triad vectors are unspecified, except that they are null and orthogonal to $c_a$. With $c_a$ fixed, we still have freedom to boost along $l^a$. The equivalent of  the Lorentz transformations for the 3D triad are discussed fully in Appendix~\ref{sec:UsefulResults}. Here we consider the effect of a boost of the form  
\begin{align}
\tilde{l}^a &= A l^a, & \tilde{n}^a& = A^{-1} n^a \,. 
\label{boosta}
\end{align}
Using the definitions in Eqs.~\eqref{ninerot}, we find that under such a boost, six of the coefficients are simply multiplied by factors of $A$, while three have nontrivial transforms,
\begin{align}
\label{BoostTransforms}
\tilde{\alpha}&=A \alpha -A_{,1}& 
\tilde{\eta} 
&= \frac{ A \eta-A_{,3} }{A} \,,&
\tilde{\delta}&= \frac{  A\delta - A_{,2} }{A^2}\,.
\end{align}
The full transforms are given in Eq.~\eqref{BoostTransforms2}; interestingly, the above coefficients with a nontrivial transform do not enter into the expressions for the Weyl scalars in Eqs.~\eqref{p0psi}.

We can always use our boost freedom to set at least one of $\tilde \alpha,\,\tilde \eta,$ or $\tilde \delta$ to zero. Note that if a boost exists that can set $\tilde \alpha = \tilde \eta = \tilde \delta = 0$, then it can be shown that $R=0$ and that the resulting spacetime is flat. Also, if one triad leg $c_a$ is chosen according to Eq.~\eqref{eqCA}, it is not possible to apply a boost to render the null vector $l^a$ geodesic, or equivalently to set the coefficient $\beta$ to zero. An example which illustrates this fact is in the asymptotic region of a radiating spacetime, where our choice of $c^a$ would point along a cylindrical radius; meanwhile, the outgoing null geodesics define a radial direction, and it is clear that these two directions are not orthogonal. Rather, we would need to locally choose some other null direction to define $l^a$. 

We now ask whether it is possible to find a coordinate $t$ whose gradient is timelike and orthogonal to $c_a$, i.e. that $t_{,a} c^a=0$. The first step is to define a timelike unit vector $T_a$ as
\begin{align}
T_a= \frac{1}{\sqrt{2}}(l_a+n_a). 
\label{Timelikeunit}
\end{align}
From the normalization conditions~\eqref{NullNorm} it is straightforward to verify that $T_aT^a = -1$. We would like to determine if $T_a$ is hypersurface orthogonal, so that it can be written as $T_a=- \varrho t_{,a}$. This  is possible if and only if $T_a$ is twist-free, $T_{[a}D_{b}T_{c]} = 0$. In 3D, this is equivalent to the vanishing of the scalar 
\begin{align}
W = \epsilon^{abc} T_a T_{c|b} = \frac12 (-\beta +\gamma -\zeta +2 \eta +\epsilon ).
\end{align}
For a general $l^a$ and $n^a$ this will not be true, but we can choose a boost $A$ that will transform $\eta$ such that $W=0$. By Eq.~\eqref{BoostTransforms}, we see we must choose 
\begin{align}
\eta = \frac{1}{2}(\beta -\gamma +\zeta -\epsilon ).
\end{align}
We have so far fixed our triad, and selected the harmonic coordinate $\psi$ and the coordinate $t$ whose gradient lies parallel to $T_a$. Let us call the third coordinate $s$.  On the coordinate basis $(t, \ s, \ \psi)$  the assumptions thus far imply that in all generality the we can express the covariant components of the triad as
\begin{align}
l_i &= ( -l_t,\ h_s,\ -h_\psi)\, , &
n_i &=(-n_t,\  -h_s,\ h_\psi)\notag\\
c_i &= (0 , \ 0, \sqrt{2/R}), 
\label{psitriaddown}
\end{align}
where $l_t$, $n_t$, $h_s$ and $h_\psi$ are free functions of $(t,\ s, \ \psi)$.
The factor $\varrho$ in the definition of $T_a$ is $\varrho = (l_t+n_t)/\sqrt{2}$. 
The metric on the coordinate basis is constructed using Eq.~\eqref{Hmn}. To see if any further metric functions can be set to zero, consider a coordinate transformation that leaves the coordinates $t$ and $\psi$ unchanged but chooses a new coordinate $s'$, such that $s =f(t,s',\psi)$. We find that the metric can be expressed in the same form except with the functions $l_t, \, n_t, \ h_s, \ h_\psi$ transformed as 
\begin{align}
h'_{\psi }&= h_{\psi} -h_s f_{,\psi }\, ,&
h'_s&= h_s
  f_{,s'} \, , \notag\\
l'_t&=  l_t -h_s f_{,t}\, \,, \ &
n'_t&= n_t+h_s f_{,t} \,.
\label{cortranspsi}
\end{align}
It is thus always possible to choose a gauge in which $h'_\psi=0$. Dropping the primes, the resulting metric on the coordinate basis is
\begin{align}
h_{ij} &= 
\left(
\begin{array}{ccc}
 -2 l_t n_t & -h_s \left(l_t-n_t\right) & 0 \\
 -h_s \left(l_t-n_t\right) & 2 h_s^2 & 0 \\
 0 & 0 & \frac{2}{R} \\
\end{array}
\right) .
\label{Metpsi}
\end{align}
For the rest of this section we make this coordinate choice. The covariant components of the triad vectors are 
\begin{align}
l^i &= \frac{1}{ \sqrt{2}  \varrho h_s}( h_s,\ l_t,\ 0)\, , &
n^i &=\frac{1}{ \sqrt{2}  \varrho h_s }(h_s,\  -n_t,\ 0) \,, \notag\\
c^i &= (0 , \ 0, \sqrt{R/2}). 
\label{psitriadup}
\end{align}

The choice of $\psi$ as a coordinate is an unfamiliar one,  and to help build some intuition we present the Minkowski metric, triad, and rotation coefficients in this coordinate system in Appendix~\ref{sec:Minkpsi}. The rotation coefficients in general axisymmetric spacetimes can be expressed in terms of the functions entering Eqs.~\eqref{psitriaddown} and~\eqref{psitriadup}, and are listed in Appendix~\ref{sec:ROTCOEFPSICOORD}. 

The expression in~\eqref{ROTcoefsMet} for the coefficient $\epsilon$,
\begin{align}
 \epsilon = -\frac{\sqrt{R} \left[ \ln(h_s  \varrho)\right]_{,\psi }}{2 \sqrt{2}}\, ,
\end{align}
can be integrated using the Bianchi identity~\eqref{BIANCI3b}, 
\begin{align}
\epsilon = \frac{\sqrt{R}\ (\ln R)_{,\psi}}{4 \sqrt{2}} \,.
\end{align}
Combining these equations shows that the metric functions obey $ \left[ \ln R (h_s \varrho)^2 \right]_{,\psi } =  0$, which after integration provides
\begin{align}
(h_s \varrho)^2 R = g(t,s). 
\label{epsint}
\end{align}
There is still some residual coordinate freedom in Eq.~\eqref{Metpsi} in that we can apply a coordinate transformation to the $s$ and $t$ coordinates without changing the form of the metric. In particular by using the coordinate transformation $s= f_2(t,s_2')$  and using a restricted version of Eq.~\eqref{cortranspsi} it is possible to choose a gauge in which $g(t,s)=1$ so that we have $(h_s \varrho)^2 R=1$. We will not necessarily make this specialization in the rest of the text.

\subsubsection{Field equations adapted to the  $\psi$  coordinate choice}

In this subsection we specialize the field equations~\eqref{Bondi0}--\eqref{BondiI3} to the case where we use $\psi$ as a coordinate and where $c_a = \sqrt{2/R} \psi_{,a}$. Recall that this choice  implies that $\gamma= - \epsilon$, $\psi_{,1}=\psi_{,2} = 0$ and $\psi_{,3} = \sqrt{R/2}$. With this specialization, we re-order the general field equations~\eqref{Bondi0}--\eqref{BondiI3} augmented by the commutation relations~\eqref{Rcomm}. One of the field equations is redundant with one of the commutation relations, while the remaining 11 equations can be split into a subset of four equations that contain directional derivatives in the $l^a$ and $n^a$ directions only,  
\begin{align}
\epsilon _{,1}&=+\beta _{,2}+2 \beta  \delta , \notag\\
\zeta _{,1}&=-\epsilon _{,2}+2 \alpha  \zeta
   ,\notag\\
\iota _{,1}&=-\theta _{,2}+\alpha  \iota -\delta  \theta , \notag\\
\delta _{,1}&=\alpha
   _{,2}-R/2+2 \alpha  \delta +\beta  \zeta +\epsilon ^2
\label{psiFeq12}
\end{align}
and a group of seven equations that fix the directional derivatives of certain rotation coefficients in the $c^a$ direction,
\begin{align}
\epsilon _{\text{,3}}&= -\delta _{,1}-\iota _{,1}+\alpha _{,2}+2 \alpha  \delta +\alpha  \iota
   +\theta  \iota +2 \epsilon ^2,\notag\\
\beta _{\text{,3}}&= \theta _{,1}+\theta  (\alpha -\theta )+2 \beta  (\epsilon -\eta
   ), \notag\\
\alpha _{\text{,3}}&= \eta _{,1}-\alpha  \eta +\alpha  \epsilon +\beta  (\delta -\iota )-\eta  \theta -\theta 
   \epsilon , \notag\\
\delta _{\text{,3}}&= \eta _{,2}-\alpha  \zeta +\delta  (\eta +\epsilon )+\zeta  \theta +\eta  \iota
   -\iota  \epsilon , \notag\\
\zeta _{\text{,3}}&= \iota _{,2}-\delta  \iota +2 \zeta  \eta +\iota ^2+2 \zeta  \epsilon , \notag\\
\theta
   _{\text{,3}}&= 2 \epsilon _{,1}-\beta  \iota -\eta  \theta +\theta  (-\epsilon ), \notag\\
\iota _{\text{,3}}&= -2 \epsilon
   _{,2}+\zeta  \theta +\iota  (\eta -\epsilon ) .
\label{psiFeq3}
\end{align}
 We showed that in the coordinate basis $(t, \, s, \, \psi)$ the metric can be written in the form~\eqref{Metpsi}.  With the choice of $l^a$ and $n^a$ in \eqref{psitriadup}, all the equations~\eqref{psiFeq12} contain only derivatives with respect $t$ and $s$, and effectively constitute a set of constraint equations that have to be satisfied for every constant $\psi$ surface.

As can be seen from the above set of  equations, choosing $\psi$ as a coordinate does not greatly simplify the field equations. For this coordinate and triad choice the major simplifications occur in the Bianchi identities~\eqref{BIANCI3b}, the form of the metric~\eqref{Metpsi}, and the simple form of the corresponding Weyl scalars. An additional advantage of this coordinate and triad choice that will be discussed in the next section is the easy identification of the axis.

\subsubsection{Axis conditions as $\lambda \rightarrow 0$}

On the axis, which for the three metric is denoted by the boundary conditions $\psi\rightarrow - \infty$ or $\lambda \rightarrow 0$, we now explore the conditions on the triad quantities required for the elementary flatness condition to hold.

The first step is to observe that working in a coordinate system where $\psi$ is a coordinate makes it easy to prove the equivalence of the two forms of the axis conditions, $K_D= 1$ in Eq.~\eqref{deficitangle} and the coordinate invariant expression in Eq.~\eqref{ElemFlat}. Assuming the metric  $h_{ij}$  can be written in the form~\eqref{Metpsi}, the metric on the space orthogonal to the axis $W_2$ is merely $ds^2_0 = g_{\psi\psi}\ d\psi^2+ \lambda d\phi^2 $, where the 4D metric component $g_{\psi \psi} = 2/(\lambda R)$. The elementary flatness condition~\eqref{deficitangle} now reads
\begin{align}
\lim_{\lambda \rightarrow 0} \frac{\sqrt{\lambda}}{\int_{-\infty}^\psi \sqrt{g_{\psi\psi}} d\psi} =1 ,
\end{align}
where use has been made of the fact that $\lambda$ is not a function of $\phi$. Applying l'H\^{o}pital's rule and differentiating above and below the line with respect to $\psi$, the elementary flatness condition becomes
\begin{align}
\lim_{\lambda \rightarrow 0} \frac{ e^\psi}{\sqrt{g_{\psi\psi}} }= \lim_{\lambda \rightarrow 0}\sqrt{ \frac{  e^{4\psi}R}{2}} = 1,
\end{align}
or equivalently $R\rightarrow 2 e^{-4\psi}$. By definition, $R=2\psi_{,a} \psi^{,a}$, showing that the covariant expression~\eqref{psiAxis} and thus  Eq.~\eqref{ElemFlat} are equivalent to the elementary flatness condition.
Note that the elementary flatness condition, in conjunction with the  condition found when examining the rotation coefficient $\epsilon$,  Eq.~\eqref{epsint}, implies that the determinant of metric on the subspace normal to the axis also remains finite as we approach the axis. To see this explicitly, observe that $\det[h_{\tilde{i}\tilde{j}}]=-2h_s^2 \varrho^2$, where  $\tilde{i},\tilde{j}\in\{s,\ t\}$. By the condition found in Eq.~\eqref{epsint} in the gauge where $g(s,t)=1$ we have  $\det[h_{\tilde{i}\tilde{j}}]=-2/R$. The determinant associated with the corresponding part of the four metric becomes $\det[g_{\tilde{i}\tilde{j}}]=-2/(R e^{4\psi})$, which by the elementary flatness condition approaches the value $-1$ on the axis as expected. 

Symmetry dictates that a null vector on the axis remains on the axis when it is sent out to infinity or toward the origin. Thus on the axis $\hat{l}^\mu$ and $\hat{n}^\mu$ are geodesic, provided they are chosen to lie along the in-going and out-going directions. In terms of the NP scalars~\eqref{4spins}, this translates into  $\hat{\kappa }= \beta e^{-\psi }/\sqrt{2} \rightarrow 0$, and $\hat{\nu }= \zeta e^{3 \psi }/\sqrt{2}  \rightarrow 0$.

In Appendix~\ref{sec:ROTCOEFPSICOORD}, explicit formulas for the expansions of the metric quantities about the axis are given and discussed. The special case of the static Schwarzschild black hole is examined in Sec.~\ref{staticpsi} where the scaling of the solution, the 3-curvature $R$ and all the rotation coefficients are explicitly computed.

\subsection{Geodesic null coordinates}
\label{sec:Choice1}

We now examine the equations in a coordinate system adapted to asymptotic null infinity, where the concept
of emitted radiation is well defined. Akin to the standard methods used in the NP formulation (see e.g. \cite{NewmanPenrose1962, Sachs1962, Tamburino1966}), this coordinate system and triad choice is tied to the tangent vectors of null geodesics. We begin with a  family of null hypersurfaces in $\mathcal S$, and we label these by a coordinate $u$, so that $h^{ij} u_{,i} u_{,j} =0$. We  then choose the covariant representation of one null triad vector to be the gradient of the coordinate $u$, setting $l_i = -u_{,i}$. Since $l_i$ is the gradient of a coordinate, it has vanishing curl. The intrinsic derivative of $l_i$ on the triad basis is
\begin{align}
l_{a|b}=\left\{
\begin{array}{ccc}
 0 & 0 & 0 \\
 \alpha  & \delta  & \eta  \\
 -\beta  & -\epsilon  & -\theta 
\end{array}  
\right\} \,.
\label{Derivlnull}
\end{align} 
The fact that $l_{a|b}$ is symmetric immediately sets $\beta=\alpha = 0$ and $\eta = -\epsilon$. Note that $\beta=\alpha = 0$ implies that $l_a$ is geodesic and affinely parametrized on $\mathcal S$, by Eqs.~\eqref{ninerot}. Also, recall that null geodesics are conformally invariant, and we can verify that here Eqs.~\eqref{4spins} imply that if $\beta = 0$ then $\hat{\kappa}=0$ in $\mathcal M$. Thus if $l^a$ is the generator of a geodesic null congruence on $\mathcal{S}$, the  corresponding null congruence in the physical manifold is also geodesic. The above conditions on the rotation coefficients further imply that the field equation~\eqref{BondiI2} is trivially satisfied. If we choose as another coordinate the affine parameter $p$ along the geodesic that $l^a$ is tangent to, we have $l^i f_{;i} = f_{,1} = \partial_p f$. Lastly, we label our third coordinate $\chi$. Expressing the null vectors on the $(u,\, p,\, \chi)$  coordinate system, we have
\be
\label{Choice1set1}
l_i = (-1,0,0),  \qquad l^i = (0,1,0) \,.
\ee
The normalization conditions~\eqref{NullNorm} allow us to restrict some of the components of the remaining triad vectors, giving  $n^u = 1$ and $c^u = 0$. Using the expression for the metric in terms of the triad vectors~\eqref{Hmn}, we can see that $h^{uu} = h^{u\chi} = 0$ and $h^{up} = -1$ follows. Three more metric functions fully determine $h_{i j}$. We parametrize these remaining metric components following the convention of \cite{Bondi1962,Bonanos2007} so that the contravariant form of the metric becomes
\be
\label{InverseBondiMetric}
h^{ij} =  \left( \begin{array}{ccc}
0&-1&0  \\
-1 & 2v_1+ v_2^2&v_2 e^{-v_3}\\
0&v_2 e^{-v_3}&e^{-2v_3}
\end{array} \right) \, ,
\ee
where $v_i$ are free functions of the coordinates.
The covariant form of the  metric on $\mathcal{S}$ is then given by 
\be
\label{BondiMetric}
h_{ij} =  \left( \begin{array}{ccc}
-2v_1 & -1&v_2 e^{v_3}\\
-1& 0  & 0   \\
v_2 e^{v_3} & 0 & e^{2 v_3}
\end{array} \right) \,.
\ee
This metric holds for any null foliation of the manifold $\mathcal{S}$, where constant $u$ surfaces denote the null hypersurfaces, the affine parameter $p$ serves as a coordinate along a particular geodesic and the coordinate $\chi$, usually associated with an angular coordinate, labels the geodesics within the hypersurface. 

We further need to fix the triad legs $n^i$ and $c^i$. One such choice that satisfies the normalization condition~\eqref{NullNorm} and gives the correct form of the metric~\eqref{InverseBondiMetric} is
\be
\label{Choice1set2}
n^i = (1, -v_1, 0), \qquad c^i = (0, v_2, e^{-v_3}) \,.
\ee
The corresponding covariant vectors are 
\be
\label{eq:C1nc}
n_i = (-v_1,-1, v_2 e^{v_3}), \qquad c_i = (0,0,e^{v_3}) \,.
\ee
On this triad, the directional derivatives applied to  a function $f$ are
\begin{align}
\label{C1Ders}
f_{,1} & =  f_{,p} \,, &f_{,2} & =  f_{,u} - v_1 f_{,p} \,, &
f_{,3} & =  v_2 f_{,p} + e^{-v_3} f_{,\chi} \,.
\end{align}
If the chosen coordinates $(u,\, p,\, \chi)$ are to be valid, they must satisfy the commutation relations given in Eq.~\eqref{COMM}. Applying the commutation relations to each successive coordinate
provides a simple way of relating the rotation coefficients to derivatives of the metric functions of Eq.~\eqref{BondiMetric}. The commutators acting on $\chi$ yield the coefficients
\begin{align}
\label{chicomm1}
\gamma &= -\epsilon ,&\theta &= -v_{3,1},&\iota &=
   v_{3,2}.
\end{align}
Applying the commutation relations to $u$ reiterates that $\alpha=\beta=0$ and $\eta = -\epsilon$.
Finally, applying the commutation relations to $p$ fix
\begin{align}
\delta &= -v_{1,1},
\label{rcomm1} \\
\epsilon &= \frac{1}{2}
   \left(v_2\, \theta-v_{2,1}\right), 
\label{rcomm2}  \\ 
\zeta
   &= v_2
   v_{3,2 }+v_{2,2}+v_{1, 3} .
\label{rcomm3}
\end{align}

\subsubsection{Field equations adapted to the geodesic null coordinate choice}

When working with geodesic null coordinates, where $\gamma=\eta=-\epsilon$ and $\alpha=\beta=0$,
the field equations \eqref{Bondi0}--\eqref{BondiI3} can be expressed in simplified form in terms of the five remaining rotation coefficients as
\ba
\label{Bondi5Sim}
\theta_{,1} & = & \theta^2 +\phi_5 \,, \\
\label{Bondi3Sim}
\epsilon_{,1} & = & \phi_3 \,, \\
\label{BondiSimC1}
  \delta_{\text{,1}}
&=&   \epsilon ^2  -\frac{\phi_0}{2} -\phi_4 \,,\\
\label{BondiSimC2}
 \theta_{,2} -\epsilon_{,3}  
&=& -\delta 
   \theta - \epsilon ^2  -  \theta  
   \iota-\frac{\phi_0}{2} 
  \,,
\\
\label{BondiI1Sim}
\iota_{,1} + \theta_{,2} & = & - \theta \delta \,, \\
\label{BondiSim4}
\zeta_{,1}  +  \epsilon_{,2} & = &    \phi_1 \,, \\
\label{Bondi2Sim}
\zeta_{,3} - \iota_{,2} & = & \iota(\iota - \delta) + \phi_2 \,, \\
\label{BondiI3Sim}
\zeta_{,1} + \delta_{,3} + 2 \epsilon_{,2} & = & \zeta \theta - 2 \iota \epsilon \,.
\ea
One of the equations is trivially solved and has been omitted. The remaining equations have been reordered, and some are linear combinations of the original set. These combinations are~\eqref{BondiSimC1} = \eqref{BondiI1}/2 $-$ \eqref{Bondi0}/2 $-$ \eqref{Bondi4} describing the derivative $\delta_{,1}$; the combination~\eqref{BondiSimC2}=\eqref{BondiI1}/2 $-$ \eqref{Bondi0}/2, yielding an expression for the combination $\theta_{,2}-\epsilon_{,3}$; and finally the combination~\eqref{BondiSim4} = \eqref{BondiI3} $-$ \eqref{Bondi1} to obtain an expression for $\zeta_{,1}+\epsilon_{,2}$. The remaining equations are simplified analogues of their counterparts in  Eqs.~\eqref{Bondi0}--\eqref{BondiI3}.

The reordering makes apparent the fact that a hierarchy exists in the  reduced system of equations, which in turn makes it possible to formally integrate the field equations in a systematic way. Suppose we begin on a null hypersurface of constant $u$ on which the directional derivatives of the function $\psi$ are given, so that $\phi_i$, \ $i=0\dots 5$ are known. Equations~\eqref{Bondi5Sim} and~\eqref{chicomm1} can be  integrated with respect to $p$ to obtain the rotation coefficient $\theta$, and subsequently the metric function $v_3$. In a similar fashion Eqs.~\eqref{Bondi3Sim} and~\eqref{rcomm3} yield $\epsilon$ and $v_2$, and subsequently~\eqref{BondiSimC1} and~\eqref{rcomm1} give $\delta$ and $v_1$. The metric functions $v_3$, $v_2$ and $v_1$  are thus determined within the null hypersurface up to boundary terms. The requirement that $D^2 \psi =0$ determines $\iota= v_{3, 2}$ using Eq.~\eqref{LapOp}. Thus the manner in which the metric function $v_3$ changes away from the initial null hypersurface is 
known. Equation~\eqref{BondiSimC2} then serves as a consistency condition which restricts some of the six integration constants that arise while integrating Eqs.~\eqref{Bondi5Sim}--\eqref{BondiSimC1}.  The other integration constants are determined by boundary conditions that will be discussed more fully in Sec.~\ref{AsflatJD}. Equation~\eqref{BondiSim4} implicitly determines $v_{2,2}$. Equation~\eqref{Bondi2Sim}, in conjunction with the condition  $D^2\psi=0$ provides an evolution equation of $\psi$. The remaining two equations,~\eqref{BondiI1Sim} and~\eqref{BondiI3Sim}, are eliminant relations that are trivially solved when the metric functions are substituted into the field equations. 

The hierarchy of field equations that arise when they are expressed on a coordinate system adapted to a null hypersurface has been extensively studied in the four dimensional context. It is known, for instance, that the equivalent equations on $\mathcal M$ are formally integrable on a constant $u$ surface \cite{Sachs1962,Bondi1962,Hobill1987,Hayward1993}. The asymptotic behavior of the metric in geodesic null coordinates, and the associated boundary conditions are further discussed in Sec.~\ref{AsflatJD}, where the relationship to the Bondi formalism is explored.  In Appendix~\ref{AsflatJDApp} we give an explicit example of how the field equations are systematically integrated in an asymptotic region far from a gravitating system, although there $l^a$ is affinely parametrized with respect to $\mathcal M$.

The gauge and triad choice  discussed in this section has the advantage of eliminating four of the nine rotation coefficients. This reduction in complexity makes apparent a hierarchy in the field equations that hints at the possibility of finding an analytic solution to the problem. In Sec.~\ref{NewSln} we carry out an example calculation in which the field equations~\eqref{Bondi5Sim}--\eqref{BondiI3Sim} are systematically solved in a special case. It should be noted however that this simplicity comes at a cost. Unlike the case where the triad was adapted to the coordinate $\psi$, and the Ricci tensor only had one non-zero component, the Ricci tensor on this triad is constructed from the three independent quantities $\psi_{,a}$. It has two degenerate eigenvectors with zero eigenvalues, and a single normalized eigenvector with nonzero eigenvalue $2 \psi_{,a} \psi^{,a}$.

The analysis performed in this section assumes that $l_i$ is affinely parametrized in $\mathcal S$. If we adjust the parameter along each geodesic, $p \to p'(u,p,\chi)$, this results in $h_{up'} \neq -1$, but otherwise preserves the form of the metric. A physically motivated alternative to affine parametrization in $\mathcal S$ is to boost $l^i$ so that it is affinely parametrized in the physical 4D spacetime, and then to use the affine parameter $\tau$ as the second coordinate instead of the parameter $p$. The field equations that result form this choice of parametrization are detailed in Appendix~\ref{app:affinereal}.

Some gauge freedom remains when $l^i$ is affine in $\mathcal S$, and $p$ is used as a coordinate. Shifting the origin of the affine parameter along each geodesic separately, $p' = p + f (u,\chi)$, transforms the metric function of~\eqref{BondiMetric} according to
\begin{align}
\label{affineshift}
v_1'&=v_1-f_{,u} \,, & v_2'&= v_2+ f_{\chi}e^{-v_3} \,,   & v_3'&=v_3 \,.
\end{align}
Relabeling the individual geodesics within a spatial slice, $\chi' = g(u,\chi)$, transforms the metric functions of~\eqref{BondiMetric} to~\cite{CarmeliBook}
\begin{align}
\label{relabeling}
v_1'& = v_1+ e^{v_3}v_2 \frac{g_{,u}}{g_{,\chi}} -\frac{1}{2} e^{2 v_3} \left( \frac{g_{,u}}{g_{,\chi}}\right)^2\,, &e^{v_3'}& =\frac{e^{v_3} }{g_{,\chi}}\,, \notag\\
v_2'&= v_2 -e^{v_3} \frac{g_{,u}}{g_{,\chi}} \,.
\end{align}
Finally, it is also possible to relabel the null hypersurfaces, setting $u'=h(u)$,  $ p'=p/ h_{,u}$. The metric components transform as
\begin{align}
\label{AffineRescale}
 v_1' &= \frac{v_1}{(h_{,u})^2} + p \frac{h_{,uu}}{(h_{,u})^3}, &
v_2' &= \frac{v_2}{h_{,u}},&
v_3' &= v_3.
\end{align}

\subsection{Asymptotic flatness and the peeling property}
\label{AsflatJD}

We will complete our discussion of useful coordinate systems on  $\mathcal S$ by discussing the asymptotic limit of the metric far from an isolated, gravitating system.
We will consider only spacetimes that are asymptotically flat and therefore admit the peeling property~\cite{Sachs1961,Sachs1962,Penrose1963,Penrose1965,PenroseRindler2}. According to the peeling property, the Weyl scalars expressed on an affinely parametrized out-going null geodesic  tetrad admit a power series expansion at  future null infinity (denoted $\mathcal{I}^+$) of the form
\begin{align}
\hat \Psi_i = \tau^{i-5}\sum_{n=0} \tau^{-n}\hat \Psi_i^{(n)} \,,
\label{eq:PeelingpowerSeries}
\end{align}
where $\tau$ is the affine parameter along the out-going null geodesics in $\mathcal{M}$ and
$\hat \Psi_i^{(n)}$ are constant along an out-going geodesic, i.e. $\hat \Psi_i^{(n)}(u,\chi)$. 
The work of  Bondi, van der Burg and Metzner~\cite{Bondi1962},  as well as Tamburino and Winicour's approach~\cite{Tamburino1966}, indicate that the  the metric functions also admit a power series expansion if expressed in terms of geodesic null coordinates. 

Appendix~\ref{AsflatJDApp} details a triad-based derivation of the asymptotic series expansions of the metric functions, in the restricted context of axisymmetric spacetimes. In this derivation, the ``Bondi news function'' is identified with the derivative of the dominant coefficient in the expansion of the shear of the out-going null tetrad leg. The calculation is performed assuming that the out-going null geodesic is affinely parametrized in $\mathcal M$, and the corresponding field equations given in Appendix~\ref{app:affinereal} are used.

The results obtained in Appendix~\ref{AsflatJDApp} for the asymptotic expansion of the metric
can be summarized as follows. In terms of affinely parametrized null coordinates, the 4D line element can be expressed as 
\begin{align} \label{affineMmet}
ds^2 &=-2e^{2\psi}w_1\ du^2 -2 du\ d\tau + e^{w_3}w_2\ du\ d\chi \notag\\
&+ e^{2w_3-2\psi}\ d\chi^2+e^{2\psi}\ d\phi^2\,,
\end{align}
where, according to Eqs.~\eqref{psiexpand0},~\eqref{w3expand},~\eqref{w2expand},~\eqref{w1expand},~\eqref{direcdirpsi3}, and~\eqref{w12def}, the metric functions  admit the following asymptotic expansion as the affine parameter \mbox{$\tau \rightarrow \infty$}:
\begin{align}
e^{2\psi}&=(1-\chi^2) \left(\tau^2  - 2\tau \sigma^{(0)}+\sigma^{(0) 2} + \frac{\hat \Psi_0^{(0)}}{3\tau} \right)+ O(\tau^{-2}), \notag\\
e^{w_3}&= \tau^2- \sigma^{(0)2}  +\frac{\sigma^{(0)} \hat \Psi_0^{(0)}}{6\tau^2} +  O(\tau^{-3}),  \notag\\
w_1 &=  \frac{1}{(1-\chi^2)}\left[ \frac{1} {2 \tau^2}
  + \frac{ \sigma^{(0)}+  \hat \Psi_2^{(0)}   }{ \tau^3}\right] + O(\tau^{-4}), \notag\\
w_2 &=
 \frac{\left[(1-\chi^2) \sigma^{(0)}\right]_{,\chi }}{  (1-\chi^2) \tau ^2}-\frac{2 \sqrt{2}  \hat \Psi _1^{(0)}}{3 \sqrt{1-\chi^2}\ \tau ^3} + O(\tau^{-4})\, .
\label{metExpand}
\end{align}
Note that the coordinate $\chi$ is chosen here to be $\chi = \cos \theta$, and $\theta$ is the usual polar angle. The axis occurs as \mbox{$\chi \rightarrow \pm 1$}.

The free functions that enter into the metric are $\sigma^{(0)}(u,\chi)$ and the dominant terms associated with the Weyl scalars $\hat \Psi_i^{(0)}(u,\ \chi)$, $i\in\{0,1,2\}$. The dominant terms of $\hat \Psi_3^{(0)}$ and $\hat \Psi_4^{(0)}$ are fixed by these free functions through Eqs.~\eqref{NPWeylexpansion} and~\eqref{psi30def} or equivalently
\begin{align}
\hat \Psi_3^{(0)}&= 
 -\frac{ \left[ (1-\chi^2) \sigma^{(0)}{}_{,u} \right]_{,\chi}  }{\sqrt{2}  
\sqrt{1-\chi^2} } \,, & \hat \Psi_4^{(0)}&=\sigma^{(0)}_{,uu} \,.
\label{psi3defpre}
\end{align}
The field equations determine the evolution of $\hat \Psi_i^{(0)}(u,\ \chi)$, $i\in\{0,1,2\}$ from one null hypersurface to another via Eqs.~\eqref{p00evo},~\eqref{p10evo}, and~\eqref{p20evo}. As can be observed from \eqref{psi3defpre}, the free function $\sigma^{(0)}_{,u}$ carries the gravitational wave content of the spacetime and is often referred to as the ``Bondi news function.''

A solution that settles down to a Schwarzschild black hole in its final state requires that in the limit $u \to \infty$ the scalars behave as
\begin{align}
\{ \hat \Psi_0, \hat \Psi_ 1,\hat \Psi_2 ,\hat \Psi_3, \hat \Psi_4, \hat \sigma^{(0)}, \hat \sigma^{(0)}_{,u} \} \to \{ 0,0,-M,0,0,0,0\}\,,
\end{align}
where the constant $M$ is the mass of the final black hole. 
For $u<\infty $, $\hat \Psi^{(0)}_{i}$ , $i\in \{0,1,2\}$ are then determined by these final conditions, provided that $\sigma^{(0)}(u,\chi) $ is given, using the evolution equations~\eqref{p00evo},~\eqref{p10evo}, and~\eqref{p20evo}. 
For easy reference these equations are repeated here :
\begin{align}
\label{p00evopre}
\hat \Psi _0^{(0)}{}_{,u}&= 3 \sigma ^{(0)} \hat \Psi_2^{(0)}
+(1-\chi^2)\left(\frac{\hat \Psi_1^{(0)} }{\sqrt{2}\sqrt{1-\chi^2}} \right){}_{,\chi } \,,
\\
\hat \Psi^{(0)}_{1 \ ,u} &= 2\sigma^{(0)} \hat \Psi_3^{(0)} + \frac{ \hat \Psi _2^{(0)}{}_{,\chi }}{\sqrt{2}\sqrt{1-\chi^2}}
\label{p10evopre} \,,  \\
\hat \Psi _2^{(0)}{}_{,u}&= 
 -\frac{ \left[ (1-\chi^2) \sigma
   ^{(0)}{}_{,u} \right]_{,\chi\chi}  }{2} -\sigma ^{(0)} \sigma^{(0)}{}_{,uu} \,,
 \label{p20evopre}
\end{align}

We now  examine how the metric and Weyl scalars behave on the axis in the limit of large distance from the isolated source. As noted in Sec.~\ref{sec:AxisCond}, and explored further in Appendix~\ref{sec:ROTCOEFPSICOORD}, the metric functions have a power series expansion in $\lambda=e^{2\psi}$ near the axis of symmetry. In addition, these expansions are such that the metric functions vanish sufficiently quickly in the approach to the axis so that there are no ``kinks'' at the axis~\cite{Bondi1962}. Note that  from~\eqref{metExpand}, the coefficient of the $du\, d\chi$ term in the metric, namely $e^{w_3} w_2$, is only regular on the axis if both  $\sigma^{(0)}$ and $\hat \Psi_1^{(0)}$ vanish on the axis, and respectively scale like
\begin{align}
 \sigma^{(0)}& =(1-\chi^2)  \tilde{\sigma}^{(0)} , & \hat \Psi_1^{(0)} =\sqrt{1-\chi^2} \tilde{\Psi}_1^{(0)} ,
\end{align}
where $  \tilde{\sigma}^{(0)}$ and $ \tilde{\Psi}_1^{(0)}$ need not vanish on the axis. Substituting these scalings into the evolution equations for the dominant expansion terms for the Weyl scalars, Eqs.~\eqref{p00evopre}--\eqref{p20evopre} and  \eqref{psi3defpre}, shows that on the axis 
\begin{align}
\hat \Psi_0 & = \hat \Psi_1 =\hat \Psi_3 =\hat \Psi_4 = 0 \,,
\end{align}
indicating that the spacetime is Type D on the axis, and there is no radiation to infinity along the axis. This is to be expected, since spin-2 transverse radiation cannot propagate along the axis and still obey axisymmetry. The only nonzero Weyl scalar is $\hat \Psi_2$, and the dominant coefficient can depend only on $u$,
\begin{align} 
\hat \Psi_2 & = - M(u) \tau^{-3} \, . 
\end{align}

The metric functions in the near-axis, large $\tau$ limit are
\begin{align}
g_{uu} & = - \left(1 - \frac{2 M(u)}{\tau} \right) + O(\tau^{-2}) \,, &
g_{u\chi} &= O(\tau^{-2}) \,,
\end{align}
with the $g_{\chi\chi}$ term becoming singular at the poles simply due to our coordinate choice. Changing from $\chi$ to the coordinate $\theta$ gives $g_{\theta \theta} = \tau^2 + O(\tau)$ while fixing $g_{u\theta} = 0 $ on the axis.
 Asymptotically, the only dynamics present are the variation of the multipole moments with changing $u$, where $M(u)$ clearly gives a monopole mass moment.

The results given thus far are for a metric whose $\tau$ coordinate coincides with the affine parameter of the geodesic null vector $\hat l^\mu$ on the physical manifold $\mathcal{M}$.
In order to convert to affine geodesic null coordinates on the manifold $\mathcal S$, and so read off the asymptotic behavior of the metric~\eqref{BondiMetric}, we need to consider the effect of the transformation between $(u,p,\chi)$ and $(u,\tau,\chi)$ coordinates, where $p = p(u, \tau,\chi)$. Expanding $dp$ in \eqref{BondiMetric} in terms of $du$, $d\tau$ and $d\chi$ and equating the result with  \eqref{BONDIMET2} yields the following relationship between the metric functions and the derivatives of the affine parameter $p$,
\begin{align} 
p_{,\tau}&= e^{2\psi}, & v_3&=w_3,\notag\\ 
v_2 &= e^{2\psi}w_2 + p_{,\chi} e^{-w_3}, & v_1&= e^{4\psi}w_1- p_{,u}. 
\label{reparam}
\end{align}
Integrating the first equation of~\eqref{reparam} with respect to $\tau$ yields
\begin{align}
\label{eq:geodesicp}
p  =& (1-\chi^2)\left( \frac{\tau^3}{3} - \sigma^{(0)} \tau^2 +  \sigma^{(0) 2}  \tau   +  \frac{\hat \Psi_0^{(0)}}{3} \ln \tau  \right) \notag\\
&+ O(\tau^{-1}) \,,
\end{align}
Inverting the series~\eqref{eq:geodesicp} to obtain an explicit expression for $\tau$ in terms of $p$ is complicated by the logarithmic term. The leading order expression can however easily be found and is 
\begin{align}\label{taup}
\tau = \left(\frac{  3 p}{1-\chi^2}\right)^{1/3} + \sigma^{(0)} + O(p^{-2/3} \ln p).
\end{align}
By working out the series expansions of Eq.~\eqref{reparam} in terms of $\tau$ and then substituting Eq.~\eqref{taup} into the result, the asymptotic behavior of the metric ~\eqref{BondiMetric} can be found to be 
\begin{align}
v_1 =& \frac{ (3p)^{2/3}}{2}(1-\chi^2)^{1/3} (1+2\sigma^{(0)}_{,u})
\notag\\ &+(3p)^{1/3}(1-\chi^2)^{2/3} \hat \Psi_2^{(0)}+O(\ln p) \notag\\
v_2 =&  -\frac{2\chi}{3}\left(\frac{3p}{1-\chi^2}  \right)^{1/3}-\frac{2}{3}\chi\sigma^{(0)} + O(p^{-1/3}) \notag\\
e^{v_3}  =& \left(\frac{3p}{1-\chi^2}\right)^{2/3} + 2\sigma^{(0)}  \left(\frac{3p}{1-\chi^2}\right)^{1/3} +  O(p^{-1/3}\ln p)\notag\\
e^{2 \psi}  =& (3p)^{2/3} ( 1 - \chi^2)^{1/3}  + O(p^{-1/3}\ln p) \,.
\end{align}

\section{Solutions to the twist-free axisymmetric vacuum field equations}
\label{sec:Solutions}

In this section we will characterize the properties of known twist-free solutions to the axisymmetric vacuum field equations within the framework that was established in the previous sections. The aim is to identify existing solutions and to catalog the assumptions made in finding them.  With the exception of the Schwarzschild solution, none of the existing asymptotically flat solutions have physical significance. The hope is that this characterization will help establish the necessary properties  a new dynamical solution, such as the head-on collision, must posses. 

A number of insights that can be gleaned by relating the four dimensional physical quantities to the three dimensional rotation coefficients are discussed in Sec.~\ref{sec:TriadEqns}. This section thus relies heavily on Sec.~\ref{sec:3dto4d}, and in particular Eqs.~\eqref{4spins} and Eqs.~\eqref{p0}, which give the 4D NP spin coefficients and associated Weyl tensor in terms of the 3D rotation coefficients discussed in Sec.~\ref{sec:TriadEqns}. Wherever possible we will also express the properties of the known solutions in terms of the two geometrically motivated triad and coordinate choices of Sec.~\ref{sec:Choice2} and Sec.~\ref{sec:Choice1}. 

We begin the discussion of analytic solutions with an example of the systematic solution of the field equations mentioned in Sec.~\ref{sec:Choice1}. We consider the special case where the spacetime admits a coordinate choice in which we can simultaneously choose $\psi$ as a coordinate with $c_{a}=\sqrt{2/R} \psi_{,a}$ and find a null coordinate $u$ such that the geodesic null vector $l_a=-u_{,a}$ is orthogonal to $c_a$.  This example has the benefit that it draws on our general results for both coordinate choices discussed in Sec.~\ref{sec:Choice2} and Sec.~\ref{sec:Choice1}. 

Having found one solution, we then place it in context with known solutions using a classification scheme based on the optical properties of the geodesic null congruence that $l^a$ is tangent to. It should be noted that the scope of many of the known solutions discussed in this section often extends beyond the restricted arena of twist-free axisymmetry, but we will restrict our discussion to this realm. 

\subsection{Special case: spacetimes that admit the coordinate choice $(u,\ p,\ \psi)$}
\label{NewSln}

Any three metric can be expressed on an affinely parametrized geodesic null coordinate basis $(u,\ p,\ \chi)$ as in Eq.~\eqref{BondiMetric}.  In this section we will consider the special case where the third coordinate $\chi = \psi$.  Making the  triad choice defined in Eqs.~\eqref{Choice1set1} and~\eqref{eq:C1nc} we thus require that $\psi_{,i}$ be spacelike and orthogonal to $l_i$.  From the results in  Sec.~\ref{sec:Choice1} on the geodesic null coordinate choice we have that $\alpha = \beta = 0$, $\epsilon = - \eta = -\gamma$ and that the simplified field equations presented in Eqs.~\eqref{Bondi5Sim}--\eqref{BondiI3Sim} hold. As discussed in Sec.~\ref{sec:Choice2}, the choice of $\psi$ as a coordinate naturally sets $\psi_{,1} = \psi_{,2} = 0$ and the metric function $e^{-2v_3}=R/2$. Furthermore the only non-zero curvature scalar is $\phi_0=R$ which  also simplifies Eqs.~\eqref{Bondi5Sim}--\eqref{BondiI3Sim}. We now proceed to solve this set of field equations.

First, we note that Eq.~\eqref{Bondi5Sim}, $\theta_{,p}=\theta^2$ can be solved by setting  $\theta = -[p+f(u,\psi)]^{-1}$. We use the coordinate freedom  discussed in Sec.~\ref{sec:Choice1} to relabel the origin of the affine parameter $p$ by an arbitrary function of $f(u,\psi)$, to give
\be
\theta = - p^{-1} \,.
\ee
Performing one more integration using the commutation relation \eqref{chicomm1}, $v_{3,p} =p^{-1}$, allows us to obtain the metric function $v_3$; equivalently, the scalar curvature $R = 2 e^{-2v_3}$ is
\begin{align}
R&=c_1(u,\psi)p^{-2} \,.
\end{align}
The next field equation \eqref{Bondi3Sim}, $\epsilon_{,p}=0$ indicates that  $\epsilon = \epsilon(u,\psi)$ only. Once more a commutation relation can be integrated to obtain the metric function $v_2$. In this case Eq.~\eqref{rcomm2}, $(v_2 p)_{,p} = -2\epsilon p$ implies that 
\begin{align}
v_2 = c_2(u,\psi)p^{-1} - \epsilon \, p \,.
\end{align}

Before proceeding, let us use the fact that $\psi$ has been chosen as a coordinate and examine the simplified Bianchi identities~\eqref{BIANCI3b}. The first equation is trivially satisfied, while the third equation $(\ln R)_{,3}=4\epsilon$ places restrictions on the integration constants already obtained. Writing out the directional derivative in terms of coordinate derivatives and substituting in the solutions for $R$, $v_3$, $\epsilon$ and $v_2$ we have
\begin{align}
2\epsilon  - \frac{  c_{1,\psi} } {2\sqrt{c_1(u,\psi)}} p^{-1} + 2 c_2(u,\psi)p^{-2}  =0 \,.
\end{align}
This expression must vanish for all powers of $p$, which implies that $\epsilon=c_2=c_{1,\psi}=0$.
A consequence of this result is that $v_2=0$, and in addition $\eta=\gamma= \epsilon=0$, and $c_1=c_1(u)$ is a function of $u$ only. The final Bianchi identity gives the coefficient $\iota$,
\begin{align}
\iota =  - \frac{c_{1,u}}{2 c_1(u)} - v_1 p^{-1} \,.
\end{align}

Substituting the results obtained thus far into the third field equation, \eqref{BondiSimC1} we obtain $\delta_{,p}= -R/2$, and thus the rotation coefficient $\delta = c_1(u)/(2p) + c_\delta(u,\psi)$. The integration constant $c_\delta(u,\psi)$ is fixed to the value $c_\delta = c_{1,u}/(2c_1)$ by evaluating the field equation~\eqref{BondiSimC2}. The commutation equation~\eqref{rcomm1}, $v_{1,p}=-\delta$, yields an expression for the final metric function $v_1$,
\ba 
v_1 &=& - \frac{c_1(u)}{2} \ln p - \frac{c_{1,u}}{2c_1} p + c_3(u,\psi) \,.
\ea
 Since $v_2=0$, the final commutation equation~\eqref{rcomm3} sets $\zeta=c_{3,\psi}\sqrt{c_1/2}\,p^{-1}$ but the field equation~\eqref{BondiSim4} implies that $\zeta_{,p}=0$. Thus we have that $c_{3}=c_3(u)$ is a function of $u$ only and $\zeta=0$. All metric functions now depend on the variables $p$ and $u$ only. 

It is useful to observe that we still have the freedom to relabel the null hypersurfaces of constant $u$ as discussed in Eq.~\eqref{relabeling}. If we transform to a new  set of coordinates $(u',\ p', \psi)$ such that $u'=\int du \sqrt{c_1(u)/2}$ and $p'=p\sqrt{2/c_1(u)}$, the metric function $v_1'$ expressed on  the new coordinate basis can be written in the form $v_1' = -\ln p'+c_3'(u')$. It is always possible to choose a coordinate $u$ that labels the null hypersurfaces in a manner such that $c_1(u)=2$.  For the rest of the section we make this choice (omitting the primes). 

The final field equation~\eqref{Bondi2Sim} reduces to $\iota_{,2}  =  -\iota(\iota - \delta)$. Upon substituting in $\iota = -v_1 p^{-1}$, $\delta=p^{-1}$ and $v_1= -\ln p + c_3$ we find that $c_{3, u} = 0$, so that if we define $\ln A = c_3$, $A$ is merely a constant. 

In summary, when $\psi_{,i}$ is orthogonal to  a geodesic null vector $l_i = u_{,i}$ the metric functions are
\begin{align}
\label{eq:psiandnull}
v_1 &= \ln\left(\frac{A}{p}\right) ,& v_2 &=0,& v_3 &= \ln p , & R&=2p^{-2},
\end{align}
 and the rotation coefficients take on the values 
\begin{align}
\alpha=\beta=\epsilon=\eta=\gamma=\zeta =0 ,\notag\\
\theta = -\frac{1}{p},\, \, \iota = -\frac{v_1}{ p}, \, \, \delta = \frac{1}{p} \,.
\label{Triad3Sln}
\end{align}

Having successfully solved the 3D field equations in this special case, let us examine some of the implications the solution has for the 4D spacetime associated with the original axisymmetric problem. The 4D NP spin coefficients for the solution found in this section are easily obtain from Eqs.~\eqref{4spins} 
\begin{align}
\hat{\kappa }&= 
\hat{\nu }= \hat{\beta}=\hat{\epsilon }=  0,\, \, \, \, \, 
\hat{\tau }=  - \hat{\pi} =  -\hat{\alpha } = \frac{e^{\psi }}{\sqrt{2}p} ,  \notag\\
\hat{\rho  }&= \hat{\sigma }= -\frac{1 }{2p}, \, \, \, \, \,
\hat{\lambda }=\hat{\mu }  = -\frac{v_1}{2p} e^{2 \psi }  , \, \, \, \, \, 
\hat{\gamma }= -\frac{ e^{2 \psi }  }{2p}.  
\end{align}
These expressions for the spin coefficients show the corresponding 4D null congruence is also geodesic and affinely parametrized. By writing down the Weyl scalars using Eq.~\eqref{p0psi}
\begin{align}
\hat \Psi_0&= \hat \Psi_4= 0 , &
\hat \Psi_1&=-\frac{e^{ \psi } }{\sqrt{2}p^2 }, \notag\\
\hat \Psi_2&= \frac{e^{2 \psi }}{p^2} , & 
\hat \Psi_3&=\frac{e^{3 \psi }}{\sqrt{2}p^2 } \ln \left(\frac{A}{p}  \right), 
\label{p0psinull}
\end{align}
we observe that $\hat l^\mu$ is a principal null direction. 
A general classification scheme  using the spin coefficients will be discussed more fully in Sec.~\ref{sec:PNDClassification}.  For now it is useful to observe that  the fact that $\hat \rho = \hat \sigma$ and that $\hat l^\mu$ is a geodesic principal null vector indicates that this spacetime is a cylindrical-type Newman-Tamburino solution \cite{NewmanTamburino1962}. These spacetimes do not depend on any free functions of $u$. In fact, the metric of Eq.~\eqref{eq:psiandnull} corresponds to the particular case of a cylindrical-type Newman-Tamburino spacetime with one of the two arbitrary constants that parametrize these solution set to zero~\cite{Stephani2003}.

The axis conditions offer no additional constraints to this solution. As we approach the axis, we have $e^{2 \psi} \to 0$.  In order for the axis to be free of singularities, the elementary flatness condition, Eq.~\eqref{ElemFlat} must hold. In this particular case, we would have $h^{ij} \psi_{,i} \psi_{,j} = p^{-2}$, and so elementary flatness would require
\begin{align}
 \lim_{\lambda \rightarrow 0} e^{2 \psi} = p \,,
\end{align}
as we approach the axis. However, we then have that $p \to 0$ as we approach the axis, and we can see from the Weyl scalars~\eqref{p0psinull} that the spacetime is singular as $p \to 0$. We can conclude that this solution possesses a curvature singularity along the axis.

\subsection{Spacetimes with special optical properties}
\label{sec:SpecialSpaces}

In the next subsection we review known, special solutions to the axisymmetric, vacuum field equations, classifying them according to their optical properties. The classification will be made according to the properties of the null congruence that the tetrad vector $\hat l^\mu$ is tangent to in $\mathcal M$, and by extension the congruence that the triad vector $l^i$ is tangent to in $\mathcal S$. One of the benefits of the NP formalism is that the spin coefficients are directly related to the optical properties of a given spacetime. By seeking solutions with specified optical properties, many simplifications become possible, and the assumptions made are physically transparent. As detailed in Sec~\ref{sec:3dto4d}, by choosing to work in twist-free axisymmetry we have at least halved the complexity of the problem of solving the 4D field equations. The  4D spin coefficients computed from the 3D rotation coefficients are all real, which has immediate implications for the null congruence they describe in 4D, 
and we will discuss these implications here.

Consider a general  axisymmetric spacetime whose axial KV is twist-free and explore the behavior of the congruence of null curves that $\hat{l}^\mu$ is tangent to in $\mathcal{M}$. The expansion and twist of this congruence are described by the real and imaginary parts of the spin coefficient $\hat{\rho}$, respectively, and constitute the first two optical scalars. From Eqs.~\eqref{4spins} we know that 
\begin{align}
\label{4rho}
\hat \rho =  \frac{\theta}{2}, \,
\end{align}
and is manifestly real, and this shows that a spacetime with a twist-free, axial KV admits a twist-free null congruence (c.f. \cite{Kramer1968,Stephani2003}).  Furthermore, a twist-free congruence is hypersurface orthogonal, and so the fact that $\hat \rho$ is real implies that $\hat l_u$ is proportional to the gradient of some potential function $u$. As in Sec.~\ref{sec:Choice1}, in this case $\hat l^\mu$ is geodesic, since $l^\nu l_{\mu;\nu} = u^{,\nu} u_{;\mu \nu} = (u^{,\nu} u_{,\nu})_{,\mu}/ 2 = 0$, and we have
\begin{align}
\hat{\kappa} =e^{-\psi}\beta/\sqrt{2}= 0.
\end{align} 
The final optical scalar that characterizes the geometrical properties of the null congruence is the shear, which measures the distortion of the congruence and is given by
\be
\label{4sigma}
\hat \sigma =  \frac{\theta}{2} + \psi_{,1} \,.
\ee
An interesting property that arises from restricting the discussion to twist-free axisymmetric spacetimes is that, if the spacetime further has a null direction along which the derivative of $\psi$ vanishes, the associated congruence has $\hat \rho = \hat \sigma$. The vanishing of any of the other directional derivatives of $\psi$ gives analogous reductions to the 3D rotation coefficients, as can be seen by studying Eqs.~\eqref{4spins}. 

\begin{figure}[t]
\includegraphics[width=1.0\columnwidth]{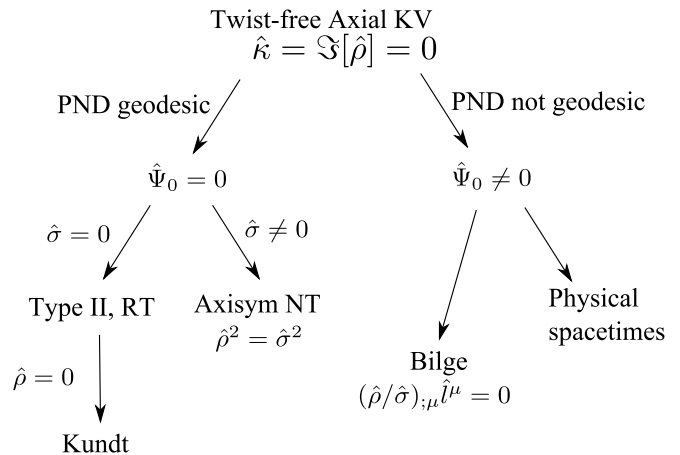}
\caption{Classification of spacetimes possessing a twist-free, axial Killing vector.  The abbreviations used in this figure can be interpreted as follows: Killing vector (KV), Principal null direction (PND), Robinson-Trautman (RT), and Newman-Tamburino (NT).}
\label{fig:AxiSymClass}
\end{figure}

A number of solutions to the field equations have been found which admit a geodesic, hypersurface orthogonal null congruence (where $\Im [\hat \rho] = 0$). We will discuss these spacetimes and their relation to the form of the field equations developed in this paper in the next subsection. Our focus will be on asymptotically flat solutions, which can represent isolated systems, and we will reserve our discussion of stationary, axisymmetric spacetimes until Section~\ref{sec:Split}. Figure~\ref{fig:AxiSymClass} gives a summary of the solutions we will be considering, along with the reductions and assumptions employed to yield the known results.

\subsection{Principal null geodesic congruences}
\label{sec:PNDClassification}

We have showed that in any twist-free, axisymmetric spacetime there exists a geodesic, hypersurface-orthogonal null congruence. If in addition, the tangent to this congruence is also assumed to be a principal null direction, so that $\hat \Psi_0=0$, the field equations can be solved and the exact metric expressions are known. To see this, it is easiest to work in a triad where $l^a$ is affinely parametrized with respect to the physical manifold $\mathcal M$. In this case, we have $\alpha = -2 \psi_{,1}$ on $\mathcal S$, $\hat \epsilon =0$ on $\mathcal M$, and Eqs.~\eqref{p0} and~\eqref{Bondi5} become
\begin{align}
\hat{\Psi}_0 &= 0 =\psi _{,1}^2+\psi_{,11},\label{poCon1}\\
\theta_{,1} &=(\theta +\psi_{,1})^2  +\psi_{,1}^2 \,. \label{poCon2}
\end{align}
Note that the directional derivative in these equations can be interpreted as a derivative with respect to the affine parameter $\tau$ in $\mathcal{M}$ and expressed as $f_{,1} = f_{,\tau}$.   For Eq.~\eqref{poCon1} the two solutions are
\begin{align}
\label{psiSolve}
\psi_{,1} = 0 && {\rm or}  && \psi_{,1}= \frac{1}{\tau + c_{1}(u,\chi)}\,.
\end{align}
Using Eq.~\eqref{poCon1}, Eq.~\eqref{poCon2}  can be rewritten as 
$\left(\theta+\psi_{,1}\right)_{,1} = \left(\theta+\psi_{,1}\right)^2$. The two solutions to this equation are
\begin{align}
\label{thetaSolve}
\theta+\psi_{,1} = 0 && {\rm or} &&\theta + \psi_{,1}= -\frac{1}{ \tau + c_{2}(u,\chi)}.
\end{align}

By substituting the solutions~\eqref{thetaSolve} and~\eqref{psiSolve} into the 4D spin coefficients, Eqs.~\eqref{4spins}, several distinct conditions on the optical scalars $\hat \rho$ and $\hat \sigma$ can be identified.  The expansion free case, where $\theta=\psi_{,1} = 0$, implies that $\hat{\rho}=\hat{\sigma}= 0$ and is known as the Kundt solution. On the other hand if $\psi_{,1}= 0$, or if $\theta=-\psi_{,1}$, we have that $\hat{\rho}=\pm\hat{\sigma}\neq 0$ which characterizes a cylindrical-type Newman-Tamburino spacetime.  In the case that $\psi_{,1}\neq 0$ and $\psi_{,1} + \theta \neq 0$ the transformation $\tau \to \tau' + f(u, \chi)$ can always be used to set $c_1=-c_2$. The optical scalars thus become $\hat \rho =- \tau/(\tau^2-c_1^2)$ and $\hat \sigma = c_1/(\tau^2-c_1^2)$. This case can be split into two distinct scenarios: If $c_1=0$, $\hat{\sigma} =0$ and the spacetime can be classified as a Robinson-Trautman spacetime. If on the other hand $c_1\neq 0$, then $\hat \sigma$ is nonzero and the 
result would again be a Newman-Tamburino spacetime, but of spherical type. However, in the case of axisymmetry, we can solve the field equations explicitly for a nonzero $c_1$, by integrating the hierarchy of field equations and matching powers (and transcendental functions) of $\tau$ at each step; the resulting spacetime has vanishing curvature and so is actually flat. This conforms to the known fact that the (non-trivial) spherical Newman-Tamburino solution can have at most only a single ignorable coordinate, namely the parameter labeling the 
null hypersurfaces $u$ \cite{Collinson1967,Stephani2003}; in other words, the spherical-type solutions are incompatible with axisymmetry. 

In the subsequent subsections we examine  the properties of the each of the spacetimes mentioned here in greater depth.

\subsubsection{Newman-Tamburino spacetimes}
\label{sec:NTspacetimes}

Newman-Tamburino spacetimes are characterized by the properties
\be
\label{NTReduction}
\hat \kappa = \hat \Psi_0 = 0 \,, \qquad \Im [\hat \rho] = 0 \, \quad \mbox{and} \quad \, \hat{\sigma} \neq 0
\ee
The metric for these solutions can be found explicitly \cite{NewmanTamburino1962,CarmeliBook} and except for special cases, the spacetimes  are of the generic Petrov Type I. The Newman-Tamburino solutions are divided into two classes, ``spherical type'' and ``cylindrical type'' solutions. The spherical type is the more general, requiring $ \hat \rho^2 \neq  \hat \sigma \hat \sigma^*$. The cylindrical type requires $ \hat \rho^2 = \hat \sigma \hat \sigma^*$. Only the cylindrical type  solutions admit a spatial KV \cite{Collinson1967}, and so are the case of interest for our study. Since all of the 4D spin coefficients are real in twist-free axisymmetry with our tetrad choice, these solutions require that the more restrictive condition $\hat{\sigma}= \pm \hat \rho$ holds, or equivalently in terms of the 3D quantities, 
\begin{align}
\mbox{either}\qquad \psi_{,1}=0  \qquad \mbox{or} \qquad \theta = -\psi_{,1}.
\end{align}

In the case with $\psi_{,1}=0$, the proper circumference of the orbits of the axial KV is unchanging along the geodesic null congruence. Therefore,  these solutions represent a spacetime that expands in the direction of the congruence. Note that the congruence is not simply frozen at a constant parameter $\tau$, since the expansion $\hat \rho$ is nonzero for these solutions. 

The Newman-Tamburino solutions do not correspond to spacetimes of physical interest. 
In addition, the metric functions have a simple polynomial dependence on
the coordinate $u$, which shows that the dynamics of these spacetimes are very simple. 
Since the properties of these solutions are well understood \cite{NewmanTamburino1962,CarmeliBook} and the general derivation of the metric functions is lengthy, we do discuss these spacetimes further. Instead, recall that the solution found Sec.~\ref{NewSln} is a special case of the cylindrical type Newman-Tamburino solutions, where in addition to $\psi_{,1}=0$ we assumed that $\psi_{,2}=0$. The solution found in Sec.~\ref{NewSln} is parametrized by one constant $A$, while the general cylindrical-type metric contains two arbitrary constants~\cite{NewmanTamburino1962, CarmeliBook, Stephani2003}.

\subsubsection{Robinson-Trautman spacetimes}
\label{sec:RTspacetimes}

The second class of solutions where the congruence is geodesic, principal null, shear-free, ($\hat{\kappa}=\hat{\sigma}=\hat{\Psi}_0=0$) and expanding ($\hat \rho \neq 0$) is known as the Robinson-Trautman spacetimes \cite{RobinsonTrautman1960,RobinsonTrautman1962}.  The solutions to the field equations in such a case can be reduced to a single nonlinear partial differential equation and have been well studied, see e.g. \cite{Jaramillo2011} and the references therein. While these equations have been used to study radiating sources in an exact, strong field setting, they do not represent physical systems, such as a stage of head-on collision of black holes. 

In terms of the 3D rotation coefficients, the conditions for the Robinson Trautman solutions are
\begin{align}
\beta&=0, & \theta &= -2\psi_{,1} \neq 0. \label{RTreduction}
\end{align}
Note that since these vacuum spacetimes admit a shearfree geodesic null congruence, the Goldberg-Sachs theorem  \cite{GoldbergSachs2009,NewmanPenrose1962,Stephani2003} states that they are algebraically special, so that $\hat \Psi_0= \hat \Psi_1=0$.  

We can verify this directly from our the 3D equations. To do so, we use an affine parametrization with respect to the physical manifold $\mathcal M$, as discussed in Appendix~\ref{app:affinereal}. This sets $\alpha=-2\psi_{,1}$. Substituting $\theta=-2\psi_{,1}$ into the field Eq.~\eqref{poCon2}  immediately gives $\hat \Psi_0 = 0$. 
Showing that $\hat \Psi_1 = 0$ can also be made to vanish requires more finesse. For this, we use Eq.~\eqref{p0} with $\beta=0$ to obtain an expression for $\hat \Psi_1$,
\begin{align}
\hat \Psi_1
&= \frac{e^{\psi }}{\sqrt{2}} \left[\psi _{,1} \left(\psi _{,3}+\eta \right)+\psi
   _{,13}\right]. 
\label{p1comm}
\end{align}
In Eq.~\eqref{p1comm} the commutation relation~\eqref{COMM} has been used to interchange to order of differentiation on $\psi$. The field equation~\eqref{Bondi3}, specialized to the case where $\theta = \alpha=-2\psi_{,1}$ and $\beta=0$, can be written as
\begin{align}
\eta _{,1}  &= -2 [\psi _{,1} (\psi _{,3} + 2\eta)+ \psi _{,13}] .
\label{etaRT}
\end{align}
Observe that with the simplifications so far $(\hat \kappa = \hat \sigma = \hat \epsilon = \hat \Psi_0 = 0)$, the two 4D Bianchi identities ~\cite{MathTheoryofBlackHoles, Stephani2003} that govern only the 
directional derivatives of $\hat \Psi_0$ and $\hat \Psi_1$ can be expressed in our notation as
\begin{align}
\hat \Psi _{1,\hat{1}} &=4 \hat{\rho } \hat \Psi _1 
& \hat \Psi _{1, \hat{3}} &=(2 \hat{\beta } +4 \hat{\tau })\hat \Psi _1
\label{Bianchi4DPND}
\end{align} 

The corresponding expressions in terms of the 3D quantities associated with the triad discussed in Appendix~\ref{app:affinereal}, where $\epsilon=-\gamma=-\eta -2\psi_{,3}$,  is
\begin{align}
\label{eq:Bianchi3DPND}
\ln (e^{4\psi}\hat \Psi_1)_{,1} &= 0,& \ln ( e^{4\psi}\hat \Psi_1)_{,3} =-5\eta.
\end{align}
If we apply the directional derivative $l^i \partial_i$ to the second of Eqs.~\eqref{eq:Bianchi3DPND}, use the commutation relations~\eqref{COMM} to switch the directional derivatives, and repeatedly use Eqs.~\eqref{eq:Bianchi3DPND}, we obtain the equation $\eta_{,1}=-2 \eta \psi_{,1}$. Substitution of this into Eq.~\eqref{etaRT} implies that $ \psi_{,1} \left(\psi _{,3}+\eta \right)=-\psi_{,13}$, and thus $\hat \Psi_1 =0$.

To complete the discussion of the Goldberg-Sachs theorem for twist-free axisymmetric spacetimes,  note that substituting the condition $\hat \Psi_0= \hat \Psi_1=0$ into the 4D Bianchi identities immediately implies that $\hat \kappa=\hat \sigma=0$, and thus the existence of a geodesic shear-fee null congruence.

\subsubsection{Expansion-free spacetimes}
\label{sec:Kundt}

We now consider the last case in our class of spacetimes that admit a geodesic principal null congruence, namely spacetimes that are both shear-free and expansion-free ($\hat \kappa = \hat \sigma = \hat{\rho}=\hat \Psi_0 = 0$). 
These metrics are Kundt solutions  and have been extensively studied~\cite{Kundt1961,McIntosh1990,Stephani2003}.
In terms of the 3D quantities, the Kundt metrics have the properties  $\beta=\theta = \psi_{,1} =0$.  Since $\psi_{,1} = 0 $, setting $\alpha=0$ implies that the geodesics can be affinely parametrized in both $\mathcal{M}$  and $\mathcal{S}$ simultaneously.
The fact that $\psi_{,1}=0$ also greatly simplifies the 3D curvature scalars, setting 
$\phi_5=\phi_4=\phi_3 =0$. Observe that choosing $l_i=-u_{,i}$ and setting  $\theta=0$ in addition to $\alpha=\beta =0$ and $\eta=-\epsilon$ implies that $D^2 u =0$, so that $u$ is a harmonic coordinate. This can be seen by taking the trace of Eq.~\eqref{Derivlnull}.

In the Kundt metrics, many of the rotation coefficients and metric functions are independent of the affine parameter $p$, which simplifies the calculations. The solution of the field equations in this case provides another simple and illustrative example of the integration of the 3D field equations. We now proceed to solve the hierarchy of field equations~\eqref{Bondi5Sim}--\eqref{BondiI3Sim} in Sec.~\ref{sec:Choice1}, in conjunction with the commutation relations applied to coordinates.  
 
In addition to $\psi_{,1}=0$, which implies that $\psi$ is independent of the affine parameter, Eqs.~\eqref{Bondi3Sim} and~\eqref{BondiI1Sim} give $\epsilon_{,1} = \iota_{,1} = 0$. Further, the commutation relation~\eqref{chicomm1} yields $v_{3\ ,1} = 0$. Thus $v_3$, $\psi$, $\epsilon$ and $\iota$ are functions of only two coordinates $u$ and $\chi$. Note that it is always possible to use the coordinate freedom to define a new coordinate $\chi' = g(u,\chi)$ such that  $v_3'=0$  in the new coordinate system, by choosing $g_{,\chi} = e^{ v_3}$. Since $\iota =v_{3, 2}$ by Eq.~\eqref{chicomm1}, it is possible to set 
\begin{align}
v_3=\iota=0.
\end{align}
The next metric function, $v_2$, can be found using the commutation relation~\eqref{rcomm1}, $v_{2,\ p} =-2\epsilon $. Since $\epsilon = \epsilon(u,\chi)$ only we can integrate the equation to yield \mbox{$v_2 = -2\epsilon p + c_1(u,\chi)$}. The ability to shift the origin of the affine parameter by a function of $u$ and $\chi$ by defining a new parameter  $p'=p+g(u,\chi)$ allows us to set $c_1=0$, and thus
\begin{align}
v_2 = -2\epsilon(u,\chi) p \,.
\end{align}
In obtaining the expressions for $v_3$ and $v_2$ we have used up most of the coordinate freedom with respect to the $\chi$ and $p$ coordinates, except for transformations of the type $\chi' =\chi + f_1(u)$ and  $p'=p+f_2(u)$ which do not spoil any of the simplifications so far.

In order to learn more about the function $\epsilon$ consider the harmonic condition on $\psi$, Eq.~\eqref{LapOp}. With the reductions employed thus far, Eq.~\eqref{LapOp} reduces to $\psi_{,33}  = 2\epsilon \psi_{,3}$. Furthermore the field equation~\eqref{BondiSimC2} simplifies to the expression $\epsilon_{,3} = \epsilon^2+\psi_{,3}^2$. Taking the sum and the difference of these two equations we obtain
\begin{align} 
(\epsilon + \psi_{,3})_{,3} &=  (\epsilon + \psi_{,3})^2,& (\epsilon-\psi_{,3} )_{,3} &=  (\epsilon-\psi_{,3})^2.   \label{epsKundt}
\end{align}
For  functions of the form $f=f(u,\chi)$ the fact that $v_3=0$, implies that the directional derivative $f_{,3}$ becomes the coordinate derivative $f_{,\chi}$, for these functions it is further true that $f_{,2}=f_{,u}$. 

Before solving Eq.~\eqref{epsKundt}, note that Eq.~\eqref{BondiSimC1} can be  reduced to
 $\delta_{,1} = \epsilon ^2- \psi_{,3}^2$, indicating that $\delta_{,1}$ is independent
of $p$ and allowing us to integrate the equation to find an explicit expression for $\delta$,
\begin{align}
\delta &=  (\epsilon ^2- \psi_{,3}^2) p - w_\delta(u,\chi) \,,
\label{delepss}
\end{align} where $w_\delta$ is an integration constant.
In addition, $\delta$ has to satisfy  the difference of Eqs.~\eqref{BondiI3Sim} and~\eqref{BondiSim4}, $ \delta_{,3} +  \epsilon_{,2}  =  -2\psi_{,2}\psi_{,3}$. Substituting in Eq.~\eqref{delepss} and evaluating, we obtain the following constraint on $w_{\delta}$:
\begin{align}
w_{\delta,\chi} =    \epsilon_{,u}  +2\psi_{,u}\psi_{,\chi}.
\end{align}
Integrating the commutation relation~\eqref{rcomm1}, $\delta=-v_{1,1}$, the metric function  $v_1$ can found to have the form 
\begin{align}
v_1= -\frac{1}{2} (\epsilon ^2- \psi_{,3}^2) p^2 + w_\delta(u,\chi)p + w_1(u,\chi) \,,
\label{v1eps}
\end{align}
where $w_1$ is a new integration constant. To see what additional constraints are to be imposed on the integration constants by the field equations, we express the only remaining coefficient $\zeta$ in terms of the metric functions using Eq.~\eqref{rcomm3} with $v_3=0$, 
\begin{align}
\zeta = v_{2, 2}+v_{1, 3} =  ( 2  \psi_{ ,u} \psi_{ ,\chi}- \epsilon_{,u})p + w_{1, \chi} +2 \epsilon w_1.
\end{align}
Substituting this expression for $\zeta$ into Eq.~\eqref{Bondi2Sim} and expressing the result using coordinate derivatives  yields the constraint
\begin{align}
(w_{1,\chi}+ 2 \epsilon w_{1,\chi} )_{,\chi}   = 2\psi_{,u}^2.
\end{align}
This is the final condition that has to be satisfied. 

All that remains now is to explicitly integrate Eq.~\eqref{epsKundt}. There are three possible cases:
\begin{align}
\epsilon &=0& {\rm and} && \psi_{,3}&=0 \label{psieps1} \,,\\
\epsilon &= \pm \psi_{,3}& {\rm and} && 2\psi_{,3} &=\mp \frac{  1  }{\chi-c_2(u)} \label{psieps2} \,,\\
\epsilon + \psi_{,3} &= -\frac{1}{\chi - c_2(u)}, & {\rm and} && \epsilon- \psi_{,3}& = -\frac{1}{\chi - c_3(u)} \,,
\label{psieps3}
\end{align}
where the functions $c_i(u)$ are arbitrary functions of $u$ only.
For the remainder of the discussion we shall concentrate on the most generic case, Eqs.~\eqref{psieps3}.
Taking the difference of the Eqs.~\eqref{psieps3} and integrating one more time gives an expression for $\psi$,
\begin{align}
\psi = \frac{1}{2} \ln \left|\frac{\chi-c_3(u)}{\chi-c_2(u)}  \right| + c_4(u). 
\label{PSISOL}
\end{align}
The Ricci scalar $R=2 \psi_{,\chi}^2$ and the coefficient $\epsilon$ are 
\begin{align}
R &= \frac{1}{2} \left[\frac{c_3-c_2  }{(\chi-c_2)(\chi-c_3)    }   \right]^2 \,, \label{Rkundt}\\
\epsilon &=  \frac{1}{2} \left[\frac{c_3+c_2 -2\chi  }{(\chi-c_2)(\chi-c_3) }   \right] \,.
\end{align}
It is straightforward to verify that the solution given here  matches that presented in the original derivation of Kramer and Neugebauer \cite{Kramer1968}, who also use the conformal 2+1 decomposition. The solution provided by Hoenselaers using the triad method \cite{Hoenselaers1978b} excludes the twist-free case, but McIntosh and Arianrhod \cite{McIntosh1990} show (after making some corrections) that it matches the above form.

We now investigate the condition of elementary flatness to see if it provides any additional constraints on the free functions. The axis is located at those points where
\begin{align}
e^{2 \psi} = e^{2 c_4} \frac{\chi -c_3}{\chi - c_2} \to 0 \,,
\end{align}
which occurs when $\chi \to c_3$. The condition for elementary flatness, Eq~\eqref{ElemFlat} or equivalently $ e^{4\psi}R\rightarrow 2 $, can be expressed as 
\begin{align}
\lim_{\lambda\to 0}\frac{e^{4 c_4}}{4(c_3-c_2)^2}= 1 \,.
\end{align}
The requirement that the axis be free of conical singularities does thus provide a further constraint on the free functions, relating $c_4$ to $c_3-c_2$. 

Let us further explore the properties of the spacetime by computing the Weyl scalars. Since the Kundt spacetime is geodesic and shear free, it is algebraically special, with  $\hat \Psi_0=\hat \Psi_1 = 0$.
The other Weyl scalars can be obtained in terms of the triad variables by making use of Eqs.~\eqref{p0},  
\ba
\label{kp2}
\hat \Psi_2 & = &e^{2\psi}   \psi_{,\chi} \left(  \psi_{,\chi} +  \epsilon  \right)  \,, \\
\label{kp3}
\hat \Psi_3 & = &  \frac{e^{3\psi}}{\sqrt{2}}\left( \psi_{,\chi u} + \epsilon \psi_{,u} + 3 \psi_{,u}\psi_{,\chi} \right) \,, \\
\label{kp4}
\hat \Psi_4 & = & e^{4 \psi}\left( \psi_{,u u} + 3 (\psi_{,u})^2 - \delta \psi_{,u} - \zeta \psi_{,\chi} \right) \,,
\ea
This shows that the metric is in general of Petrov Type II. In the case where $c_2 = c_3$, we can see from our solution in Eq.~\eqref{PSISOL} that $\psi = c_4(u)$, and so using  Eq.~\eqref{kp2} we have that $\hat \Psi_2 = 0$. In this case the metric is of Type III. In case 2 mentioned in Eq.~\eqref{psieps2},  $\epsilon=-\psi_{\chi}$ also implies $\hat \Psi_2 =0$ and that the spacetime is also of Type III. Finally, we consider the special case 1 of Eq.~\eqref{psieps1} where \mbox{ $\epsilon=\psi_{,\chi}=0$}, for which the vanishing of the Weyl scalars \mbox{$\hat \Psi_2 = \hat \Psi_3 = 0$} implies that the metric is Type N. This completes the full classification of axisymmetric spacetimes that admit a geodesic principal null congruence.

\subsection{Other axisymmetric, twist-free spacetimes}
\label{sec:GeneralSln}

Given the complete classification of spacetimes with special optical properties, it is clear that dynamical spacetimes of physical interest are not represented in this class of solutions. We must necessarily consider spacetimes whose geodesic, hypersurface-orthogonal congruence is not a principal null congruence, and $\Psi_0 \neq 0$. 

Known exact solutions that do not admit a geodesic principal null congruence include the twist-free solutions with the restriction $ (\hat \rho/\hat \sigma )_{;\mu} \hat l^\mu = 0$ which have been found by Bilge~\cite{Bilge1989}. However, Bilge and G\"{u}rses also showed that this class of spacetimes, though generally of Type~I, is not asymptotically flat \cite{Bilge1986}. The class of twist-free spacetimes that obey $(\hat \rho/\hat \sigma )_{;\mu} \hat l^\mu = 0$ includes the vacuum Generalized Kerr-Schild (GKS) metrics, when they have twist-free congruences. GKS metrics are of the form
\be
\label{GKS}
 g_{\mu \nu} = \tilde g_{\mu \nu} + H \hat l_\mu \hat l_\nu \,,
\ee
where $g_{\mu \nu}$ and $\tilde g_{\mu \nu}$ are both solutions to the vacuum field equations; $\hat l_\mu$ is a geodesic null vector with respect to both metrics (it forms the twist-free congruence); and $H$ is some function on spacetime \cite{Bilge1989}. Gergely and Perj\'{e}s showed that the GKS spacetimes which admit twist-free congruences are the homogeneous, anisotropic Kasner solutions \cite{GergelyPerjes1994} (although two of the three constants that classify the Kasner spacetime must be set equal in order for there to be a single rotational symmetry and so an axial KV).

Another method to find solutions with one spatial KV, which we mention for completeness, is to use a different triad choice than those discussed here. The idea, as developed by Perj{\'e}s in the study of stationary spacetimes \cite{Perjes1970} where the KV is timelike and the conformal 3D quotient space of the Killing orbits is spacelike, is to orient one of the triad legs along an {\it eigenray}.  The eigenray is a curve defined such that if its spatial tangent vector is geodesic in the 3D quotient space, it is the projection of a null ray in the full 4D space. The triad formalism of Perj\'{e}s adapts naturally to the case of a spatial KV rather than a timelike KV, in which case the eigenray is timelike and the triad is chosen such that
\begin{equation}
\psi_{,1} = \psi_{,2}\,, \qquad \psi_{,3} = 0 \,.
\end{equation}
If the eigenray vector is, in fact, assumed to be geodesic, then solutions to the field equations can be found. In stationary spacetimes these were found in~\cite{Perjes1970,KotaPerjes1972}, and in the case of a spacetime with a spatial KV they were found by Luk{\'a}cs \cite{Lukacs1983}. When the geodesic eigenray is shearing, the solutions are Kasner (as in the case of the twist-free GKS spacetimes), and when it is not shearing they are Type D, and so very restricted.

Thus, future studies which aim to extract physical information about isolated dynamical, axisymmetric spacetimes will have to focus on general spacetimes, where none of the principal null directions are geodesic, and which do not fall within Bilge's class of metrics.

\section{Stationary Axisymmetric Vacuum Spacetimes}
\label{sec:Split}

As discussed in Sec.~\ref{sec:Intro}, the SAV field equations have been completely solved, in the sense that techniques exist that can generate any SAV solution. In this section, we briefly discuss the case of SAV spacetimes in the context of the conformal 3D metric and Ernst equation of Sec.~\ref{sec:ErnstPot}. We then specifically focus 
on  the twist-free case, and catalog the simplifications to the triad formalism discussed in Sec.~\ref{sec:TriadEqns} when static spacetimes are considered. We recast the static metric in the form discussed in Sec.~\ref{sec:Choice2} with $\psi$ chosen as a coordinate. Finally, as an illustrative example, we explore the properties of the Schwarzschild metric re-expressed using $\psi$ as a coordinate, and we derive the scaling of the rotation coefficients as the axis is approached.  We compare the results to the general expansions derived in the time-dependent case in Appendix~\ref{sec:ROTCOEFPSICOORD}.

\subsection{SAV spacetimes with twist}
\label{sec:SAVtwist} 

A stationary, axisymmetric spacetime possesses two KVs. One is spacelike, with closed orbits, which we denote $\xi^\mu$ as in previous sections. The other is timelike, and we call it $\eta^\mu$. The ignorable coordinate associated with $\eta^\mu$ we denote $t$, and we will later work in a gauge were $\eta^i = \delta^i_t$.  The KV $\eta^\mu$ places additional restrictions on the spacetimes considered up until now.

Carter showed that for axisymmetric spacetimes which are asymptotically flat, the two KVs commute~\cite{Carter1970}. In addition, in vacuum, the pair of KVs are surface forming, which allows us to use coordinates $t$ and $\phi$ in order to separate the metric into a pair of two dimensional blocks. Since any two metric is conformally flat, we can further choose coordinates $\rho$ and $z$ such that these coordinates are isotropic on their block~\cite{Wald1984}, and express the metric in the Weyl canonical form,
\begin{align}
\label{WeylForm}
ds^2 =&e^{-2U} \left[ e^{2k} \left( d \rho^2 + dz^2 \right) + \rho^2 d\phi^2 \right]  - e^{2U} (dt - A d\phi)^2  \,,
\end{align}
where the functions $U,\, A,\, k$ are functions of $\rho$ and $z$ only.

This form of the metric is  extensively used for exploring the SAV field equations. Associated with  the metric functions is an Ernst potential, which is often used in constructing solutions to the field equations. It should be noted that the Ernst potential usually employed in the discussion of SAV spacetimes in the literature $\mathcal E_{(\eta)}$ is associated with the timelike KV $\eta^{\mu}$, and is not the Ernst potential $\mathcal E$ associated with the axial KV $\xi^\mu$ introduced in Eq.~\eqref{EQERN}. In the SAV context, $\mathcal E_{(\eta)}= e^{2U} + i \varphi$ where $e^{2U}=- \eta^\mu \eta_\mu$ and $\varphi_{,\mu} = \epsilon_{\mu \nu \sigma \rho} \eta^\nu \eta^{\rho;\sigma}$. In order to make a direct connection with the notation used in this paper,  we redefine the metric functions and cast the metric of Eq.~\eqref{WeylForm} in the form
\begin{align}
\label{AxialForm}
ds^2 = & e^{2 \psi} \left( d\phi - B dt \right)^2 + e^{-2 \psi}\left[ e^{2 \gamma} \left(d\rho^2 + dz^2  \right) - \rho^2 dt^2 \right] \,.
\end{align}
with $e^{2 \psi} = \lambda = \xi^\mu \xi_\mu$ as before. The metric functions of the two forms are related by
\begin{align}
B & = A e^{2 U - 2 \psi} , & e^{2 \gamma} = &  e^{2 k + 2 \psi - 2 U} , \notag \\
 e^{2 \psi} & = \rho^2 e^{-2U} - A^2e^{2U} .
\end{align}
Using the metric \eqref{AxialForm}, the conformal 3-metric is given by
\begin{align}
h_{ij} = & {\rm diag} [ -\rho^2, e^{2 \gamma}, e^{2\gamma} ] \,.
\label{Hcanonical}
\end{align}
It turns out that for the line element~\eqref{AxialForm}, the Ernst equation for $\mathcal E = \lambda + i \omega$ and field equations for $\psi, \gamma,$ and $B$ can be solved in an identical manner to the more usual SAV case, where  $U, \ k,$ and $A$ are sought and when $\mathcal E_{(\eta)} = f + i \varphi$. As derived previously, the field equations reduce to a single nonlinear equation for $\mathcal E$, namely the first equation in Eqs.~\eqref{EQERN}. Written out in the coordinates associated with metric~\eqref{AxialForm}, the equation for $\mathcal E$ becomes
\begin{align}
\label{SAVErnstEq}
\nabla^2 \mathcal E & \equiv \rho^{-1} (\rho \mathcal E_{,\rho})_{,\rho} + \mathcal E_{,zz} = 2\frac{(\mathcal E_{,\rho})^2 + (\mathcal E_{,z})^2}{\mathcal E + \mathcal E^*} \,.
\end{align}
Here we have defined $\nabla^2$ as the usual flat space Laplace operator, in cylindrical coordinates.
When working with SAV spacetimes it is often useful to introduce the complex coordinate $\zeta = (\rho + i z)/\sqrt{2}$ to express the equations more compactly on the complex plane.  In these coordinates $\partial_\zeta = (\partial_\rho - i \partial_z)/\sqrt{2}$.
Once  $\mathcal E$ is known the metric functions for $\gamma$ and $B$ can be found by making use of 
the line integrals 
\begin{align}
\label{GammaEq}
\gamma_{,\zeta} & = \frac{\sqrt{2} \rho(\mathcal E)_{,\zeta} (\mathcal E^*)_{,\zeta} }{(\mathcal E + \mathcal E ^*)^2}  \,,\\
\label{BEq}
B_{,\zeta} & = -\frac{2 \rho  (\mathcal E - \mathcal E^*)_{, \zeta}}{(\mathcal E + \mathcal E ^*)^2} \,.
\end{align}

\subsection{Twist-free SAV spacetimes}
\label{sec:SAVnotwist}

We now specialize to the case of static axisymmetric spacetimes, which are twist-free. The KV $\xi^\mu$ is hypersurface orthogonal, and thus  $\omega=A=B=0$. The Ernst equation~\eqref{SAVErnstEq} reduces to the 3D cylindrical Laplacian applied to $\psi$,
\begin{align}
\nabla^2 \psi =  \rho^{-1}(\rho \psi_{,\rho})_{,\rho} + \psi_{,zz}= 0 \,. 
\label{lappsi}
\end{align}
The two metric functions $\gamma$ and $\psi$ are related to the metric functions in Weyl canonical metric~\eqref{WeylForm} by
\begin{align}
\psi =& \ln \rho - U \,, & \gamma &= \ln \rho + k - 2U \,.
\end{align}
Since $\ln \rho$ is a homogeneous solution to the cylindrical Laplacian, we see that if $U$ obeys the cylindrical Laplace equation, then  $\psi$ does also, and vice versa. All homogeneous solutions to the cylindrical Laplace equation are known (for example in terms of Legendre polynomials). The only difficulty is in specifying boundary conditions whose corresponding solution gives a spacetime with the desired physical interpretation. We will examine this issue using specific examples.

\begin{figure}[t]
\includegraphics[width=0.95 \columnwidth]{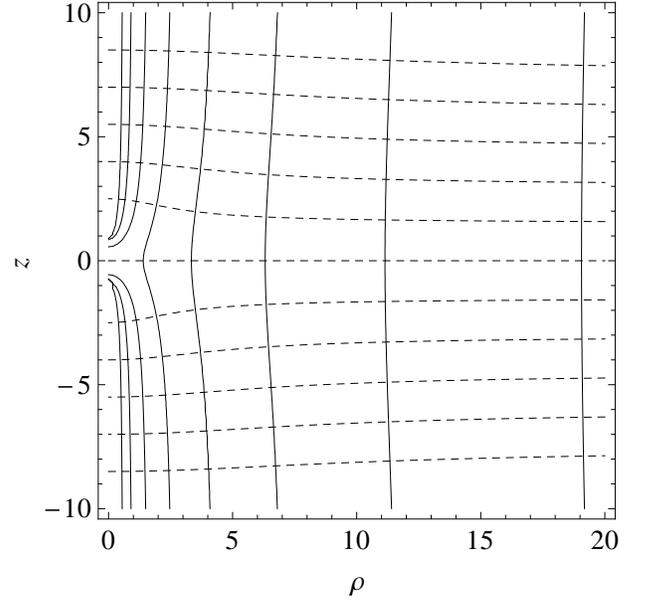}
\caption{Orthogonal coordinates $\psi$ (solid lines) and $s$ (dashed lines) plotted for a single black hole of mass $M=1$ in canonical Weyl coordinates $(\rho,z)$.}
\label{fig:ONEBH}
\end{figure}

In the context of the line element~\eqref{WeylForm}, a single Schwarzschild black hole is generated by a line charge of length $2M$ placed on the axis. With this as the boundary condition, the Laplace equation for $U$ can be solved; in Weyl coordinates, the solution is
\begin{align}
\label{WeylSchw}
e^{2 U} & =  \frac{r_+ + r_- - 2 M}{r_+ + r_- + 2M} \,, \\
 e^{2 \gamma}  & = \rho^2 \frac{(r_++ r_- + 2 M )^3}{4 (r_+ + r_- - 2M)r_+ r_-}\,, 
\end{align}
where $r_\pm^2  = \rho^2 + ( z \pm M)^2$ and $e^{2\psi}=\rho^2 e^{-2U}$. The function $\psi$ for the Schwarzschild metric in $(\rho, z)$ coordinates is plotted in  Fig.~\ref{fig:ONEBH}. The black hole lies on the axis between $z/M=\pm 1$, where the equipotential lines of $\psi$ meet the axis at almost right angles. Further from the axis the surfaces of constant $\psi$ rapidly approach surfaces of constant cylindrical radius. For the single black hole, as we will discuss in the next section, the appropriate axis conditions guaranteeing elementary flatness are satisfied at $\rho = 0$ outside the line charge.

Multiple static black holes solutions can be found by placing multiple line charges along the axis~\cite{Israel1964}. Mathematically this corresponds to the superposition of multiple $U_i$ potentials given in Eq.~\eqref{WeylSchw}, centered at positions $z_i$ and with masses $M_i$. The corresponding $\psi$ potential is then constructed using $\psi =  \ln \rho- \sum_{i=1}^m U_i$. Given the $\psi$ potential of a superimposed set of black holes, we can always construct a solution for the metric function $\gamma$ satisfying Eq.~\eqref{GammaEq}, which for static spacetimes becomes 
\begin{align}
\gamma_{,\rho} & =\rho [(\psi_{,\rho})^2 - ( \psi_{,z})^2] \,, & \gamma_{,z} & = 2\rho \psi_{,\rho} \psi_{,z}\,. \label{EQLine}
\end{align}
Despite the fact that a solution can be found, it is not possible to find a solution for which  elementary flatness holds along every connected component of the axis. From Eq.~\eqref{WeylForm} we see that if $k\neq0$ along a component of the axis, then elementary flatness does not hold there. Similarly, comparison of the line element~\eqref{AxialForm} with Minkowski space written with $\psi$ as a coordinate [see Eq.~\eqref{eq:MinkMetric} in Appendix~\ref{sec:Minkpsi}], shows that elementary flatness requires $\gamma = 2\psi - \ln \rho$ along the axis. The conical singularities that result when the elementary flatness  condition 
is not  met are interpreted as massless strings or struts which hold the black holes apart, keeping them stationary. The fact that these singularities always appear in static, axisymmetric black hole solutions is in line with our intuition that black holes should attract each other, and can never remain stationary at a fixed separation. Approximate models based on using the tension on the strut to evolve binaries in a head-on collision scenario can be found e.g. in~\cite{PhysRevD.52.816, PhysRevD.10.3171, Kates86}. In the time-dependent case the harmonic equation governing $\psi$ once again suggests that a generalized superposition principal could hold, at least on the initial time slice. 

In the time-dependent case, however the counterpart of equations~\eqref{EQLine} defining the gradient of the potential $\gamma$ do not exist. Instead the second order elliptic equations for $\gamma$ associated with the  initial value problem must be solved. This  allows the freedom to impose the boundary conditions guaranteeing elementary flatness. We thus expect that in dynamical spacetimes the conical singularities can be removed. We have yet to fully explore the time-dependent equations in the framework provided by this paper. 

\subsection{Static Axisymmetric spacetimes with $\psi$ chosen as a coordinate }
\label{staticpsi}

In order to explore SAV spacetimes within the framework of more general axisymmetric spacetimes, we transform to a coordinate and triad system adapted to the scalar field $\psi$. To do this, consider the coordinate system $(t,\  s,\ \psi)$, where $t$ remains the ignorable coordinate associated with the timelike KV, $\psi$ is a potential that obeys Eq.~\eqref{lappsi} and the coordinate $s$  is orthogonal to $\psi$ and defined by
\begin{align}
\partial_{\rho}s &=  \rho \partial_z \psi & \partial_{z} s &= - \rho \partial_{\rho} \psi. \label{sGrad}  
\end{align}
The integrability condition for $s$  is guaranteed by the vanishing of the Laplace equation~\eqref{lappsi} for $\psi$. Changing coordinates from $(t,\, \rho,\, z)$ to $(t,\, s,\, \psi)$ yields a metric in the form
\begin{align}
 h_{ij}={\rm diag}[ -\rho^2,\, 2/S,\, 2/R] \,, 
\label{hsav}
\end{align} 
where the functions $S$ and $R$ are the normalization factors defined by $S = 2 s_{;i}s^{;i}$ and $R =2 \psi_{;i}\psi^{;i}$. Note that from the definitions in Eq.~\eqref{sGrad} and the metric~\eqref{Hcanonical}, it follows that $S = \rho^2 R$. This is consistent with the result~\eqref{epsint} obtained by integrating the equations for the coefficient $\epsilon$, in the special case where $g(t,s)=1$. This means that once $R$ and $\rho(s, \psi)$ are known the entire metric in $(\psi,\, s)$ coordinates is known. In terms of the Weyl canonical coordinates $(\rho,\, z)$, we have that $R= 2(\psi_{,\rho}^2+\psi_{,z}^2)e^{-2\gamma}$, and in addition from the static field equations for $\gamma$, Eq.~\eqref{EQLine}, we also have the identity $R^2 = 4 e^{-4\gamma}(\gamma_{,\rho}^2+\gamma_{,z}^2) / \rho^2  $.

Let us now take a closer look at the rotation coefficients associated with the metric~\eqref{hsav}. When we consider the more general metric in Eq.~\eqref{Metpsi}, make the 
triad choice~\eqref{psitriadup}, and substitute both that $l_t=n_t$ and that the metric functions are independent of time into Eqs.~\eqref{ROTcoefsMet}, we find  
\begin{align}
\alpha &=\delta, &  \gamma &=-\epsilon , & \beta &=-\zeta ,&\eta&=0 ,
&\theta&= \iota. \label{statRot}
\end{align}
Furthermore note that for the choice of triad vectors~\eqref{psitriadup}, the directional derivatives  obey the relation $f_{,2}=-f_{,1}$ for any time-independent function $f$.

Noting that $h_s^2 = 1/(\rho^2 R) $ and that $\varrho = \rho$ in the static case, 
the remaining independent rotation coefficients can be expressed in terms of only the functions $R$, $\rho$ and their derivatives as follows,
\begin{align}
\epsilon &= \frac{R_{,\psi}   }{4\sqrt{2R}  },
&\beta &=-
\frac{\sqrt{R} \rho_{,\psi}}{\sqrt{2}\rho } - \epsilon,  \notag\\
 \theta &=
\frac{\rho R_{,s}}{4\sqrt{R}}, 
& \alpha &=
-\frac{\sqrt{R} \rho_{,s}  }{2}.
\label{ROTStat2}
\end{align}

To further illustrate the implications of choosing $\psi$ as a coordinate, we now turn to the concrete case of the Schwarzschild metric. Computationally it is useful to  introduce prolate spheroidal coordinates $(x,\, y)$ related to the Weyl coordinates by the transformation
\begin{align}
\rho^2 &= M^2(x^2-1)(1-y^2),\, &z&= M xy.
\end{align}
In spheroidal coordinates,  $D^2\psi=0$ is equivalent to requiring that
\begin{align}
 \partial_{x}\left[ (x^2-1)\partial_x  \psi\right]  + \partial_y \left[ (1-y^2)  \partial_{y} \psi  \right]=0 .
\label{lapxy}
\end{align}
In terms of the $(x,\, y)$ coordinates, the norms of the Schwarzschild azimuthal and timelike KVs are 
\begin{align}
e^{2\psi} &= (x+1)^2(1-y^2), & e^{2U} =  \frac{x-1}{x+1} \,. \notag
\end{align}
Note that the upper and lower segments of the axis are identified by the coordinate values $y=1$ and $y=-1$, respectively. The event horizon of the black hole is indicated by $x=1$.
By direct substitution it is easy to verify that both $\psi$ and $U$ satisfy Eq.~\eqref{lapxy}.

It is further possible to verify that the potential $s = (x-1)y $ has a gradient orthogonal to the gradient of $\psi$ and obeys Eq.~\eqref{sGrad}. It can thus be used as the second spatial coordinate. 
The lines of constant $\psi$ and $s$ coordinates are plotted in the $(\rho, \ z)$ plane in Fig.~\ref{fig:ONEBH}. Since in Weyl coordinates the $(\rho, z)$ plane is conformally flat, the fact that the curves intersect orthogonally in Fig.~\ref{fig:ONEBH} indicates that their gradients are orthogonal to each other. The strong warping influence of the black hole at $\rho = 0$, $-1 \leq z \leq 1$ on constant $\psi$ surfaces in these isotropic coordinates is clearly visible. This behavior can be ascribed in large part to  the coordinates, which compress the black hole horizon onto the axis. As $e^\psi \rightarrow 0$ away from the hole, contours of constant $\psi$ approach the axis,  but on the black hole $\psi$ acts as an angular coordinate, changing as the surface is traversed.

The 3D Ricci curvature scalar $R$ for the Schwarzschild metric is
\begin{align}
R &=
\frac{2 \left(x+2 y^2-1\right)}{(x+1)^5 \left(1-y^2\right)^2}
=2e^{-4 \psi } \left(1-2 \frac{e^{2 \psi
   }}{(x+1)^3} \right) \,,
\label{SchRICCS}
\end{align}
and obeys the axis condition~\eqref{psiAxis}. Also note that Eq.~\eqref{SchRICCS} is written is the same form  as the more general series expansion of $R$ about the axis given in Eq.~\eqref{seriesRpsi}. In Schwarzschild case, the series truncates after the first order. To facilitate compact notation later on, let us define a function $R_0$ that is finite on the axis by 
\begin{align} 
R_0 =  e^{4\psi}\frac R 2 = \frac{2y^2+x-1}{x+1} \,.
\end{align}

The nontrivial rotation coefficients for the Schwarzschild metric are
\begin{align}
\label{Schrot}
\beta 
&=-\frac{3 e^{2 \psi } (x-1)}{2 R_0^{3/2} (x+1)^6} \,,
\notag\\
\theta 
&=-\frac{3 e^{\psi } \sqrt{x-1} \sqrt{(x+1)^2-e^{2 \psi }}}{\sqrt{2} R_0^{3/2} (x+1)^{11/2}}\,,
\notag\\
\epsilon &= \frac{ e^{-2 \psi   }}{R_0^{3/2}} 
\left( \frac{e^{4 \psi } (3 x-7)  -1 }{2  (x+1)^6}  + \frac{3  e^{2 \psi }}{ 2(x+1)^9}
\right),
\notag\\
\alpha
&=\frac{e^{-\psi } \sqrt{(x+1)^2-e^{2 \psi }}}{\sqrt{2} R_0^{1/2} \sqrt{x-1} (x+1)^{5/2}}\, .
\end{align}
From the above expressions it is clear that on the axis, $e^{2\psi}=0$ ($y=\pm 1$), we have that $\beta = \theta=0$ while $\epsilon$ and $\alpha$ diverge. The fact that $\beta = 0$ is an indication that the null vector $l^a$ is geodesic on the axis, as is expected from symmetry. The divergence of $\alpha$ is an indication that this geodesic has been poorly parametrized.  A better choice would be to boost the null vector so that it is affinely parametrized. 
Note that in the boosted frame the vector $(\tilde{l}^a+\tilde{n}^a)/\sqrt{2} = (A l^a+ A^{-1} n^a)/\sqrt{2}$ is no longer hypersurface orthogonal. Choosing $A$ so that the component of the $l^a$ vector along the $t$ coordinate direction corresponds to that in the Kinnersley frame~\cite{Kinnersley1969} results in an affine parametrization on the axis. This boost transformation is achieved by setting  $A =\rho(x+1)/[\sqrt{2} (x-1)] $. The transformation of the rotation coefficients into the boosted, affinely parametrized frame are given in Eq.~\eqref{BoostTransforms2}, and is straightforward to compute. 
 
In this frame, the Weyl scalars computed from Eq.~\eqref{p0psi} are
\begin{align}\label{Schweyl} 
\hat \Psi _0 &= -\frac{3 e^{2 \psi }}{4 R_0 (x+1)^5}\,, 
\qquad
\hat \Psi _1=-
   \frac{3 e^{\psi } y}{2 \sqrt{2} R_0 (x+1)^4} \, , \notag\\ 
\hat \Psi _2&= -
   \frac{1 }{ R_0 (x+1)^3}+
 \frac{e^{2 \psi } (3 x+1)}{2 R_0 (x+1)^6} \, , 
\notag\\
\hat \Psi _3&=
   \frac{3 e^{\psi } (x-1) y}{\sqrt{2} R_0 (x+1)^5}\, , 
\qquad 
\hat \Psi _4 = -\frac{3 e^{2 \psi } (x-1)^2}{R_0 (x+1)^7}\, .
\end{align}
On the axis, the only nonzero Weyl scalar is \mbox{$\hat \Psi_2=-(x+1)^{-3}$}, where 
$x+1$ can be associated with the standard Schwarzschild radius along the axis.  As we move off the axis 
the other Weyl scalars take on non-zero values.  This is expected because our choice of triad is adapted to the gradient of $\psi$ rather than being a geodesic null triad. In fact, anywhere off the axis in the Schwarzschild spacetime, a triad adapted to the $\psi$ coordinate can never have its null vector $l^a$ be geodesic, as can be verified using Eq.~\eqref{Schrot}. This provides an explicit example of the general arguments regarding the boost transforms in~\eqref{BoostTransforms}. 

When the Schwarzschild metric is recast into a form with $\psi$ as a coordinate, many of the equations appear to be unwieldy and offer little additional insight, but they do allow for an explicit verification that a frame exists in which the near-axis scaling of the rotation coefficients and metric functions computed in Appendix~\ref{sec:ROTCOEFPSICOORD} hold. The main motivation for choosing a triad adapted to the $\psi$ coordinate is that the results obtained in this analysis of the SAV case generalize to dynamic spacetimes. The Weyl scalars and rotation coefficients should give some indication of the expected behavior of their counterparts in dynamical spacetimes, near the axis and near black holes. It should be noted that many of the features that make the standard SAV analysis elegant are due to the availability of isotropic $(\rho, z)$ coordinates, and the ability to linearly superimpose multiple solutions in the static case. These properties do not generalize to time-dependent spacetimes.
One feature that is common to both the SAV and the time-dependent analysis is that $\psi$ is a harmonic function that plays a crucial role in determining the configuration of the spacetime. It is hoped that a generalized superposition principal by which a new solutions for $\psi$ can be formed by the ``sum'' of two or more existing solutions can be found. 
Such a superposition of solutions, although straightforward to achieve on an initial value slice, will  subsequently be complicated by the fact that the potential $\psi$ influences the evolution of the metric functions that determine its own evolution in a possibly nonlinear way, making a ``sum'' of two solutions nontrivial.

\section{Conclusions}
\label{sec:Conclusions}

In this paper we have reviewed the reduction of the Einstein field equations in the case where the spacetime admits an axial Killing vector.  The problem of finding solutions to the field equations in 4D then reduces to a problem of finding solutions to the field equations in 3D with an additional scalar field source term. We specialized to the case of vacuum spacetimes, and then to twist-free spacetimes, where the field equations become especially simple, but are still capable of describing spacetimes of physical relevance. Of particular interest is the case of a head-on collision of non-spinning black holes. In order to recast the equations into a form that seems especially amenable to investigation and intuition, we have presented a triad-based formulation of the equations due to Hoenselaers. We have expanded upon the original work of Hoenselaers, linking this formalism to the Newman-Penrose formalism and discussing two triad and coordinate choices that help to simplify the equations. We have also 
reviewed the known twist-free axisymmetric solutions which are not necessarily captured by the SAV equations, classifying them according to their optical properties, which correspond to certain simplifications in the field equations.

We have  introduced and explored the use of a harmonic coordinate $\psi$ on the 3 manifold $\mathcal S$. Recall that $\mathcal S$ is conformally related to the manifold $\bar{ \mathcal{S}}$ of orbits of the KV. The function $\psi$ corresponds to a scalar field source for the Einstein field equations on $\mathcal S$, and its use as a coordinate simplifies the curvature on $\mathcal S$ considerably. There is a natural expansion about the axis of symmetry in terms of $\lambda = e^{2\psi}$, which provides inner boundary conditions for the field equations on $\mathcal S$, and we have provided these series expansions and their connection to the elementary flatness condition. We have further revisited the case of static, axisymmetric spacetimes using $\psi$ as a coordinate, in order to concretely illustrate the near-axis behavior of the triad and curvature quantities. Meanwhile, the assumption of asymptotic flatness provides the usual outer boundary behavior for the metric functions and curvature quantities, in 
terms of a Bondi expansion in geodesic null coordinates far from the 
sources. In order to develop a unified notation, we have given explicit expressions for the Bondi expansion in the triad formalism, eventually arriving at series expansions for the metric $h_{ij}$ on $\mathcal S$.

We intend this work to serve as a comprehensive and usable reference for the challenging goal of arriving at new solutions to the field equations. One immediate application of this work is to investigate the behavior of the scalar $\psi$ in numerical axisymmetric simulations, such as the head-on collision of black holes. The observed behavior of $\psi$ and the norm of its gradient $R$ in simulations may give new insights. For example, by tracking these quantities, one could clearly quantify how the nonlinear collision differs from simple linear super-position of black holes. Numerical computation of the other triad quantities can also help to succinctly quantify these simulations, once an appropriate triad is fixed. This method of recasting a numerical simulation has the advantage of identifying the variables commonly employed in most solution generation techniques as well as highlighting the role of the generalized Ernst potential ($e^{2\psi}$). Similarly, the investigation of $\psi$ and $R$ for the 
numerical studies of the critical collapse of gravitational waves in axisymmetry~\cite{lrr-2007-5,Sorkin:2010tm},  may lead to further understanding. In this case, the spacetime likely has additional symmetries which can help guide further 
analytic and numerical investigation of this solution.

In the triad formalism reviewed in this work, we have presented the asymptotic expansion of the field equations and a similar expansion near the axis of symmetry. These expansions give the boundary conditions and dynamics which would be needed in any sort of axisymmetric evolution. It is natural to consider the connection between the triad formalism and the initial data for such an evolution. This data may be in the form of quantities on an initial spatial or null slice. Previous work on null initial data \cite{1994JMP....35.4184G,1983JMP....24.1193W,1999PhRvD..60h4019H} immediately carries over to the triad quantities. Future work can examine the relationship between $\psi$ and $R$ and the momentum and Hamiltonian constraints on an initial spatial slice  as well as the subsequent evolution of these fields. Note that while the Hamiltonian and momentum constraints
do not constrain $\psi$ and $R$, it may be possible to identify a preferred choice for the $\psi$ associated with two black holes where the high frequency content is minimal.
  For example, the form that the scalars take in known initial data formulations for head-on collisions such as those in \cite{PhysRevD.47.1471} and the associated emitted junk radiation is of interest.

It is clear, though, that new techniques will still be needed in order to make analytic progress in the head-on collision. In the past, analytic methods have allowed for an exploration of curvature quantities on the horizons of the holes in a head-on merger~\cite{Lehner1999,Gomez2001}. With this and the expansions of the field equations near the boundaries, it would seem that an integration of the field quantities along null surfaces would be possible. One barrier to such an integration is the expectation that the null surfaces will caustic and become singular, especially near merger. For example, the horizon data of~\cite{Lehner1999,Gomez2001} could only be numerically evolved on null slices contacting a merged horizon, due to the lack of a null foliation for a bifurcated horizon~\cite{Gomez2002} (actually \cite{Gomez2002} used the initial data in the context of the evolution of the fission of a white hole). Ideally, a non-singular set of null coordinates could be found to cover the entire region of 
spacetime of interest, as illustrated schematically in Fig.~\ref{MANI3}. In such a coordinate system the triad formulation proves to be a powerful tool. We leave the search for such a set of null surfaces as the subject of future work.

\acknowledgements
We thank Yanbei Chen and An{\i}l Zengino\u{g}lu for valuable discussions. JB would like to thank Y. Chen and C. Ott for their hospitality while at Caltech. TH and AZ would like to thank NITheP of South Africa for their hospitality during much of this work. AZ is supported by NSF Grant PHY-1068881, CAREER Grant PHY-0956189, and the David and Barbara Groce Startup fund at Caltech. TH acknowledges support from  NSF Grants No. PHY-0903631 and No. PHY-1208881, and the Maryland Center for Fundamental Physics.

\appendix

\section{Dimensional reduction of the 4D field equations}
\label{DimensionalReduction}

In this appendix we review some of the results pertaining to the curvature of the three dimensional manifold $\bar{\mathcal{S}}$ whose  induced metric $\bar h_{\mu\nu}$ is related to the metric on the four dimensional manifold $\mathcal{M}$ by Eq.~\eqref{eq:ThreeMetric}.
In this subsection $\xi^\mu$ is not necessarily a KV, but merely assumed to be  timelike, i.e. $\xi^\mu \xi_{\mu} = \lambda > 0$.
Just as in Sec.~\ref{sec:Axisym}, Eq.~\eqref{DERIVOP}, the covariant derivative operator  $\bar D_\alpha$ is defined by the full contraction of the 4D derivative operator with the projector $ \bar h^\alpha_\nu =\delta^\alpha_\nu-\lambda^{-1} \xi_{\nu} \xi^{\alpha}$. For convenience, the definition of $\bar D _\alpha$,  Eq.~\eqref{DERIVOP} is repeated below,
 \begin{align}
\bar D_\alpha \bar T_{\beta \gamma} & = \bar h_\alpha^\mu\bar h_\beta^\nu \bar h_\gamma^\rho (\nabla_\mu \bar T_{\nu \rho}) \,. 
\label{DERIVOPAPP}
\end{align}

\subsection{Generalized Gauss-Codazzi equations for a timelike projected manifold}
\label{GaussCODAZZI}

As in the  case of the 3+1 split in numerical relativity~\cite{BaumgarteShapiro2010}, the  Gauss-Codazzi equations  describe the relationship between the 3D and 4D curvature tensors associated with the metrics
$\bar h_{\mu\nu}$ and $g_{\mu\nu}$  respectively. The Gauss equation can be derived by  considering the 3D Ricci identity, which defines the contraction of the 3D Riemann tensor with an arbitrary covector $\bar {V}_\alpha $ on $\bar{ \mathcal{S}}$,
\begin{align}
\label{RICCI3G}
\bar R_{\alpha \beta \gamma \delta} \bar V^\beta & = 2\bar D_{[\gamma }\bar D_{\delta]}
\bar V_\alpha \,.
\end{align}
The derivation proceeds by  writing out the derivative operators on the right hand side in terms of the 4D quantities $g_{\mu \nu}, \nabla_\mu$ and $ \xi^\mu$ using Eqs.~\eqref{eq:ThreeMetric} and~\eqref{DERIVOPAPP}. As an intermediate step we define the quantity $\bar K_{\beta \alpha}$ to be
\begin{align}
\bar K_{\beta \alpha} 
= -\lambda^{-1/2}\bar h^{\mu}_\alpha \bar h^{\nu}_\beta \nabla_\mu \xi_\nu ,
 \label{ExtrinsicK}
\end{align}
and expand the double covariant derivative operator applied to $\bar V_\gamma$ as
\begin{align}
\bar D_\alpha \bar D_\beta \bar V_\gamma  =& \bar h^\mu_\alpha \bar h^\nu_\beta \bar h^\rho_\gamma \nabla_\mu  \left(\bar h^\sigma_\nu \bar h^\tau_\rho \nabla_\sigma \bar V_\tau \right) \notag \\
=& \bar h^\mu_\alpha \bar h^\nu_\beta \bar h^\rho_\gamma \nabla_\mu \nabla_\nu \bar V_\rho +\bar K_{\gamma \alpha}\bar K_{\delta \beta} \bar V^\delta
\notag \\ 
+& \bar K_{\beta\alpha} \bar h^\tau_\gamma \lambda^{-1/2}\xi^{\sigma} \nabla_\sigma \bar V_\tau .
\label{doubleCov}
\end{align}
To derive the second line of Eq.~\eqref{doubleCov} we repeatedly use the definition of $\bar K_{\alpha\beta}$ given in Eq.~\eqref{ExtrinsicK},  the fact that $ \bar h^\mu_\alpha \bar h^\alpha_\beta = \bar h^\mu_\beta $, and the identity $\bar V_\tau \xi^\tau =0$ .

Substituting Eq.~\eqref{doubleCov}  into  Eq.~\eqref{RICCI3G} results in  an expression relating the Riemann tensor on $\bar{ \mathcal{S}}$  to the Riemann tensor on $\mathcal{M}$,
\begin{align}
\label{Riem3dpre}  
\bar R_{\alpha \beta \gamma \delta} \bar V^\beta 
=& \left( \bar h^\mu_\alpha \bar h^\nu_\beta \bar h^\rho_\gamma \bar h^\sigma_\delta R_{\mu \nu \rho \sigma}  
+ \bar K_{\alpha \gamma}\bar K_{\beta \delta}    -\bar K_{\alpha \delta}\bar K_{\beta \gamma} \right)\bar V^\beta  \notag\\
& + 2\bar K_{[\delta\gamma]}       \bar h^\tau_\alpha  \lambda^{-1/2}\xi^{\sigma} \nabla_\sigma \bar V_\tau. 
\end{align}
This result holds for any vector $\xi^\mu$ with norm $\lambda$.
The top line of Eq.~\eqref{Riem3dpre} resembles the usual Gauss equation often encountered in a 3+1 split of spacetime if the tensor $\bar K_{\alpha \beta}$ is identified with the extrinsic curvature of the embedded hypersurface (there is a relative sign change in front of the terms containing quadratic products in the tensor $\bar K_{\beta\alpha}$ that results from the fact that we are considering a timelike rather than spacelike 3 manifold).
 The tensor  $\bar K_{\alpha \beta}$  defined in Eq.~\eqref{ExtrinsicK} can be identified with the extrinsic curvature of a hypersurface embedded in $\mathcal M$ only if $\xi^\mu$ is hypersurface orthogonal.  
The second line of Eq.~\eqref{Riem3dpre} contains a term with the prefactor $\bar K_{[\delta \gamma]}$. In general if $\xi^\mu$ has twist, this term is non-vanishing and must be retained. 

For all vectors $\bar V^\alpha $ whose Lie derivative with respect to $\xi^\mu$ vanishes,  $\mathcal{L} _\xi \bar V_\alpha = 0 $, we can simplify the second term in Eq.~\eqref{Riem3dpre} using  $ \bar h^\tau_\alpha \lambda^{-1/2} \xi^{\sigma} \nabla_\sigma \bar V_\tau = \bar K_{   \beta \alpha} \bar V^\beta$ to yield     
\begin{align}
\label{Riem3dpre2}  
\bar R_{\alpha \beta \gamma \delta} \bar V^\beta 
=& \left( \bar h^\mu_\alpha \bar h^\nu_\beta \bar h^\rho_\gamma \bar h^\sigma_\delta R_{\mu \nu \rho \sigma}  
+ \bar K_{\alpha \gamma}\bar K_{\beta \delta}    -\bar K_{\alpha \delta}\bar K_{\beta \gamma} \right)\bar V^\beta  \notag\\
& + 2\bar K_{[\delta\gamma]}  \bar K_{   \beta \alpha} \bar V^\beta.
\end{align}
Since the vector $\bar V^\beta$ is arbitrary provided  $\mathcal{L} _\xi \bar V_\alpha = 0 $ and $\bar V_\mu \xi ^\mu=0$, it can be dropped from Eq.~\eqref{Riem3dpre2} to give an expression for the curvature on the 3 manifold in terms of projected quantities. An important consequence of the 3D Riemann tensor so obtained is that in order for it to have the correct symmetries the tensor $\bar K_{\beta\alpha}$ must be antisymmetric, $\bar K_{\beta\alpha} = \bar K_{[\beta\alpha]}$. The condition  $\bar K_{(\alpha\beta)} =0$ is the same as requiring that the projection of the 4D Killing equation hold. 

A concise way of expressing the generalized Gauss equation Eq.~\eqref{Riem3dpre2} in terms of the vector $\xi^\mu$ in the case where $K_{(\alpha \beta)}=0$ is 
\begin{align}
\label{Riem3d}  
\bar R_{\alpha \beta \gamma \delta} = & \bar h^\mu_\alpha \bar h^\nu_\beta \bar h^\rho_\gamma \bar h^\sigma_\delta R_{\mu \nu \rho \sigma} \notag \\
& +  \frac{4}{\lambda}  \bar h^\mu_{[\alpha} \bar h^\nu_{\beta]} \bar h^\rho_{[\gamma} \bar h^\sigma_{\delta]} \left(\nabla_\mu \xi_{(\nu} \right) \left(\nabla_{\rho)} \xi_\sigma\right) \,.
\end{align}
Note that the since the curvature tensor $\bar R_{\alpha \beta \gamma \delta}$ is defined on a 3D manifold which has a vanishing Weyl tensor,  $\bar R_{\alpha \beta \gamma \delta}$   can be constructed solely from the Ricci tensor $\bar R_{\alpha \beta}$. As a result only the contracted Gauss equation
\begin{align}
\label{Riem3dcon}  
\bar R_{ \beta  \delta} = &   \bar h^\nu_\beta \bar h^\sigma_\delta \left(
 R_{ \nu  \sigma}- \lambda^{-1}\xi^\mu \xi^\rho R_{\mu \nu \rho \sigma}
\right) \notag \\
& +  \frac{4}{\lambda} \bar h^{\alpha \gamma}  \bar h^\mu_{[\alpha} \bar h^\nu_{\beta]} \bar h^\rho_{[\gamma} \bar h^\sigma_{\delta]} \left(\nabla_\mu \xi_{(\nu}\right) \left( \nabla_{\rho)} \xi_\sigma\right) \,,
\end{align}
needs to be considered.

The components of the 4D curvature tensor where one index has been projected onto $n^\mu$, the unit normal in the $\xi^\mu$ direction (and not to be confused with the null triad or tetrad vectors used elsewhere in this text), are related to quantities defined on the 3 manifold $\bar{\mathcal{S}}$ via the Codazzi equations. These equations can be derived  by applying the 4D Ricci identity to the unit vector $n^\mu= \lambda^{-1/2}\xi^\mu$ and projecting the result onto the 3D manifold $\bar{\mathcal{S}}$ 
\begin{align}
\label{RICCI4D}
 \bar h^\mu_\alpha  \bar h^\rho_\gamma \bar h^\sigma_\delta
n^\nu R_{\mu \nu \rho \sigma}& = 2\bar h^\mu_\alpha   \bar h^\rho_\gamma \bar h^\sigma_\delta \nabla_{[\rho} \nabla_{\sigma]}  n_\mu \,.
\end{align}
When expanding the right hand side of Eq.~\eqref{RICCI4D} in terms of 3D quantities, 
it is useful to observe that $\bar K_{\beta \alpha}$ can be expressed 
as the gradient of the unit vector $n^\mu$ which has been twice contracted with the
the projection operator. 
Expanding this relation and using the fact that $n^\mu n_\mu =1$ yields the identity
\begin{align} 
 \nabla_\mu n_\nu=-\bar{K}_{\nu\mu} + n_\mu \bar a_\nu ,
\label{covn}
\end{align}
where $\bar a_\beta = n^\nu \nabla_\nu n_\beta$ is a measure of how the unit vector $n^\nu$ is changing when parallel propagated. Note that $\bar h^\mu_\alpha \bar a_\mu = \bar a_\alpha$ since $\bar a_\mu n^\mu=0$.
Substituting Eq.~\eqref{covn} into Eq.~\eqref{RICCI4D} yields a generalized Codazzi equation
\begin{align}
\label{CodazziGen}
\bar h^\mu_\alpha  \bar h^\rho_\gamma \bar h^\sigma_\delta
n^\nu R_{\mu \nu \rho \sigma} = \bar D _\delta \bar K_{\alpha \gamma}-\bar D_\gamma \bar K_{\alpha \delta}+2\bar a_\alpha \bar K_{[\gamma \delta]}.
\end{align}
The last term once again vanishes in the case where $\xi^\mu$ is 
hypersurface orthogonal, but has to be retained if we consider
vectors $\xi^\mu$ with twist.
Contracting Eq.~\eqref{CodazziGen} on the indices $\alpha$, $\gamma$ 
with the metric $\bar h^{\alpha \gamma}$ yields the contracted Codazzi relation
\begin{align}
\label{CodazziGenCon} 
\bar h^\sigma_\delta
n^\nu R_{ \nu  \sigma}
=  \bar D_\delta \bar K -\bar D^\alpha \bar K_{\alpha \delta}+2\bar a^\alpha \bar K_{[\alpha \delta]},
\end{align}
where $\bar K = \bar K^\alpha_\alpha$.

There is one more nonzero contraction of the 4D Riemann tensor with the projector $\bar h^\mu _\alpha$ and the unit vector $n^\mu$, which is computed by contracting the second and third  indices of the Riemann tensor with $n^\mu$ and the remaining indices with the projection operators $\bar h _\alpha ^\mu$.
\begin{align}
\label{RICCI4D2}
 \bar h^\mu_\alpha  n^\rho  \bar h^\sigma_\delta
n^\nu R_{\mu \nu \rho \sigma}& =
 -   \bar h^\mu_\alpha    \bar h^\sigma_\delta n^\rho \nabla_{\rho}  \bar{K}_{\mu\sigma} +  
\bar{K}_{\alpha\rho} 
\bar {K}^{\rho}_{\ \delta}+
  \bar a_\alpha    \bar a_{\delta}
\notag\\
&-\bar h^\mu_\alpha   \bar h^\sigma_\delta
\nabla_{\sigma}
 \bar a_\mu  .
\end{align}
Equations~\eqref{Riem3dpre2},~\eqref{CodazziGen} and~\eqref{RICCI4D2} express  the 4D curvature tensor in terms of projected quantities for a projection operator based on an arbitrary spacelike vector $\xi^\mu$. When $\xi^\mu$ is hypersurface orthogonal, these expressions reduce to the usual Gauss-Codazzi equations. The case where $\xi^\mu$ is a KV is addressed in the next section.

\subsection{Field Equations expressed on the 3D quotient manifold in the case where $\xi^\mu$ is a Killing Vector}
\label{GaussCODAZZIK}

We now specialize the results of the Gauss-Codazzi equations derived in the previous section  to the case where $\xi^\mu$ is a KV obeying Eq.~\eqref{killingEQ}. When $\xi^\mu$ is a KV the derivation and results presented here are equivalent to that found in~\cite{Geroch1971}. 

The Killing equation~\eqref{killingEQ} implies that the tensor $\bar K_{\alpha\beta}$ defined in Eq.~\eqref{ExtrinsicK} is antisymmetric and thus the generalized Gauss equation~\eqref{Riem3d} holds. The Killing equations along with the identity $\xi^\mu\lambda_{,\mu} =0$ can  be used to express the covector $\bar a _\beta$ defined below  Eq.~\eqref{covn} as 
\begin{align}
\bar a_\beta = -\frac{1}{2 \lambda} \lambda_{,\beta} .
\label{axidef}
\end{align}
This result, in conjunction with  Eq.~\eqref{covn},  then allows us to write the gradient of the KV as 
\begin{align}
\nabla_\mu\xi_\nu  = -\lambda^{1/2}\bar K_{\nu\mu} -\frac{1}{ \lambda} \xi_{[\mu} \nabla_{\nu ]} \lambda.
\label{gradkillK}
\end{align}
Substituting Eq.~\eqref{gradkillK} into the definition of the twist, Eq.~\eqref{twist}, we obtain  the expression
\begin{align}
\omega_\mu & =\lambda^{1/2} \epsilon_{\mu \nu \rho \sigma} \xi^\nu \bar K^{\rho\sigma} \,,
\label{twistKK}
\end{align}
which can be inverted using the identity $\epsilon_{\mu \nu \rho \sigma} \epsilon^{\mu \tau \chi \epsilon} = - 6 \delta^\tau_{[\nu} \delta^\chi_\rho \delta^{\epsilon}_{\sigma]}$
and the fact that $\xi^\mu \bar K_{\mu\nu}=0$, to yield $\bar K_{\beta \alpha}$ in terms of the 
twist 
\begin{align}
\bar K^{\alpha\beta} =\frac{1}{2\lambda^{3/2}} \epsilon^ { \alpha \beta \epsilon \mu}\xi_\epsilon  \omega_\mu .
\label{Ktwist}
\end{align}
Finally substituting Eq.~\eqref{Ktwist} back into Eq.~\eqref{gradkillK} yields the  identity
\begin{align}
\label{gradxi}
\nabla_\mu \xi_\nu & = \frac{1}{2\lambda} \epsilon_{\mu \nu \rho \sigma} \xi^\rho \omega^\sigma - \frac{1}{\lambda}\xi_{[\mu} \nabla_{\nu]} \lambda  \,.
\end{align}

Given the expressions for $\bar K_{\alpha \beta}$, $\bar a_\beta$ and $\nabla_\mu\xi_\nu$ in terms of the twist and norm of the KV, we can begin to evaluate the Gauss-Codazzi equations. An expression for  Ricci tensor on the three manifold can be found by substituting the double contraction of the KV with the 4D Riemann tensor,  Eq.~\eqref{RICCI4D2}, into the contracted Gauss equation~\eqref{Riem3dcon} and using the relations given in this section to arrive at
\begin{align}
\bar R_{\alpha \beta}  =&  \bar h_\alpha^\mu \bar h_\beta^\nu R_{\mu \nu} + \frac{1}{2\lambda}\bar D_\alpha \bar D_\beta \lambda - \frac{1}{4 \lambda^2} \bar D_\alpha \lambda \bar D_\beta \lambda \notag
\\ &
 -\frac{1}{2 \lambda^2}\left( \bar h_{\alpha \beta} \omega_\gamma \omega^\gamma - \omega_\alpha \omega_\beta \right) \,. \notag
\end{align}
This equation used extensively in  Sec. \ref{sec:Reduction} where it is referred to as Eq.~\eqref{RICCI3}.
The first term on the right hand side of~\eqref{RICCI4D2} vanishes because of the symmetry in the indices $\alpha$, $\delta$ on the left hand side of~\eqref{RICCI4D2} and the antisymmetry of $\bar K_{\mu\alpha}$.

Since $\bar K_{\alpha\beta}$ is antisymmetric, $\bar K=0$. Substituting Eq.~\eqref{Ktwist} and subsequently 
Eq.~\eqref{gradxi} into the contracted Codazzi equation~\eqref{CodazziGenCon} we obtain
\begin{align}
\bar h^\sigma_\delta
n^\nu R_{ \nu  \sigma}
=  \frac{1}{2\lambda^{3/2}} \epsilon_ {  \delta}^{\ \ \alpha \mu \nu}\xi_\mu  \bar D_\alpha \omega_\nu ,
\end{align}
which can be rewritten to yield
\begin{align}
\bar D_{[\alpha} \omega_{\beta]}=-\epsilon_{\alpha \beta \rho \sigma}  \xi^\rho R^{\sigma}_{\nu}\xi^{\nu}. \notag
\end{align}
 which relates the twist of $\omega_\mu$ 
to the 4D Ricci curvature. This equations is reference as Eq.~\eqref{twistpotential} in Sec.  \ref{sec:Reduction} .
 
The divergence of $\omega_\alpha$ is found by considering the totally antisymmetric part of the generalized Codazzi equation, or equivalently contracting  $\xi_{\nu}\epsilon^{\nu \alpha \gamma \delta } $ with Eq.~\eqref{CodazziGen}, which becomes
\begin{align}
\xi_{\mu}\epsilon^{\mu \alpha \gamma \delta }
n^\nu R_{\alpha \nu \gamma \delta} =  2 \xi_{\mu}\epsilon^{\mu \alpha \gamma \delta }(\bar D _\delta \bar K_{\alpha \gamma}+\bar a_\alpha \bar K_{\gamma \delta}).
\label{AntiCodazziGen}
\end{align}
The first Bianchi identity,  $R_{\mu \nu \rho \sigma} + R_{ \rho \mu  \nu \sigma } - R_{ \sigma \mu \nu \rho }=0 $ sets the term on the left hand side of Eq. \eqref{AntiCodazziGen} to zero. The right hand side of Eq.~\eqref{AntiCodazziGen} can be evaluated using the following expression for the derivative of the extrinsic curvature,
\begin{align}
\bar D_\gamma K_{\alpha\beta} 
  &= -\frac{1}{\lambda} K_{\alpha\beta} \bar D_\gamma \lambda  +
\frac{1}{2\lambda^{1/2}} \epsilon_ { \alpha \beta}^{\ \ \ \  \mu \nu}\xi_\mu  \bar D_\gamma \omega_\nu .
\end{align}
The resulting expression for the divergence of the twist vector quoted in Eq.~\eqref{domega} is
\begin{align}
\bar D^\alpha \omega_\alpha =& \frac{3}{2\lambda} \omega_\alpha \bar D^\alpha \lambda \,.  \notag
\end{align}
An equation governing the harmonic operator applied to $\lambda$ can be obtained by contracting  the final projection of the Riemann tensor in Eq.~\eqref{RICCI4D2} with the three metric $\bar h^{\alpha\delta }$, and making use of the antisymmetry of $\bar K_{\mu\sigma}$ and the expressions~\eqref{Ktwist} for $\bar K_{\alpha \beta}$ and~\eqref{axidef} for $\bar a_\beta$. The result is quoted in Eq.~\eqref{laplambda} and given below,
\begin{align}
\bar D^2 \lambda = & \frac{1}{2\lambda} \bar D_\alpha \lambda \bar D^\alpha \lambda - \frac{1}{\lambda} \omega_\mu \omega^\mu - 2 R_{\mu \nu} \xi^{\mu} \xi^{\nu} \,. \notag
\end{align}
This completes the derivation of the reduced field equations on $\bar {\mathcal S}$, used in Sec.~\ref{sec:Axisym} to discuss axisymmetric spacetimes.

\section{Lorentz transforms of the 3D tetrad}
\label{sec:UsefulResults}

Here we discuss the effect of Lorentz transforms of the triad. As usual, these come in three types: boosts along the null vector $l^i$, and rotations about each of the two null vectors $l^i$ and $n^i$. First we discuss boosts. Let
\begin{align}
\tilde{l}^i &= A l^i, & \tilde{n}^i& = A^{-1} n^i \,.
\end{align}
Then $\tilde{l}_{i;j}= Al_{i;j}+A_{,j}l_{i}$ and  $\tilde{n}_{i;j}=n_{i;j}/A- A_{,j}n_{i}/A^2$. The nine rotation coefficients become
\begin{align}
\label{BoostTransforms2}
\tilde{\epsilon}&= \epsilon\,, 
&\tilde{\gamma}&=  \gamma\,, \notag\\
\tilde{\theta}&=A\theta \,, 
&\tilde{\iota} &=A^{-1} \iota \,, \notag\\
\tilde{\beta}&=A^2 \beta &\tilde{\zeta}  
&= A^{-2}\zeta \,, \notag\\
\tilde{\alpha}&=A \alpha -A_{,1}& \tilde{\eta} 
&=  \eta-A^{-1}(A_{,3}) \,,\notag\\
\tilde{\delta}&=  A^{-1}\delta -A^{-2}( A_{,2})\,,
\end{align}
where the directional derivatives are with respect to the original triad. The six curvature scalars transform as 
\begin{align}
\tilde{\phi}_5 &=A^2\phi_5,
&\tilde{\phi_2} &=A^{-2}\phi_2\notag\\
\tilde{\phi}_3 &= A\phi_3, 
&\tilde{\phi}_1 &= A^{-1}\phi_1,\notag\\
\tilde{\phi}_4 &= \phi_4,
&\tilde{\phi}_0 &= \phi_0 \,. 
\end{align}

Next, let us consider rotations about the null vectors. The usual rotations about the null vectors in a null tetrad in the 4D are restricted to those that do not mix the axial KV $\xi^\mu$ with the three vectors that span the 3D hypersurfaces which correspond to $\mathcal S$. These transforms must leave the difference and sum of the complex NP spatial vectors $\hat m^\mu$ and $\hat m^{*\mu}$ invariant (since these correspond to the normalized KV $\hat d^\mu$ and the spatial vector $\hat c^\mu$, respectively). The usual rotations by complex parameters (see e.g. \cite{Stephani2003}) are reduced to rotations by real parameters, $a$ and $b$.

For a rotation about $l^i$, we have
\begin{align}
\label{lrot}
\tilde l^i & = l^i\,, & \tilde c^i& = c^i + a l^i\,, & \tilde n^i &= n^i + a c^i + \frac{a^2}{2} l^i \,.
\end{align}
For a rotation about $n^i$ we have in complete analogy
\begin{align}
\tilde n^i & = n^i\,, & \tilde c^i& = c^i + b n^i\,, & \tilde l^i &= l^i + b c^i + \frac{b^2}{2} n^i \,.
\end{align}
We are primarily interested in a fixed null direction $l^i$, so let us consider rotations about this vector. We have the following transforms for the rotation coefficients,
\begin{align}
\tilde \alpha &\  = \alpha - a \beta \,,  &
\tilde \beta &\ = \beta \,, \notag \\ 
\tilde \gamma &\ = \gamma - a \alpha +\frac{a^2}{2} \beta + a_{,1} \,, &
\tilde \epsilon &\ = \epsilon + a \theta +\frac{a^2}{2} \beta \,,  \notag \\
\tilde \eta &\ =  \eta + a (\alpha - \theta) - a^2 \beta \,, &
\tilde \theta &\ = \theta + a \beta \,, \notag 
\end{align}
\begin{align}
\label{SpinTransforms}
\tilde \delta= &\ \delta +a (\eta - \epsilon) + \frac{a^2}{2} ( \alpha - 2 \theta)- \frac{a^3}{2}\beta \,, \notag \\
\tilde \iota = &\ \iota + a(\gamma -\eta) +\frac{a^2}{2} (\theta - 2 \alpha) + \frac{a^3}{2} \beta + a_{,3} + a a_{,1} \,, \notag \\
\tilde \zeta = &\ \zeta + a (\iota - \delta) + \frac{a^2}{2} ( \epsilon +\gamma - 2 \eta) + \frac{a^3}{2} (\theta - \alpha) + \frac{a^4}{4} \beta \notag \\
& + a_{,2} + a a_{,3} + \frac{a^2}{2}a_{,1}  \,.
\end{align}
In these expressions, the directional derivatives are with respect to the original triad vectors. The six curvature scalars transform as
\ba
\label{CurvatureRot}
\tilde \phi_5 & = & \phi_5 \,, \notag \\
\tilde \phi_4 & = &\phi_4 + a \phi_3 + \frac{a^2}{2} \phi_5 \,, \notag \\
\tilde \phi_3 & = & \phi_3 + a \phi_5 \,, \notag \\
\tilde \phi_2 & = & \phi_2 + 2 a \phi_1 + a^2(\phi_0 + \phi_4) +a^3 \phi_3 + \frac{a^4}{4} \phi_5 \,, \notag \\
\tilde \phi_1 & = & \phi_1 + a(\phi_0 + \phi_4) + \frac{3a^2}{2} \phi_3 +\frac{a^3}{2} \phi_5 \,, \notag \\
\tilde \phi_0 & = & \phi_0 + 2a \phi_3 + a \phi_5 \,.
\ea

Finally, we consider rotations about $n^i$. We first note that when interchanging $l^i$ and $n^i$, the rotation coefficients exchange identities as
\begin{align}
&\{ \alpha,\, \beta, \, \gamma,\, \delta, \,\epsilon,\, \zeta, \, \eta,\, \theta,\, \iota \}\notag \\
&\to \{ - \delta,\, -\zeta,\, -\epsilon, -\alpha,\, -\beta,\, - \eta,\, -\iota,\, -\theta \} \,.
\end{align}
Thus, the effect of a rotation around $n^i$ by a factor $b$ on the rotation coefficients can be derived from the expressions given for a rotation around $l^i$ by first applying the above relations to those transforms, and then taking $a \to b$ and swapping directional derivatives in the $l^i$ and $n^i$ directions, $f_{,1} \to f_{,2}$ and vice versa. The rotation coefficients thus transform as 
\begin{align}
\tilde \delta &= \delta - b \zeta\,, & \tilde \zeta&= \zeta \,, \notag \\
\tilde \epsilon &= \epsilon - b \delta + \frac {b^2} {2} \zeta - b_{,2}\,, & \tilde \gamma &= \gamma + b \iota + b^2 \zeta \,, \notag \\
\tilde \eta &= \eta + b( \delta - \iota) - b^2 \zeta \,, & \tilde \iota & = \iota + b \zeta \,, \notag 
\end{align}
\begin{align}
\tilde \alpha  =& \alpha + b (\eta - \gamma) + \frac{b^2}{2} ( \delta - 2 \iota) - \frac{b^3}{2} \zeta \,, \notag \\
\tilde \theta  =& \theta + b (\epsilon - \eta) +\frac{b^2}{2} (\iota - 2 \delta) + \frac{b^3}{2} \zeta - b_{,3} - b b_{,2} \,, \notag \\
\tilde \beta =& \beta + b (\theta - \alpha) +\frac{b^2}{2} ( \gamma + \epsilon - 2 \eta ) + \frac{b^3}{2}(\iota - \delta) \notag \\
&+ \frac{b^4}{4}\zeta - b_{,1} - b b_{,3} - \frac{b^2}{2} b_{,2} \,.
\end{align}
We can write similar transformations for the curvature scalars $\phi_i$ by noting that, under the exchange of $l^i$ and $n^i$ the curvature scalars transform as $\{ \phi_5, \, \phi_4,\, \phi_3, \, \phi_2,\, \phi_1,\, \phi_0\}\to \{ \phi_2, \, \phi_4,\, \phi_1, \, \phi_5,\, \phi_3,\, \phi_0\}$ and applying these transforms and $a\to b$ to Eqs.~\eqref{CurvatureRot}.

\section{Minkowski spacetime with $\psi$ as a coordinate}
\label{sec:Minkpsi}
To gain a better intuition into the choice of $\psi$ as a coordinate, let us consider Minkowski space in cylindrical coordinates $(t, z, \rho, \phi)$. In these coordinates $\rho = e^{\psi}$ and so $d\rho = e^\psi d \psi$. Inserting this gives the line element
\begin{equation}
\label{eq:MinkMetric}
ds^2 = - dt^2 + dz^2 + e^{2 \psi} ( d\psi^2 + d \phi^2) \,.
\end{equation}
Next, consider the metric on the conformal space $\mathcal S$. We have
\begin{equation}
h_{ij} = {\rm diag} [ - e^{2\psi}, e^{2 \psi} , e^{4 \psi}]\,.
\end{equation}
We can see immediately that $R = 2 e^{4\psi}$, which trivially obeys the axis condition in $\mathcal S$ as $\psi \to - \infty$. Looking at how the metric functions enter the triad in Eqs.~\eqref{Metpsi} and~\eqref{psitriadup}, we see that an appropriate choice for a null triad adapted to the timelike gradient $T_a$ is
\begin{align}
\label{eq:FlatSimpTriad}
l^i & = ( e^{-\psi}, e^{-\psi},0)/\sqrt{2} \,, \qquad n^i = (e^{-\psi}, -e^{-\psi} ,0)/\sqrt{2}\,, \nonumber \\
c^i & = (0,0,e^{-2\psi}) \,.
\end{align}
Using this triad, we can compute the rotation coefficients and begin to get a sense of the way each coefficient should behave as we approach the axis. However, first let us note that the triad chosen above has some troubling features. Comparing these to the corresponding tetrad vectors in $\mathcal M$, as given by Eq.~\eqref{physicalvec}, we see that
\begin{align}
\hat l^\mu & = (e^{-\psi}, e^{-\psi}, 0, 0)/\sqrt{2}\,, \qquad \hat n^\mu = (e^{\psi}, -e^{\psi},0,0)/\sqrt{2} \,, \nonumber \\
\hat c^\mu &= (0,0,e^{-\psi},0)\,.
\end{align}
Near the axis, we see that our chosen $\hat l^\mu$ and $\hat n^\mu$ vectors are poorly behaved; $\hat l^\mu$ blows up on the axis, and $\hat n^\mu$ vanishes. We must boost the triad vectors by a factor of $A= e^{\psi}$ in order for the corresponding physical tetrad to the be well behaved on the axis. Computing the rotation coefficients on $\mathcal S$ with the boosted triad legs $l^i = (1,1,0)/\sqrt{2}$ and $n^i = e^{-2\psi}(1,-1,0)/\sqrt{2}$ leads to
\begin{equation}
\gamma = - \epsilon =-\eta = e^{-2\psi} \,, 
\end{equation}
with all others vanishing. Note that under a null boost, $\gamma$ and $\epsilon$ do not change; we cannot prevent the pathological behavior of these coefficients on $\mathcal S$ as $\psi \to -\infty$. Meanwhile, $\eta$ does transform, and using a boost $A = e^{m \psi}$ we find $\tilde \eta = \eta - m e^{-2 \psi}$, which can be used in this case to make $\tilde \eta = 0$ with the choice $m = -1$. This infinite boost at the axis has no effect on the vanishing of the other coefficients, and returns us to the triad we originally considered in Eq.~\eqref{eq:FlatSimpTriad}.

\section{Rotation coefficients  associated with a triad adapted to the  $\psi$  coordinate}
 \label{sec:ROTCOEFPSICOORD}

The rotation coefficients associated with the triads in Eqs.~\eqref{psitriaddown} (with $h_\psi =0$) and~\eqref{psitriadup} and the corresponding  metric Eq.~\eqref{Metpsi} are now computed using equations~\eqref{lambdaabc} and~\eqref{rotrecon}. Recall that for the case under consideration
$\eta = \frac{1}{2}(\beta -\gamma +\zeta -\epsilon )$ and   $\gamma = -\epsilon$. The remaining coefficients are
\begin{align}
\alpha &= -\frac{l_{t,s}+ h_{s,t}}{\sqrt{2} \varrho h_s}\, , 
& \delta &=
   \frac{ h_{s,t} - n_{t,s}}{\sqrt{2} \varrho h_s}\, , \notag\\
\beta& = \frac{\sqrt{R} \left(l_t h_{s,\psi }-h_s l_{t,\psi }\right)}{2  \varrho h_s}, 
& \zeta &= \frac{\sqrt{R} \left( h_s n_{t,\psi } - n_t
   h_{s,\psi }\right)}{2 \varrho h_s}\,,
\notag\\
\theta &=
   \frac{\left(l_t R_{,s}+ h_s R_{,t} \right)}{2\sqrt{2}  \varrho h_s R}\, ,
&
\iota &=
   \frac{ \left(n_t R_{,s}- h_s R_{,t}\right)}{2\sqrt{2}  \varrho h_s R} ,\notag\\
\epsilon &= -\frac{\sqrt{R} \left( \varrho h_s \right)_{,\psi }}{2 \sqrt{2}  \varrho h_s}\,  .
\label{ROTcoefsMet}
\end{align}
From Eq.~\eqref{ROTcoefsMet} the dominant scaling of the rotation coefficients near the axis can now be obtained by requiring that the physical metric expressed in terms of $(t,s,\lambda,\phi)$ coordinates is regular as the axis is approached, $\lambda \rightarrow 0$. 

To examine the behavior of the metric components as we near the axis, we will quote a result of Rinne and Stewart \cite{rinnestewart}. Consider a local Lorentz frame in a neighborhood near a point on the axis, $p \in W_2$, and let us use Cartesian coordinates $(x,y)$ on the space orthogonal to $W_2$, so that the KV can be represented as
\begin{align}
\xi^\mu \partial_\mu = - y \partial_x + x \partial_y \,.
\end{align}
If we insist that scalar quantities have a regular expansion in $(x,y)$ about the axis, 
\begin{align}
\label{eq:AxisExpansion}
f(x,y) = \sum_{m,n =0} f^{(m, \,n)} x^m y^n \,,
\end{align}
and that their Lie derivative with respect to $\xi^\mu$ vanishes, then it can be shown that the expansion must in fact be of the form 
\begin{align}
f(x^2 + y^2) = \sum_{n=0} f^{(n)} \lambda^n \,,
\end{align}
noting that $\lambda = x^2 + y^2$ near the axis if our coordinates are appropriately normalized. By applying the same Lie derivative argument to the metric, it can be shown \cite{rinnestewart} that the $(t,s)$ block of the metric admits expansions in $\lambda$ as if the metric functions are scalar quantities,
\begin{align}
g_{ss} &= \sum_{n=0} g_{ss}^{(n)}(t,s) \lambda^{n},&g_{tt} &= \sum_{n=0} g_{tt}^{(n)}(t,s) \lambda^{n}, \notag\\
 g_{ts} &= \sum_{n=0} g_{ts}^{(n)}(t,s) \lambda^{n}. 
\label{scalingg}
\end{align}
 It is always possible to choose $s$ and $t$ coordinates on the axis to be orthogonal to each other.
This choice sets the first term in the off-diagonal metric function $g_{ts}^{(0)}$ to zero. 

The series expansions about the axis in Eq.~\eqref{scalingg}  also set the series expansion for the functions entering into the metric $h_{ij}$ in Eq.~\eqref{Metpsi}.  Explicitly we have that the functions $h_s$, $n_t$ and $l_t$  admit the following expansions near the  symmetry axis:
\begin{align}
n_{t} &= e^{\psi}\sum_{n=0} n_{t}^{(n)}(t,s) \lambda^{n},&l_{t} &= e^{\psi}\sum_{n=0} l_{t}^{(n)}(t,s) \lambda^{n},\notag\\
h_{s} &= e^{\psi}\sum_{n=0} h_{z}^{(n)}(t,s) \lambda^{n} & \varrho&= e^{\psi}\sum_{n=0} \varrho^{(n)}(t,s)\lambda^n  , \label{scalingg2}
\end{align}
where the coefficients $\varrho^{(n)}=(l_t^{(n)} + n_t^{(n)} )/\sqrt{2}$.   If the metric is chosen to be diagonal on the axis, we further have that $l_t^{(0)} = n_t^{(0)}$ . In Eq.~\eqref{epsint} we integrate the equations describing the rotation coefficient $\epsilon$ by means of one of the Bianchi identities to yield $(h_s \varrho)^2 = g(t,s)/R$. This result remains valid  in the neighborhood of the axis and allows us to find an expression for the function $g(t,s)$ in terms of the expansion coefficients  $h_s^{(0)}$ and $\varrho^{(0)}$, namely 
\begin{align}
g(t,s) = 2( h_s^{(0)} \varrho^{(0)})^2.
\end{align}
This result, together with the expansions given in Eq.~\eqref{scalingg2} and the series expansion of $R$ implied by the axis condition,  
\begin{align}
R = 2 e^{-4\psi}\left( 1+  \sum_{n=1} R^{(n)} \lambda^n \right) ,
\label{seriesRpsi}
\end{align}
allows us to determine that on the axis the rotation coefficients scale as 
\begin{align}
\alpha &\to -\frac{
   \left(h_{s,t}^{(0)}+l_{t,s}^{(0)}\right)}{\sqrt{g(t,s)}
   \sqrt{\lambda }}+O\left(\sqrt{\lambda }\right),\notag\\
\delta &\to \frac{
   \left(h_{s,t}^{(0)}-n_{t,s}^{(0)}\right)}{\sqrt{g(t,s)}
   \sqrt{\lambda }}+O\left(\sqrt{\lambda }\right), \notag\\
\theta &\to
   \frac{  \left(R^{(1)}_{,t} h_s^{(0)}+l_t^{(0)}
   R^{(1)}_{,s}\right) \sqrt{\lambda }}{2\sqrt{g(t,s)}}+O\left(\lambda
   ^{3/2}\right), \notag\\
\iota &\to \frac{   \left(n_t^{(0)} R^{(1)}_{,s}-h_s^{(0)}
   R^{(1)}_{,t}\right) \sqrt{\lambda }}{2\sqrt{g(t,s)}}+O\left(\lambda
   ^{3/2}\right), \notag\\
\beta &\to \frac{2  \left(h_s^{(1)}
   l_t^{(0)}-h_s^{(0)} l_t^{(1)}\right)}{\sqrt{g(t,s)}}+O\left(\lambda
   ^1\right), \notag \\
\zeta &\to \frac{2  \left(h_s^{(0)}
   n_t^{(1)}-h_s^{(1)} n_t^{(0)}\right)}{\sqrt{g(t,s)}}+O\left(\lambda
   ^1\right) , \notag\\
\epsilon &\to -\frac{1}{ \lambda }
 - R^{(1)} -2 \left( \frac{ \varrho^{( 1 )} }{ \varrho^{( 0 )}} +  \frac{ h_s^{( 1 )} }{ h_s^{( 0 )}} \right)+O\left(\lambda ^1\right).
\label{axisRot}
\end{align}
While $\beta$ and $\zeta$ have $O\left(\lambda^0\right)$ terms, we know that null rays which remain on the axis must be geodesic, and thus these terms must vanish. We then have 
\begin{align}
 l_t^{(1)}&= \frac{h_s^{(1)}
 l_t^{(0)}}{h_s^{(0)}}, 
&  n_t^{(1)}&= \frac{h_s^{(1)} n_t^{(0)}} {h_s^{(0)}}, 
&\frac{\varrho^{(1)}}{\varrho^{(0)}} &= \frac{h_s^{(1)}}{h_s^{(0)}}. 
\label{l1n1}
\end{align}
Recall that when working with a triad adapted to $\psi$ as a coordinate, Eq.~\eqref{p0psi} states that the Weyl scalar $\hat \Psi_0 = \sqrt{R/2}\beta$, and as a result~\eqref{l1n1} implies that  the geodesics along the axis are principal null. It also implies that if the metric is chosen to be diagonal on the axis so that $l_t^{(0)} = n_t^{(0)}$, then this property persists to order $O(\lambda^2)$, since $l_t^{(1)} = n_t^{(1)}$. 

We now substitute the expansions into the field equations~\eqref{psiFeq12} and~\eqref{psiFeq3}, and begin to solve them order by order in $\lambda$. We start by looking at the subset of equations that have directional derivatives only in the $l^a$, $n^a$ directions, namely Eqs.~\eqref{psiFeq12}, and choose to set the metric diagonal on the axis. The dominant terms that arise from the sum and difference of the first two equations in Eqs.~\eqref{psiFeq12} give
\begin{align}
\left( R^{(1)} +\frac{ 4 h_s^{(1)}}{h_s^{(0)}}\right)_{,s}   &=0, &
\sqrt{2}\left( R^{(1)} + \frac{4 h_s^{(1)}}{h_s^{(0)}}\right)_{,t} &=0, \notag 
\end{align} 
respectively. These imply that 
\begin{align}
\frac{h_s^{(1)}}{h_s^{(0)}} = -\frac{1}{4}R^{(1)} +k_1 \,, \label{Rhzk}
\end{align}
with $k_1$ a constant.
Together Eqs.~\eqref{axisRot},~\eqref{l1n1} and~\eqref{Rhzk} give 
\begin{align} 
\epsilon = -\lambda^{-1} - 4 k_1 + O(\sqrt{\lambda}) \,.
\end{align}
The dominant $\lambda^{-1}$ term in the fourth equation of \eqref{psiFeq12} and the first equation in \eqref{psiFeq3} governing $\epsilon_{,3}$ can only vanish if both $k_1 = 0$ and the equation 
\begin{align}
\frac{2   h^{(0)}_{s ,tt}}{(\varrho^{(0)})^ 2 h_s^{(0)}}& -\frac{2 \varrho^{(0)}_{,t} h^{(0)}_{s ,t}}{(\varrho^{(0)})^{3} h_s^{(0)}}+
\frac{\varrho^{(0)}_{,s}   h^{(0)}_{s, s}}{\varrho^{(0)} (h_s^{(0)})^{3}}
-\frac{\varrho^{(0)}_{,ss}}{\varrho^{(0)}  (h_s^{(0)})^{ 2}} \notag \\
+2 R^{(1)}&=0
\end{align} 
holds. With this, the equations in~\eqref{psiFeq3} that govern $\delta_{,3}$,   $\alpha_{,3}$,  $\theta_{,3}$,  and $\iota_{,3}$ are satisfied to $O(\sqrt{\lambda})$ and the third equation in~\eqref{psiFeq12} to $O(\lambda)$. Setting the $O(\lambda^0)$ term to zero in the equation governing $\epsilon_{,3}$ and the $O(\lambda ^0)$ term to zero in those for  $\zeta_{,3}$, $\beta_{,3}$ in~\eqref{psiFeq3}, and furthermore setting the $O(\lambda^0)$ term to zero in the fourth equation of~\eqref{psiFeq12} fixes the $R^{(2)}$, $n^{(2)}_t$, $l^{(2)}_t$ and $h^{(2)}_s$ coefficients in terms of $\varrho^{(0)}$, $h^{(0)}_s$, $R^{(1)}$, and  their derivatives. These expressions are lengthy, and we will only give the off diagonal term here,
\begin{align}
l_t^{(2)}-n_t^{(2)} = \frac{\varrho^{(0)}_{,s} R^{(1)}_{,t}}{16 \varrho^{(0)} h_s^{(0)}}+\frac{h^{(0)}_{s ,t}
   R^{(1)}_{,s}}{16 (h_s^{(0)})^2}-\frac{R^{(1)}_{,ts}}{16 h_s^{(0)}} \,.
\label{Seriesnodiag}
\end{align} 
Equation~\eqref{Seriesnodiag}, indicates that for a time-dependent metric the off-diagonal term $l_t-n_t$ is $O(\lambda^{3/2})$. It is also clear that when the spacetime is dynamic, the diagonalization of the $(t,s)$ block cannot be maintained off the axis. This off-diagonal term can be interpreted as a shift governing the motion of the coordinates along constant $\psi$ slices as time progresses. Continuing on, it appears that the expansion coefficients of the metric functions are fixed at each higher order by the lowest order terms $\varrho^{(0)}$ , $h^{(0)}_s$, and $R^{(1)}$. The same behavior occurs far from gravitating sources in asymptotically flat spacetimes, where the expansion is in orders of inverse affine distance. This is the asymptotic expansion of the Bondi formalism, which we discuss in Sec.~\ref{AsflatJD} and Appendix~\ref{AsflatJDApp}.

\section{Geodesic null coordinates affinely parametrized with respect to the 4D manifold}
 \label{app:affinereal}

In this appendix we write down the field equations adapted to a geodesic null coordinate system that is affinely parametrized with respect to the physical manifold $\mathcal M$. The results obtained in this section can be directly derived from Sec.~\ref{sec:Choice1} by applying a boost $A = e^{2\psi}$ using Eqs.~\eqref{BoostTransforms2}, and then selecting a new parameter along the geodesic. However, because this choice of coordinates is used in our discussion of the Bondi expansion, we will give the equations in full here. Let us once again choose a null coordinate $u$ such that $h^{ij}u_{,i}u_{,j} = 0$, but this time we choose the null vector $l_i= -e^{2\psi} u_{,i}$. The symmetry of  $u_{,a|b}=-(e^{-2\psi}l_a)_{|b}$, when expressed on the triad basis, leads to the following conditions on the rotation coefficients,
\begin{align}
\alpha &= -2 \psi_{,1}, & \beta &= 0, & \eta &= -(\epsilon +2 \psi_{,3}). 
\label{affinerot1} 
\end{align}
Making use of Eqs.~\eqref{4spins}, we immediately see that this choice for $l_i$ yields $\hat \kappa = \hat \epsilon = 0$, so we see that $\hat l_\mu$ is geodesic and affinely parametrized in the physical space. Let $\tau$ denote the affine parameter in $\mathcal M$, and choose the triad to be
\begin{align}
l^i &= (0,1,0),& n^i &=(e^{-2\psi},- w_1, 0), & c^i&=(0, w_2,e^{-w_3}). 
\label{laffine}
\end{align}
Then the three metric on $\mathcal{S}$ becomes 
\begin{align}
h_{ij}=\left(
\begin{array}{ccc}
 -2 e^{4 \psi } w_1 & -e^{2 \psi } & e^{w_3+2 \psi } w_2 \\
 -e^{2 \psi } & 0 & 0 \\
 e^{w_3+2 \psi } w_2 & 0 & e^{2 w_3} \\
\end{array}
\right). \label{BONDIMET2}
\end{align}
Applying the commutation relations to $\chi$, $u$, and $\tau$ respectively yield
\begin{align}
\gamma &= -\epsilon , &\theta &= -w_{3,1}, &\iota &=
   w_{3,2},\notag\\
\alpha &= -2 \psi _{,1}, & \beta &= 0,& \eta &= -2 \psi
   _{,3}-\epsilon,  \label{affinerot2}
\end{align}
and
\begin{align}
\delta &= -2 w_1 \psi
   _{,1}-w_{1,1}, \, \, \,
  \epsilon = \frac{1}{2}
   \left(w_2 \theta-w_{2,1}-2 \psi
   _{,3}\right),\notag\\
 \zeta &= w_2
   w_{3,2}+w_{2,2}+2w_1 \psi
   _{,3}+w_{1,3}\,.  \label{Tauint2} 
\end{align}
Substituting the expressions for the coefficients $\gamma$, $\alpha$, $\beta$, and $\eta$ found from these relations into the field equations~\eqref{Bondi0}--\eqref{BondiI3} and reorganizing gives
\begin{align}
\theta _{,1}= &
  2 \theta  \psi _{,1}+2 \psi
   _{,1}^2+\theta ^2, \label{BondAffineP1}\\
\epsilon _{,1} = &
2 \psi _{,1} \psi _{,3},  \label{BondAffineP2} \\
\delta _{,1}= &
\epsilon ^2-2 \psi _{,1} \left(\psi _{,2}+2 \delta \right)-2
   \psi _{,12}-\psi
   _{,3}^2,    \label{BondAffineP3}\\
\theta _{,2}-\epsilon
   _{,3}  &= -\delta  \theta -\theta  \iota -\epsilon ^2-\psi _{,3}^2  ,    \label{BondAffineP4}\\
\iota _{,1}+\theta _{,2}&=
-2 \iota  \psi
   _{,1}-\delta  \theta
   ,    \label{BondAffineP5}\\
\zeta _{,1}+\epsilon _{,2}
&=
 -4 \zeta  \psi _{,1}+2 \psi
   _{,2} \psi _{,3},    \label{BondAffineP6}\\
\zeta
   _{,3}
-\iota
   _{,2}
&=
2 \psi _{,2}^2-4 \zeta  \psi _{,3}-\delta  \iota +\iota ^2,   \label{BondAffineP7} \\
\zeta_{,1}+\delta _{,3}  +2 \epsilon _{,2}
&=\zeta  \theta -2 \iota  \epsilon 
-4 \zeta  \psi _{,1} -2 \delta  \psi
   _{,3} \notag\\
& -4 \psi _{,2} \psi _{,3}-2 \psi
   _{,23}  \,. \label{BondAffineP8}
\end{align}
As expected, the hierarchy present in  Eqs.~\eqref{Bondi5Sim}--\eqref{BondiI3Sim} (where $l^a$ is affine in $\mathcal S$) persists, which allows us to formally integrate the field equations.

We conclude this appendix by detailing how the metric functions in Eq.~\eqref{BONDIMET2} transform with the remaining coordinate freedom~\cite{CarmeliBook}. These transformations include shifting the the origin of the affine parameter along each geodesic separately, $\tau' = \tau + f (u,\chi)$, which transforms the metric function of~\eqref{BondiMetric} according to
\begin{align}
\label{affineshift2}
w_1'&=w_1-e^{-2\psi}f_{,u}\,, & w_2'&= w_2+ f_{\chi}e^{-w_3}\,,   & w_3'&=w_3 \,.
\end{align}
Relabeling the individual geodesics within a spatial slice, $\chi' = g(u,\chi)$ transforms the metric functions of~\eqref{BONDIMET2} to
\begin{align}
\label{relabeling2}
w_1'& = w_1+ e^{w_3-2\psi}w_2 \frac{g_{,u}}{g_{,\chi}} -\frac{1}{2} e^{2 w_3-4\psi} \left( \frac{g_{,u}}{g_{,\chi}}\right)^2 \notag\\
w_2'&= w_2 -e^{w_3-2\psi} \frac{g_{,u}}{g_{,\chi}}, \ \ \ \ \  \, e^{w_3'} =\frac{e^{w_3} }{g_{,\chi}} .
\end{align}
And finally, by relabeling the null hypersurfaces, setting $u'=h(u)$,  $ \tau'=\tau/ h_{,u}$ ,the metric components transform as
\begin{align}
\label{AffineRescale2}
 w_1' &= \frac{w_1}{(h_{,u})^2} + \tau e^{-2\psi}\frac{h_{,uu}}{(h_{,u})^3}, &
w_2' &= \frac{w_2}{h_{,u}},&
w_3' &= w_3.
\end{align}

\section{Derivation of the Asymptotic expansion of an affinely parametrized
metric in null coordinates}
\label{AsflatJDApp}

In this appendix we systematically solve the field equations given in Appendix~\ref{app:affinereal}
in the asymptotic regime far from an isolated, gravitating system.  Our method of solution further illustrates the integration of the hierarchy of the field equations that results when they are expressed on a null slicing. We focus only on spacetimes that admit the peeling property~\cite{Sachs1961, Sachs1962, Penrose1963, Penrose1965, PenroseRindler2}. In this case the Weyl scalars have a power series expansion at  future null infinity of the form $ \hat \Psi_i = \tau^{i-5}\sum_{n=0} \tau^{-n}\hat \Psi_i^{(n)}  $, where $\hat \Psi_i^{(n)}$ are constant along an out-going null geodesic, i.e. $\hat \Psi_i^{(n)}(u,\chi)$, and $\tau$ is the affine parameter along the geodesic. Using Eq.~\eqref{eq:PeelingpowerSeries} as a starting point, we  derive the power series expansion  of the metric functions and the rotation coefficients in terms of a series in $1/\tau$. This information makes apparent the boundary conditions that have to be imposed when solving the complete set of equations, and provides explicit 
information 
about the fall-off of all rotation coefficients at null infinity . 

Consider first the properties of the null tetrad vector $\hat l^\mu$. We will take $\hat l ^\mu$ to be the tangent to out-going, null geodesics far from the isolated system, so that $\beta=0$, and also affinely parametrized in $\mathcal M$, which sets $\alpha = -2\psi_{,1}$. Directional derivatives in the $l^i$ direction on $\mathcal S$ can thus be expressed as $f_{,1}=f_{,\tau}$. The simplified field equations and form of the metric for this choice of parametrization is given in Appendix~\ref{app:affinereal}. Specifying the series expansion for $\hat \Psi_0$  on a null hypersurface of constant $u$ gives almost all of the data required to continue the spacetime off the hypersurface. During the calculation that follows, we quantify how this information is transmitted to the metric functions.

 From Eq.~\eqref{p0} and~\eqref{eq:PeelingpowerSeries} we have that the leading order terms of the function $\psi$ are related to the  coefficient $\hat \Psi_0^{(0)}$ via   $\psi_{,11} + (\psi _{,1})^2 =   \tau^{-5} \hat \Psi_0^{(0)}  +  O(\tau^{-6})$. Solving this equation term by term, we find that
\begin{align}
\psi_{,1} = \frac{1}{\tau}+ \frac{\sigma^{(0)}}{\tau^{2}} + \frac{\sigma^{(0)\, 2}}{ \tau^{3}} + \frac{( \sigma^{(0)\,3} - \Psi_0^{(0)}/2   )}{ \tau^4} + O(\tau^{-5}) .
\label{psiexpand}
\end{align}
 Here, $\sigma^{(0)}=
\sigma^{(0)}(u,\chi)$ is a function whose properties have yet to be defined. We will see that $\sigma^{(0)}$ corresponds to the dominant term in the series expansion of the shear of $\hat{l}^\mu$. Its labeling corresponds to the choice made in~\cite{newman:891}, whose derivation we initially follow closely when working out the expansion properties of the optical scalars. In the series expansions that follow we keep terms of sufficiently high order to indicate where dominant terms of the expansions of the Weyl scalars enter into  the metric functions. 

Integrating Eq.~\eqref{psiexpand} with respect to $\tau$ adds an additional integration constant $\psi^{(0)}(u, \chi)$, and allows us to express $\psi$ as
 \begin{align}
\psi = &\psi^{(0)}+ \ln \tau  -\frac{\sigma^{(0)}}{\tau} - \frac{\sigma^{(0)\, 2}}{2 \tau^{2}}  -\frac{( \sigma^{(0)\, 3} - \hat \Psi_0^{(0)}/2   )}{3 \tau^3} \notag\\ & + O(\tau^{-4}) .
\label{psiexpand0}
\end{align}
A series expansion for $\theta$ can be found by substituting a power series ansatz for $\theta$ into  Eq.~\eqref{BondAffineP1} and using Eq.~\eqref{psiexpand} to define the series expansion for $\psi_{,1}$. It can be shown that the leading order behavior of the solution admits only two possibilities, namely $\theta = -\tau^{-1} + O(\tau^{-2})$ or  $\theta = -2\tau^{-1} + O(\tau^{-2})$. The former corresponds to a cylindrical type spacetime that is not astrophysically relevant. We will only consider the latter case and further discuss the justification for this choice after Eq.~\eqref{w3expand}. Using our coordinate freedom to relabel the origin of the affine parameter, $\tau'=\tau + f(u,\chi)$, it is always possible to set the next coefficient in the expansion to zero \cite{newman:891}. Examining the remaining coefficients in Eq.~\eqref{BondAffineP1} term by term leads to the series expansion
\begin{align}
\frac{\theta}{2}  = -\frac{1}{\tau}- \frac{\sigma^{(0)\, 2}}{\tau^{3}} - \frac{ \sigma^{(0)\,4} - \sigma^{(0)}\hat \Psi_0^{(0)}/3   }{ \tau^5} + O(\tau^{-6}) .
\label{thetaexpand}
\end{align}
With this, we can integrate Eq.~\eqref{affinerot2}, \mbox{$w_{3,1}=-\theta$}, to arrive at an expression for the metric function $w_3$,
\begin{align}
w_3 = & w_3^{(0)}+2\ln(\tau) -  \frac{\sigma^{(0)\, 2}}{\tau^{2}} - \frac{3  \sigma^{(0)\, 4} - \sigma^{(0)} \hat \Psi_0^{(0)}  }{6 \tau^4} \notag \\ 
&+ O(\tau^{-5}) ,
\label{w3expand}
\end{align}
which adds the integration constant $w_3^{(0)} (u, \chi)$ to our list of undetermined expansion coefficients. Note that the leading order term of the metric coefficient \mbox{$g_{\chi\chi}=e^{2w_3-2\psi}$}  in $\mathcal{M}$ is proportional to $ \tau^2$, which is typical for a surface of constant $\tau$ that is asymptotically spherical. Had we made the selection $\theta=-\tau^{-1} + O(\tau^{-2})$ above, the leading order term would have been independent of $\tau$.

The next set of variables to be considered in the integration hierarchy are $\epsilon$, $\hat{\Psi}_1$, and the metric function $w_2$. For our chosen triad, determining $\epsilon$ also fixes two other rotation coefficients; from Eqs.~\eqref{affinerot1} and~\eqref{affinerot2} we have that $\eta = -2\psi_{,3}-\epsilon$ and $\gamma=-\epsilon$.  The peeling property of the Weyl scalars~\eqref{eq:PeelingpowerSeries} in conjunction with the expression for $\hat \Psi_1$ given in Eq.~\eqref{p0} and the field equation~\eqref{BondAffineP2} yields
\begin{align}
  2 \epsilon  \psi _{,1}&+\left(3  - \frac{ \psi_{,11}}{\psi _{,1}^2}  \right) \epsilon_{,1}  +\frac{ \epsilon_{,11}}{\psi _{,1}} \notag \\
&= \frac{4e^{-\psi}}{\sqrt{2}}\left(\hat \Psi _1^{(0)} \tau^{-4}+ O(\tau^{-5})\right),
\end{align}
which can be systematically solved to yield a series solution for $\epsilon$,
\begin{align}
\epsilon &= \frac{\epsilon^{(0)}}{\tau}+  \frac{\epsilon^{(1)}}{\tau^2}+  \frac{\epsilon^{(2)}}{\tau^3}+  \frac{\epsilon^{(3)}}{\tau^4} +O(\tau^{-5}).
\notag
\end{align} 
Next, using Eq.~\eqref{BondAffineP2}, we find a series expansion for $\psi_{,3}$ of the form
\begin{align}
\psi_{,3}&=-\frac{\epsilon ^{(0)}}{2 \tau }+ \frac{\sigma ^{(0)} \epsilon^{(0)}-2 \epsilon^{(1)}}{2 \tau^2}+\frac{2\sigma^{(0)} \epsilon^{(1)}-3 \epsilon^{(2)}}{2\tau^3} \notag\\ 
&+\frac{6 \sigma ^{(0)} \epsilon^{(2)}- \hat \Psi _0^{(0)} \epsilon ^{(0)} - 8 \epsilon^{(3)}} {4 \tau^4}+O\left(\tau^{-5}\right).
\end{align}
The higher order coefficients in the expansion for $\epsilon$ are fixed in terms of existing quantities as follows, 
\begin{align}
\epsilon^{(2)} &=\sigma^{(0)}\left(2\epsilon^{(1)}-\epsilon^{(0)}\sigma^{(0)} \right),   \notag\\
\epsilon^{(3)} &=  \left(3\epsilon^{(1)} -2\epsilon^{(0)}\sigma^{(0)}\right)\sigma^{(0)\ 2}   -\frac{\epsilon^{(0)} \hat \Psi_0^{(0)} }{6} + \frac{\sqrt{2}\hat \Psi_1^{(0)} }{3 e^{\psi^{(0)}}} ,
\end{align}
and so far the coefficients $\epsilon^{(0)}$ and $\epsilon^{(1)}$ are unconstrained. It turns out that the leading coefficient $\epsilon^{(0)}$ can be set to zero using a gauge transform.

To see this, note that we can obtain an expansion for the metric function $w_2$ using Eq.~\eqref{Tauint2}.
The resulting expansion is
\begin{align}
w_2&=-\frac{\epsilon^{(0)}}{2} -\frac{\epsilon^{(0)}\sigma^{(0)}}{\tau} + \frac{w_2^{(2)}  }{\tau^2}\notag\\
&+ \frac{-6\epsilon^{(0)}\sigma^{(0)\ 3} - \epsilon^{(0)} \psi_0^{(0)} - 4\sqrt{2} \Psi_1^{(0)}e^{-\psi_0}    }{6\tau^3} + O(\tau^{-4}) .
\label{w2expandpre}
\end{align}
It is then possible to make use of the gauge transformation $\chi'= g(u,\chi)$ to relabel the geodesics such that $\epsilon^{(0)} = 0$, using Eq.~\eqref{relabeling2}.  The resulting expansions for $\epsilon$, $\psi_{,3}$ and $w_2$  reduce to
\begin{align}
\epsilon=& \epsilon^{(1)} \left( 
\frac{1}{\tau ^2}+
\frac{2 \sigma ^{(0)} }{\tau ^3} +
\frac{3 \sigma ^{(0)\, 2} }{\tau^4} \right) 
+   \frac{\sqrt{2}  \hat \Psi _1^{(0)}}{3\tau ^4 e^{\psi^{(0)}}  }+ O\left(\tau^{-5}\right), \\
\psi_{,3}&= -\epsilon -\frac{\sqrt{2} \hat \Psi _1^{(0)}}{3 e^{\psi ^{(0)}} \tau ^4}
 + O\left(\tau^{-5}  \right) ,
\label{psicomma3}\\
w_2 &=\frac{w_2^{(2)}}{\tau ^2}-\frac{2 \sqrt{2}
    \hat \Psi _1^{(0)}}{3 e^{\psi ^{(0)}} \tau ^3} + O(\tau^{-4}) .
\label{w2expand}
\end{align}

The next set of variables obtained via the systematic integration of the field equations includes $\delta$, the Weyl scalar $\hat \Psi_2$, the metric function $w_1$, and the derivative $\psi_{,2}$. The coefficient $\delta$ can be obtained by integrating the field equation~\eqref{BondAffineP3}, where the commutation relations, Eq.~\eqref{COMM}, the definition~\eqref{p0} of $\hat \Psi_2$ and the harmonic equation~\eqref{LapOp} for $\psi$, have been used to replace directional derivatives in the $n^i$ direction with known series expansions. The resulting equation
\begin{align}
\delta _{,1}+2 \delta  \psi _{,1}= ( \epsilon + \psi _{,3})^2-2 e^{-2 \psi } \hat \Psi _2,
\end{align}
implies that $\delta$ must admit the series expansion
\begin{align}
\delta= &\frac{\delta ^{(1)}}{\tau^2} + \frac{2 \sigma ^{(0)} \delta ^{(1)}  }{\tau^3} 
+\frac{ 3 \sigma ^{
   (0)\, 2} \,\delta ^{(1)}+e^{-2 \psi ^{(0)}} \hat \Psi_2^{(0)}   }{\tau^4}  \notag\\ &+O(\tau^{-5}) 
\end{align}
The expansion for $\delta$ can now be used to obtain the expansions for the metric function  $w_1$ using Eq.~\eqref{Tauint2}, $\delta = -2 w_1 \psi_{,1}-w_{1,1}$. The first two terms in the resulting expression are
\begin{align}
w_1&= \frac{ -\delta ^{(1)} }{\tau}+\frac{w_1^{(2)}}{\tau^2}+O(\tau^{-3}).
\end{align}
If the metric $g_{\mu \nu}$ is to be asymptotically flat, the metric function $e^{2 \psi} w_1$ must be finite  as $\tau\rightarrow \infty$, and thus the integration constant $\delta^{(1)}=0$. The resulting expansion for the metric function $w_1$ becomes 
 \begin{align}
\label{w1expand}
w_1&= \frac{w_1^{(2)}}{\tau^2}+ \frac{2 w_1^{(2)} \sigma ^{(0)} + e^{-2
   \psi ^{(0)}}  \Psi
   _2^{(0)}   }{ \tau^3} + O(\tau^{-4}) .
\end{align}
A series for the directional derivative $\psi_{,2}$ can be obtained from the field equation~\eqref{BondAffineP3} using the commutation relation to switch the order of differentiation on $\psi$. The resulting expression becomes
\begin{align}
6 \psi _{,1} \psi _{,2}+2 \psi _{,21} &= -2 \delta  \psi
   _{,1}-\delta _{,1}-\psi _{,3}^2+\epsilon ^2 ,
\end{align}
which implies that $\psi_{,2}$ has the series expansion
\begin{align}
\psi_{,2} =\frac{\psi _{,2}^{(0)}}{\tau ^3}+
\frac{3 \sigma ^{(0)} \psi
   _{,2}^{(0)}-e^{-2 \psi ^{(0)}} \hat \Psi _2^{(0)}}{\tau ^4} + O(\tau^{-5}).
\label{psi2expand}
\end{align}

Since the series expansions for $\psi$ and the three metric functions $w_1,\ w_2,\ w_3$ are  given, the directional derivatives of any function expressed as a series in $1/\tau$ can  also be expanded as a series. We begin by examining the directional derivatives of $\psi$ to determine what restrictions the resulting expressions place on the existing expansion coefficients. By considering the directional derivative $\psi_{,3}$  and the expansion~\eqref{psicomma3} we obtain the result
\begin{align}
\epsilon ^{(1)}&= -e^{-w_3^{(0)}} \psi^{(0)}{}_{,\chi },\notag\\
w_2^{(2)}&= e^{-w_3^{(0)}} \left(2\sigma^{(0)} \psi^{(0)}{}_{,\chi }+\sigma^{(0)}{}_{,\chi }\right).
\label{direcdirpsi3}
\end{align}
Examining the directional derivative $\psi_{,2}$ and the expansion~\eqref{psi2expand}, we obtain  $\delta ^{(1)}= -e^{-2 \psi ^{(0)}} \psi^{(0)}{}_{,u}=0$ which implies that $ \psi ^{(0)}=  \psi^{(0)}(\chi)$ is independent of $u$. We also have that
\begin{align}
 \psi _{,2}^{(0)}&= 
-w_1^{(2)} 
-e^{-2 \psi^{(0)}}\sigma^{(0)}_{,u}.
\end{align}

Finally, we examine the remaining field equations to obtain  further restrictions on the expansion coefficients.  Substituting the expansions and directional derivatives obtained thus far into Eq.~\eqref{BondAffineP4} yields the condition $w_3^{(0)}{}_{,u}= 2 \psi^{(0)}{}_{,u}=0,$ at order $\tau^{-3}$, which implies that  
\begin{align}
w_3^{(0)}= 2 \psi^{(0)} + f^{(0)}(\chi).
\end{align}
Note however that the coordinate transformation \mbox{$\chi'=g(\chi)$} transforms the metric function as $e^{w_3'}=e^{w_3} /g_{,\chi}$ and allows us to set $f^{(0)}(\chi)$ to any arbitrary function of our choosing. We can understand the meaning of a choice of $f^{(0)}$ by insisting that as $\tau \to \infty$, the $(\chi, \phi)$ block of the physical metric on $\mathcal M$ has the geometry of a sphere. In other words, $g_{AB} \to \tau^2 \, \Omega_{AB}$, where $\Omega_{AB}$ is a metric on the unit 2-sphere, and $\{A, B \} \in (\chi, \phi)$. With this requirement on the angular geometry of out-going null surfaces, a particular choice of $f^{(0)}$ allows us to fix both $\psi^{(0)}$ and to identify the particular angular coordinate $\chi$ corresponding to this choice of $f^{(0)}$. For example, setting  $f^{(0)}(\chi) = - 2\psi^{(0)}$ corresponds to the choice where the metric on the unit 2-sphere $\Omega_{AB}$ has unit determinant. In this case, we find that $e^{2\psi^{(0)}}=1-\chi^2$, 
where $\chi= \cos \theta$, $\theta$ is the usual angular coordinate, and the axis is located at $\chi = \pm 1$. For comparison, note that setting $f^{(0)}(\chi)=0$ corresponds to the Fubini study metric representation of the sphere used in~\cite{newman:891}. Henceforth we will set  $f^{(0)}(\chi) =  -2\psi^{(0)}$, which means that we have selected $\chi = \cos \theta$ and that $w_3^{(0)}=0$.

Evaluating Eq.~\eqref{BondAffineP4} at order $\tau^{-4}$ yields the additional condition
\begin{align}
w_1^{(2)}&= -\psi
   ^{(0) \,2}_{,\chi }
-\frac{1}{2}
   \psi ^{(0)}{}_{,\chi\chi }=  \frac{1}{2(1-\chi^2)}. 
\label{w12def}
\end{align} 
Next, at order $\tau^{-5}$ we obtain an equation that evolves $\hat \Psi_{0}^{(0)}$ from one hypersurface to the next hypersurface
\begin{align}
\label{p00evo}
\hat \Psi _0^{(0)}{}_{,u}&= 3 \sigma ^{(0)} \hat \Psi_2^{(0)}
+e^{ 2 \psi^{(0)}}\frac{\left(e^{-\psi ^{(0)}} \hat \Psi_1^{(0)}\right){}_{,\chi }}{\sqrt{2}}.
\end{align}
Using equations~\eqref{affinerot2} and~\eqref{Tauint2}, the expansions for $\iota$ and $\zeta$ can be shown to have the form
\begin{align}
\zeta &= \frac{\zeta ^{(0)}}{\tau
   ^4}+ O\left(\tau^{-5}  \right),
&\iota &= \frac{\iota
   ^{(0)}}{\tau ^3}+\frac{\iota ^{(1)}}{\tau ^4}+O\left(\tau^{-5}\right).
\label{iotazetaexpand}
\end{align}
where the expansion coefficients are related to those functions already defined by
\begin{align}
\zeta^{(0)}&= 
\frac{ \left[(1-\chi ^2) \sigma
   ^{(0)}{}_{,u}\right]_{,\chi}  }{(1-\chi^2)^2  }, \notag\\
\iota^{(0)}&= -\frac{1}{1-\chi^2}, \notag\\
\iota ^{(1)}&= -2\frac{\sigma^{(0)}\sigma ^{(0)}{}_{,u}+ \Psi _2^{(0)} }{ 1-\chi^2  },
\end{align}
Using Eq.~\eqref{p0} the dominant terms in the remaining Weyl scalars 
can be shown to be
\begin{align}
 \hat \Psi_3 &= \frac{ \hat \Psi_3^{(0)}}{\tau^2}+O(\tau^{-3}), & \hat \Psi_4  &=-\frac{\sigma ^{(0)}{}_{,uu}}{\tau} + O(\tau^{-2}),
\label{NPWeylexpansion}
\end{align}
where
\begin{align}
\hat \Psi_3^{(0)}&= 
 -\frac{ \left[ (1-\chi^2) \sigma
   ^{(0)}{}_{,u} \right]_{,\chi}  }{\sqrt{2}  
\sqrt{1-\chi^2} } \,.
\label{psi30def}
\end{align}

The evolution equations that propagate the coefficients $\hat \Psi_1^{(0)}$ and $\hat \Psi_2^{(0)}$ from one null hypersurface to the next can be obtained by examining~\eqref{BondAffineP6} and~\eqref{BondAffineP7} respectively at $O(\tau^{-6})$, yielding the expressions
\begin{align}
\hat \Psi^{(0)}_{1 \ ,u} &= 2\sigma^{(0)} \hat \Psi_3^{(0)} + \frac{ \hat \Psi _2^{(0)}{}_{,\chi }}{\sqrt{2}\sqrt{1-\chi^2}}, \label{p10evo} 
\\
\hat \Psi _2^{(0)}{}_{,u} &=
 -\frac{ \left[ (1-\chi^2) \sigma
   ^{(0)}{}_{,u} \right]_{,\chi\chi}  }{2} -\sigma ^{(0)} \sigma^{(0)}{}_{,uu}.
\label{p20evo}
\end{align}
The  field equations  \eqref{BondAffineP5}  and  \eqref{BondAffineP8} yield no additional constraints and vanish to $O(\tau^{-8})$. Also note that the harmonic equation for  $\psi$ has been satisfied.

The results obtained thus far are now briefly summarized.  The 4D line element can be expressed as 
\begin{align} \label{affineMmet2}
ds^2 &=-2e^{2\psi}w_1\ du^2 -2 du\ d\tau + e^{w_3}w_2\ du\ d\chi \notag\\
&+ e^{2w_3-2\psi}\ d\chi^2+e^{2\psi}\ d\phi^2\,,
\end{align}
where, according to Eqs.~\eqref{psiexpand0},~\eqref{w3expand},~\eqref{w2expand},~\eqref{w1expand},~\eqref{direcdirpsi3}, and~\eqref{w12def}, the metric functions  admit the following asymptotic expansion as the affine parameter $\tau \rightarrow \infty$:
\begin{align}
e^{2\psi}&=(1-\chi^2) \left(\tau^2  - 2\tau \sigma^{(0)}+\sigma^{(0) 2} + \frac{\hat \Psi_0^{(0)}}{3\tau} \right)+ O(\tau^{-2}), \notag\\
e^{w_3}&= \tau^2- \sigma^{(0)2}  +\frac{\sigma^{(0)} \hat \Psi_0^{(0)}}{6\tau^2} +  O(\tau^{-3}),  \notag\\
w_1 &=  \frac{1}{(1-\chi^2)}\left[ \frac{1} {2 \tau^2}
  + \frac{ \sigma^{(0)}+  \hat \Psi_2^{(0)}   }{ \tau^3}\right] + O(\tau^{-4}), \notag\\
w_2 &=
 \frac{\left[(1-\chi^2) \sigma^{(0)}\right]_{,\chi }}{  (1-\chi^2) \tau ^2}-\frac{2 \sqrt{2}  \hat \Psi _1^{(0)}}{3 \sqrt{1-\chi^2}\ \tau ^3} + O(\tau^{-4})\, .
\label{metExpand2}
\end{align}

To fully specify the solution we must make a choice for the functions $\hat \Psi_i^{(0)}(u,\ \chi)$, $i\in\{0,1,2\}$ on a hypersurface $u_0$ and specify the function  $\sigma^{(0)}(u,\chi)$ for all $u$ and $\chi$. A natural hypersurface  to choose is $u\rightarrow \infty$ and to specify the functions to correspond to the Schwarzschild solution. In this case
\begin{align}
\{ \hat \Psi_0, \hat \Psi_ 1,\hat \Psi_2,  \hat \sigma^{(0)}, \hat \sigma^{(0)}_{,u} \} \to \{ 0,0,-M,C e^{-2\psi^{(0)}},0\} \,.
\end{align}
The constant $M$ is the mass of the final black hole and $C$ is an arbitrary constant. The fact that the shear of the null bundle must be regular on the axis sets $C=0$.
For $u<\infty $, $\hat \Psi^{(0)}_{i}$, can now be determined provided $\sigma^{(0)}(u,\ \chi) $ is given.

The results given this section are further discussed in Sec.~\ref{AsflatJD}.

The expansion of the 4D spin coefficients can be obtained from the results  in this appendix using Eqs.~\eqref{4spins}. The series expansion of the shear $\hat{\sigma}$ is found to be
\begin{align}
\hat{\sigma} = \frac{\sigma^{(0)}}{\tau^{2}}   + \frac{( \sigma^{(0)\ 3} - \Psi_0^{(0)}/2   )}{ \tau^4} + O(\tau^{-5}), \label{sigexpand}
\end{align}
confirming that $\sigma^{(0)}$ is indeed the dominant term in the expansion of the shear. It should be noted that our expansion for the spin coefficient $\hat{\tau}$ is
\begin{align}
\hat{\tau}&=  \hat{\overline{\alpha}} + \hat{\beta}= \frac{e^{\psi}}{\sqrt{2}}(\psi_{,3} + \epsilon)
= -\frac{1}{3\tau^3} \psi_1^{(0)} + O(\tau^{-4}), \notag\\
\end{align}
and that the prefactor of $1/3$ in front of the $\tau^{-3}$ term differs from the result obtained in  \cite{newman:891}.

\bibliography{../References/MyRefs}

\end{document}